\documentclass{article}

\usepackage{arxiv}

\usepackage[utf8]{inputenc} %
\usepackage[T1]{fontenc}    %
\usepackage{hyperref}       %
\usepackage{url}            %
\usepackage{booktabs}       %
\usepackage{amsfonts}       %
\usepackage{nicefrac}       %
\usepackage{microtype}      %
\usepackage{lipsum}
\usepackage{multirow}       %
\usepackage{biblatex}       %
\usepackage{rotating}       %
\usepackage{glossaries}     %
\usepackage{subfig}         %
\usepackage{titlesec}       %
\usepackage{adjustbox}      %
\usepackage{float}
\usepackage{makecell}       %
\usepackage{enumitem}       %
\usepackage{csquotes}       %
\usepackage{afterpage}      %
\usepackage{placeins}       %
\usepackage{tabularx}       %
\usepackage{array}          %
\usepackage{geometry}       %
\usepackage{graphicx}       %
\usepackage{lscape}         %
\usepackage[table]{xcolor}  %

\usepackage[format=plain,
            labelfont={bf,it},
            textfont=it]{caption}

\title{Synthetic Image Generation in Cyber Influence Operations: An Emergent Threat? \\ 
 }

\author{
  Melanie Mathys \\
  Institute for Data Science I4DS \\
  University of Applied Sciences FHNW\\
  Brugg-Windisch AG, Switzerland \\
  \texttt{melanie.mathys@fhnw.ch} \\
    \And
  Marco Willi \\
  Institute for Data Science I4DS \\
  University of Applied Sciences FHNW\\
  Brugg-Windisch AG, Switzerland \\
  \texttt{marco.willi@fhnw.ch} \\
   \And
  Michael Graber \\
  Institute for Data Science I4DS \\
  University of Applied Sciences FHNW\\
  Brugg-Windisch AG, Switzerland \\
  \texttt{michael.graber@fhnw.ch} \\
    \And
  Raphael Meier \\
  Cyber-Defence Campus \\
  armasuisse S+T \\
  Thun BE, Switzerland \\
  \texttt{raphael.meier@armasuisse.ch} \\
}

\begin{document}

\maketitle




\begin{abstract}

The evolution of artificial intelligence (AI) has catalyzed a transformation in digital content generation, with profound implications for cyber influence operations. This report delves into the potential and limitations of generative deep learning models, such as diffusion models, in fabricating convincing synthetic images. We critically assess the accessibility, practicality, and output quality of these tools and their implications in threat scenarios of deception, influence, and subversion. Notably, the report generates content for several hypothetical cyber influence operations to demonstrate the current capabilities and limitations of these AI-driven methods for threat actors. While generative models excel at producing illustrations and non-realistic imagery, creating convincing photo-realistic content remains a significant challenge, limited by computational resources and the necessity for human-guided refinement. Our exploration underscores the delicate balance between technological advancement and its potential for misuse, prompting recommendations for ongoing research, defense mechanisms, multi-disciplinary collaboration, and policy development. These recommendations aim to leverage AI's potential for positive impact while safeguarding against its risks to the integrity of information, especially in the context of cyber influence.

\end{abstract}


\section*{Disclaimer: Personality Rights}

To protect individual personality rights, we have deliberately blurred or cropped parts of some synthetic images depicting real persons in order to prevent any potential harm to their rights. Additionally, we selectively replaced textual references to specific individuals with generic descriptions. We believe these alterations do not significantly diminish the demonstration of the models’ capabilities nor their potential use in cyber influence operations.


\section{Introduction}

Artificial intelligence (AI) has brought about a significant shift in the way we can generate and modify digital content. This technology simplifies and enhances the production of diverse, high-quality digital materials. Specifically, deep learning advancements have enabled the generation of remarkable conversational, informational, or creative text, and the crafting or modification of images from textual instructions or prompts \cite{rombach_high-resolution_2022}. These strides in AI have not only fueled advancements but have also blurred the lines between real and synthetic content, thus presenting substantial challenges to societal norms, particularly in the realm of fact verification—a cornerstone of democratic integrity \cite{kucharavy_fundamentals_2023}.
Amidst these developments, there is a growing concern over the exploitation of these advanced models for cyber influence operations \cite{cordey_cyber_2019} by threat actors. With synthetic content becoming ever more convincing and the barrier to entry lowering due to accessible software, the potential for misuse is significant. This project seeks to thoroughly examine the landscape of generative deep learning for synthetic image generation, scrutinizing the sophistication and accessibility of cutting-edge models and the tools that facilitate their use. We aim to uncover the extent and nuances of these technologies and their implications for cyber influence.

Our investigation began with a review and categorization of methods and software capable of generating synthetic images, culminating in organizing and presenting our findings at an internal workshop. This collaborative forum united experts from machine learning to cybersecurity, fostering dialogue on the impact of deep learning on cyber influence and exploring hypothetical cyber influence threat scenarios involving such technologies.

This report consolidates our efforts, detailing the implementation of selected cyber influence threat scenarios from the workshop, highlighting both achievements and constraints encountered. By dissecting these findings, we aim to illuminate the practical realities of synthetic image generation within the context of cyber influence operations, offering a crucial assessment of its potential to be weaponized by threat actors.


\section{Background}

\subsection{Overview of Applied Methods} 

This section recapitulates the most relevant synthetic image generation methods explored and reviewed in the initial phase of our research. While we forgo an in-depth technical exposition, including architectural details, we provide a succinct summary of the methodologies applied. 

\begin{figure}[ht]
\centering
\includegraphics[width=1.0\linewidth]{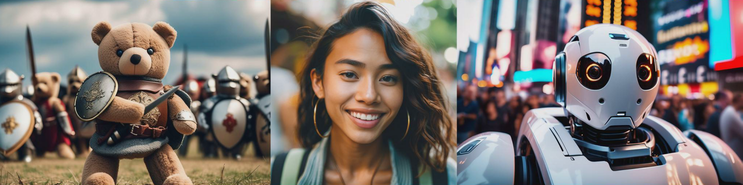}
\caption{Synthetic images generated with Stable Diffusion XL. Images adapted from \cite{podell_sdxl_2023} , licensed under \href{https://creativecommons.org/licenses/by/4.0/}{CC BY 4.0 Deed}.}
\label{fig:images:bg:sdxl}
\end{figure}

Our work relies on diffusion models, a class of probabilistic models that synthesize images by gradually denoising a random noise distribution. This technique, dating back to 2015 \cite{sohl-dickstein_deep_2015}, underpins the \textit{Stable Diffusion} family of models, which was introduced in 2022 \cite{rombach_high-resolution_2022}. These models can be used to generate images from textual descriptions as well as other inputs and are open-source. A notable addition to the Stable Diffusion model family, Stable Diffusion XL \cite{podell_sdxl_2023}, enhances image resolution and offers a supplementary model for post-processing refinement (see Figure \ref{fig:images:bg:sdxl} for example images).

\begin{figure}[ht]
\centering
\includegraphics[width=1.0\linewidth]{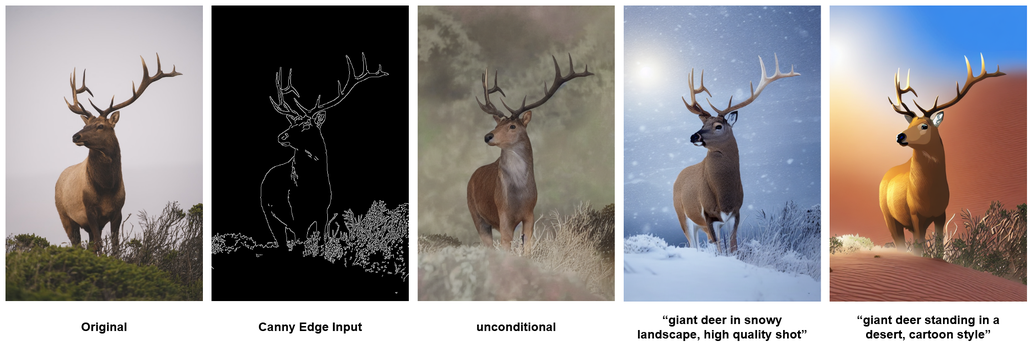}
\caption{Synthetic images generated with ControlNet conditioning. The first image on the left depicts a photograph \href{https://unsplash.com/de/fotos/braunhirsche-tagsuber-auf-grunem-gras-aJuv14zf-ZY}{(Source}, by Y S, @santonii, on Unsplash \cite{noauthor_unsplash_nodate}). The second image is the conditioning input (a canny edge map) derived from the photograph. The other images are synthetic and closely follow the canny edge map conditioning and (optionally) additional text input (the fourth and fifth images, see text below the images) which allows for more control over the generated images.}
\label{fig:images:bg:controlnet}
\end{figure}

Additionally, we utilized \textit{ControlNet} \cite{zhang_adding_2023}, a method which provides additional control over the generation process by guiding the model via visual inputs, such as sketches or edge maps (see Figure \ref{fig:images:bg:controlnet} for examples). This enables fine-grained control over the output and helps reduce unwanted artifacts.

\begin{figure}[ht]
\centering
\includegraphics[width=1.0\linewidth]{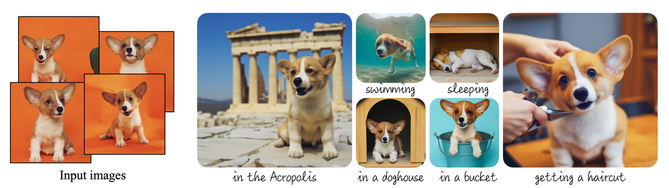}
\caption{Synthetic images generated with Dreambooth. The images on the left are the training images / photographs depicting the subject of interest. The other images are synthetic and depict the target subject in different situations. Figure from \cite{ruiz_dreambooth_2022}, licensed under \href{https://creativecommons.org/licenses/by/4.0/}{CC BY 4.0 Deed}.}
\label{fig:images:bg:dreambooth}
\end{figure}

Some use cases required the adaptation of pre-trained models to instill novel concepts into the models, for example, to generate images depicting specific persons or objects which open-source models do not know out-of-the-box. Adapting (\textit{fine-tuning}) such models requires the collection of example images (training data) depicting the concept of interest. Several innovations made it feasible to fine-tune large generative models with few computational resources (e.g. \cite{hu_lora_2021}) and small datasets. In particular, Dreambooth \cite{ruiz_dreambooth_2022} was developed to efficiently personalize diffusion models using a minimal amount of training images (often 3-5 suffice, according to \cite{ruiz_dreambooth_2022}). Dreambooth associates a subject (e.g. a person) with a unique (textual) identifier which can be used to prompt the fine-tuned model to generate images of the subject in novel contexts (see Figure \ref{fig:images:bg:dreambooth} for an example).

\FloatBarrier

\subsection{Tools and Resources}

One aspect of our work was to assess how accessible deep learning models for synthetic image generation are and what know-how is required to employ them effectively. We thus focused on models which are open-source and on tools which facilitate the interaction with such models in a user-friendly manner. We assume a threat actor, even with substantial technical skills and knowledge, will often depend on open-source models because model training requires extensive expertise and hardware which may not be feasible to attain by a threat actor.

\textit{Hugging Face} \cite{noauthor_hugging_nodate} is a platform which provides access to models, data and code to operate a larger variety of open-source models. The Hugging Face \textit{diffusers} \cite{von_platen_diffusers_2022} and \textit{transformers} \cite{wolf_transformers_2020} libraries, supplemented by comprehensive documentation and tutorials, enable quick and easy assembly of deep learning pipelines. The platform also offers scripts for model training and fine-tuning, as well as pre-trained models for download. We trained all our models using Hugging Face libraries.

\textit{Automatic1111's Stable Diffusion WebUI} \cite{automatic1111_stable-diffusion-webui_2023} offers a browser-based, graphical user interface for a large number of existing models and methods belonging to or complementing the Stable Diffusion family. It allows for easy installation and usage of such models and offers rich documentation and a large community to help get started. While deep learning models are effective, refinement is often required to address imperfections and ensure images align with user intent. Automatic1111's Stable Diffusion WebUI offers various tools, such as inpainting with mouse movements, to simplify and streamline such refinements. We used them extensively for certain generative tasks. It is built on top of Gradio \cite{abid_gradio_2019} which enables the creation of web-based user interfaces for models. The platform civitai.com\footnote{\url{https://civitai.com/}} is primarily aimed at users of Automatic1111's Stable Diffusion WebUI, providing them with a hub for sharing models and model artifacts, as well as synthetic images and artwork. It, rather surprisingly, also included a model pertaining to one of our hypothetical threat scenarios (see Section \ref{sec:scenario1_synthetic_force_multiplier}).

\FloatBarrier

\subsection{Cyber Influence Threat Scenarios}

One of the primary goals of our internal workshop on synthetic media generation and cyber influence was to develop hypothetical cyber influence operation threat scenarios that rely on content generated by deep learning models. To provide the workshop participants with structure and guidance regarding how to develop the threat scenarios, we assessed various frameworks that enable the characterization of cyber influence operations. In particular, we assessed five frameworks: \textit{4Ds} \cite{nimmo_anatomy_2015}, \textit{BEND} \cite{carley_social_2020}, \textit{ABCDE} \cite{alaphilippe_adding_2020}, \textit{DISARM} \cite{terp_disarm_2022}, and \textit{SCOTCH} \cite{blazek_scotch_2021}. To balance complexity and expressiveness, we chose the most common elements from the different frameworks and framed the characteristics in simple terms. The framework closest to what characteristics we chose is \textit{SCOTCH}. We asked the participants to define each influence operation along the following items:

\begin{itemize}%
\item Actor: Who operates the influence operation?
\item Objective(s): What are the desired effects of the operation?
\item Target(s): Who is the target audience to be influenced?
\item Content \& Methods: What content is used in the operation? (and which type of deep content generation method may be used?)
\item Channel(s): Where is the content distributed to the target audience?
\item Technique(s): Which technical mechanism of exploitation and / or psychological phenomena are being exploited?
\end{itemize}

Table \ref{tab:scenarios_large} shows an overview of the threat scenarios developed by the workshop participants.

\begin{landscape}
\begin{table}[ht]
\centering
\caption{Overview of the scenarios developed by the workshop participants.}
\label{tab:scenarios_large}
\resizebox{\columnwidth}{!}{%
\begin{tabular}{|c|l|cc|l|l|cccc|ccc|cccccccc|}
\hline
\rowcolor[HTML]{EFEFEF} 
\textbf{\#} & \multicolumn{1}{c|}{\cellcolor[HTML]{EFEFEF}\textbf{Threat Scenario}} & \multicolumn{2}{c|}{\cellcolor[HTML]{EFEFEF}\textbf{Actor}} & \multicolumn{1}{c|}{\cellcolor[HTML]{EFEFEF}\textbf{Objective}} & \multicolumn{1}{c|}{\cellcolor[HTML]{EFEFEF}\textbf{Target}} & \multicolumn{4}{c|}{\cellcolor[HTML]{EFEFEF}\textbf{Content}} & \multicolumn{3}{c|}{\cellcolor[HTML]{EFEFEF}\textbf{Channel}} & \multicolumn{8}{c|}{\cellcolor[HTML]{EFEFEF}\textbf{Technique}} \\ \hline
 &  & \rotatebox{90}{State actor} & \rotatebox{90}{Non-State Actor} &  &  & \rotatebox{90}{Meme} & \rotatebox{90}{Synthetic Image} & \rotatebox{90}{Synthetic Documents} & \rotatebox{90}{Other} & \rotatebox{90}{Social Media} & \rotatebox{90}{Legacy Media} & \rotatebox{90}{Other} & \rotatebox{90}{Inauthentic Documents} & \rotatebox{90}{Develop Memes} & \rotatebox{90}{Propagate half-truth} & \rotatebox{90}{Pollute Information Space} & \rotatebox{90}{Impersonate} & \rotatebox{90}{Reframe Context} & \rotatebox{90}{Use Hashtags} & \rotatebox{90}{Other} \\ \hline
1 & \textbf{Synthetic Force Multiplier} & x &  & \makecell[l]{Cause deterrence and\\uncertainty of the\\adversary, bind their\\resources.} & \makecell[l]{Adversary's armed\\forces and their\\allies.} &  & x &  &  & x &  &  & x &  & x &  &  &  &  &  \\ \hline
2 & \textbf{Intelligence DDoS} & x &  & \makecell[l]{Overwhelm capacity of\\adversary's intelligence\\services.} & \makecell[l]{Intelligence\\service of\\adversary.} &  & x & x &  & x & x &  & x &  &  & x &  &  &  &  \\ \hline
3 & \textbf{Tailored Military Disinformation} & x &  & \makecell[l]{Mislead the adversary's\\OSINT branch and\\kinetic forces.} & \makecell[l]{Specific vulnerable\\unit of adversary's\\armed forces\\and intelligence\\service.} &  & x & x &  & x &  & x & x &  &  &  &  &  &  &  \\ \hline
4 & \textbf{Environmentalists} &  & x & \makecell[l]{Reputational damage\\of companies with high\\CO2 emission.} & \makecell[l]{General public,\\journalists.} &  & x & x & x & x & x & x & x &  &  &  & x & x & x & x \\ \hline
5 & \textbf{Sinaloa Cartel} &  & x & \makecell[l]{Improve reputation and\\recruit new members.} & \makecell[l]{Potential and\\current members,\\local communities,\\general public.} & x & x &  & x & x & x & x &  &  &  &  & x & x & x & x \\ \hline
6 & \textbf{Prevent CO2 Regulation} & x & x & \makecell[l]{Prevent further CO2\\regulation at an\\upcoming international\\climate conference.} & \makecell[l]{Representing\\ministers and\\delegations of\\developing\\countries.} &  & x &  & x & x &  &  & x &  &  &  & x &  &  &  \\ \hline
7 & \textbf{Undermine Democracy} &  & x & \makecell[l]{Gain more political\\influence, weaken\\democratically legitimate\\state structures.} & \makecell[l]{Critical mass\\of citizens of a\\western democracy,\\own members.} &  & x &  &  & x &  &  & x &  &  &  &  &  &  &  \\ \hline
\end{tabular}%
}
\end{table}
\end{landscape}

\clearpage


\section{Threat Scenarios}

The following sections describe all threat scenarios that we fleshed out using deep generative models, which is a subset of the threat scenarios developed in the internal workshop on synthetic media generation and cyber influence operations. For each threat scenario we present background information and concepts, followed by the most convincing synthetic images that we were able to generate. Technical details about the models, as well as discussions about their potential \& limitations, as well as learnings, follow in subsequent sections.

\subsection{\textit{Synthetic Force Multiplier}}
\label{sec:scenario1_synthetic_force_multiplier}

This threat scenario involves a nation-state's armed forces aiming to enhance their kinetic weapons' perceived potency, deter and intimidate adversaries, and engage the intelligence resources of their opponents.

\begin{figure}[ht]
\centering
\includegraphics[width=0.6\linewidth]{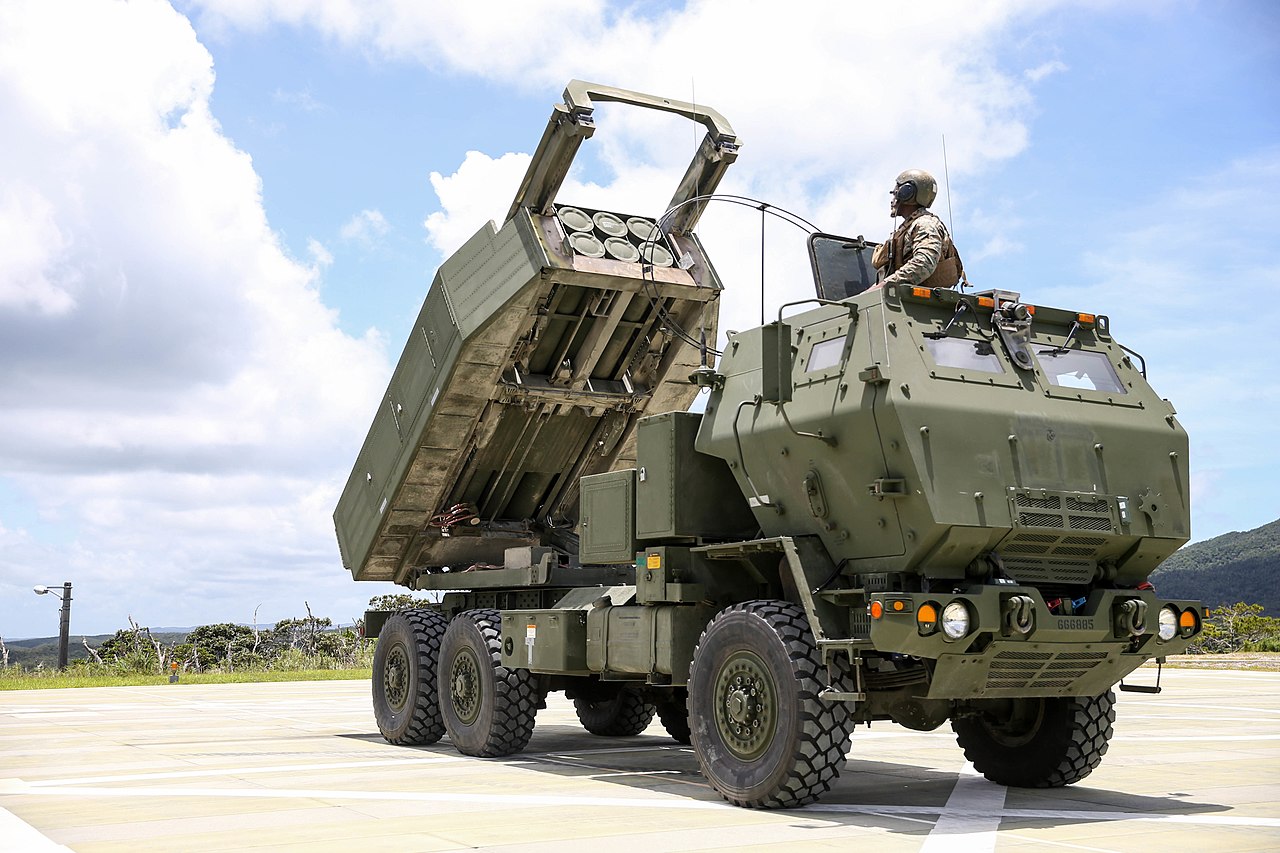}
\caption{Real image of a HIMARS vehicle. Photo credit: Cpl. Donovan Massieperez \href{https://en.m.wikipedia.org/wiki/File:3dMarineDivisionHIMARS.jpg}{(Source)}, licensed under \href{https://creativecommons.org/licenses/by-sa/4.0/deed.en}{CC BY-SA 4.0 Deed}.}
\label{fig:images:scenario_1:himars_example}
\end{figure}

To illustrate this threat scenario, we attempted to create synthetic images depicting HIMARS (M142 High Mobility Artillery Rocket System) vehicles (see Figure \ref{fig:images:scenario_1:himars_example} for an example). We chose the HIMARS system because it is a high-value asset with variable striking range capable of shaping the battlefield. The presence of a HIMARS system poses a threat to enemy force concentrations and supply routes within its range. A threat actor in possession of HIMARS systems may attempt to increase an adversary's uncertainty about the precise number and geolocations of its fielded HIMARS systems by the use of synthetic images. This can be regarded as an instance of ambiguity-increasing military deception aimed at suboptimizing the enemy's force disposition \cite{smith_military_1993}. Unlike the hypothetical actor of this scenario we do not have access to such vehicles and thus photographs of them. Therefore, we resorted to images found on the internet to train deep learning models. In order to fool a resourceful adversary, synthetic images need to appear authentic and photo-realistic.

\begin{figure}[ht]
  \centering
  \subfloat[]{\includegraphics[height=6cm]{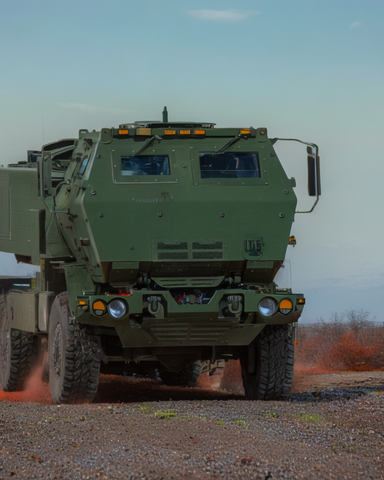}\label{capt:sc1_best_gen_example1}}
  \quad
  \subfloat[]{\includegraphics[height=6cm]{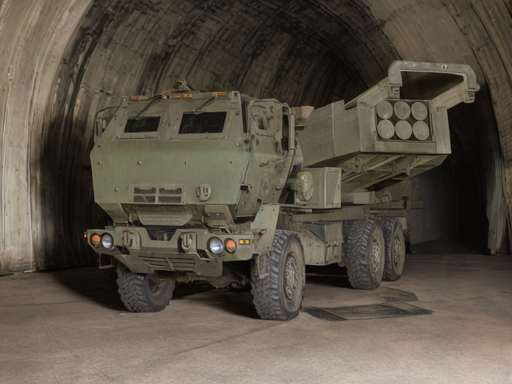}\label{capt:sc1_best_gen_example2}}
\caption{Synthetic images of HIMARS vehicles. Both images were created with ControlNet conditioning on canny edge maps from real photographs. Image  \protect\subref{capt:sc1_best_gen_example1} was created using the prompt: \enquote{photography of a HIMARS vehicle driving on a gravel road, nice weather, blue sky}.  Image \protect\subref{capt:sc1_best_gen_example2} was synthesized with \enquote{a photography of a HIMARS in a military bunker, large space, clean structures, low light}.}
\label{fig:images:scenario1:best_generations}
\end{figure}

Figure \ref{fig:images:scenario1:best_generations} shows the most convincing examples that we managed to synthesize using fine-tuned Stable Diffusion models with ControlNet conditioning. The examples were manually selected each from a sample of 16 synthetic images (using the same prompt and the same conditioning input for all of the 16 samples). The results appear realistic at first glance; however, upon close inspection, they contain clear artifacts, such as irregular tire profiles.

\FloatBarrier

\subsection{\textit{Tailored Military Disinformation}}
\label{sec:scenario2_tailored_military_disinfo}

This threat scenario envisions a foreign intelligence service seeking to mislead its adversary's Open Source Intelligence (OSINT) unit and its kinetic forces. Their goal is to induce the adversary to waste ammunition and inadvertently cause collateral damage to civilian infrastructure and population.

One possible strategy to reach this goal involves generating synthetic images of military resources as potential targets, strategically placed in specific locations. The distribution of these crafted images could take place through social media platforms or orchestrated "information leaks" such as seemingly misplaced USB sticks in an abandoned command vehicle.

The specific locations in the images must be recognizable, either by place name signs, familiar landmarks, or similar identifiable features. They must be part of an active war zone. Also, the images must appear authentic to successfully mislead the adversary.

\begin{figure}[H]
  \centering
  \subfloat[Background image by Anna Hunko on Unsplash \cite{noauthor_unsplash_nodate}.]{\includegraphics[width = 0.3\textwidth]{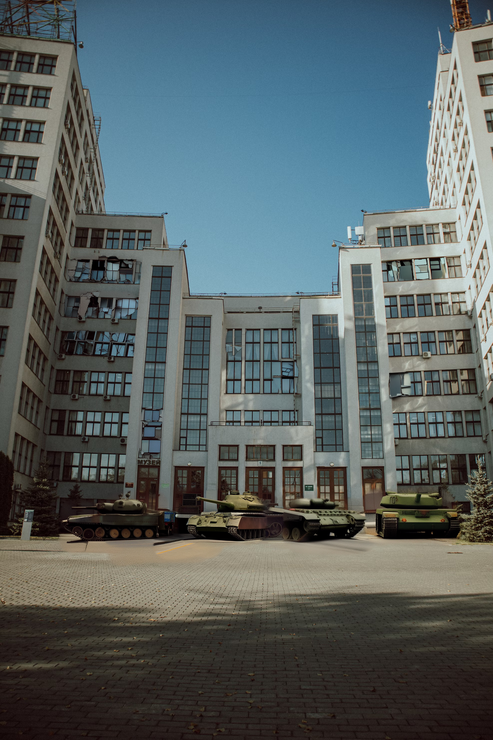}\label{fig:tailored_disinf_a}}
  \quad
  \subfloat[Background image by Kate Tepl on Unsplash \cite{noauthor_unsplash_nodate}.]{\includegraphics[width = 0.67\textwidth]{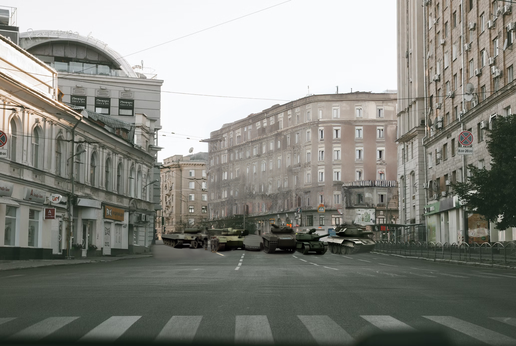}\label{fig:tailored_disinf_b}}
\quad
  \subfloat[Background image by Anna Hunko on Unsplash \cite{noauthor_unsplash_nodate}.]{\includegraphics[width = 0.29\textwidth]{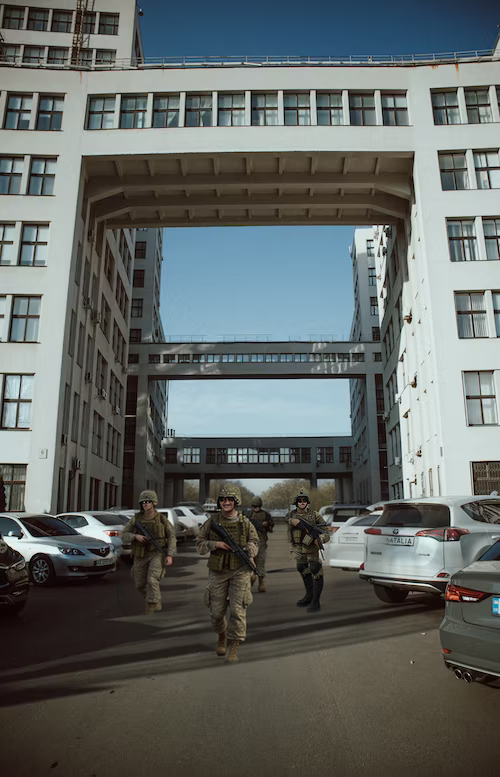}\label{fig:tailored_disinf_c}}
  \quad
  \subfloat[Background image by Julia Rekamie on Unsplash \cite{noauthor_unsplash_nodate}.]{\includegraphics[width = 0.67\textwidth]{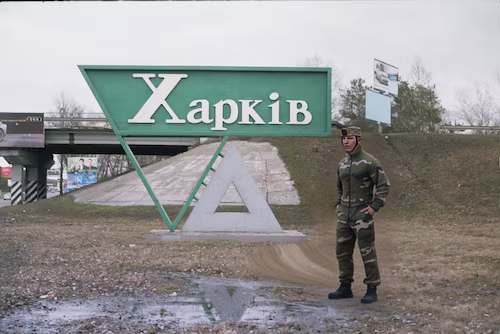}\label{fig:tailored_disinf_d}}
\caption{Tanks and soldiers inpainted into various locations in the city of Kharkiv, Ukraine. All backgrounds come from authentic photographs sourced from Unsplash \cite{noauthor_unsplash_nodate}. Building facades in images \protect\subref{fig:tailored_disinf_a} and \protect\subref{fig:tailored_disinf_b} have additionally been altered to make them appear damaged.}
\label{fig:images:scenario2:best_generations}
\end{figure}

For creating images that could meet these requirements, we decided to select real photographs of different places in the Ukrainian city of Kharkiv, and to use inpainting to add tanks and soldiers to these scenes. Figure \ref{fig:images:scenario2:best_generations} shows the most convincing synthetic examples that we managed to create.
\clearpage

\FloatBarrier
\clearpage

\subsection{\textit{Sinaloa Cartel}}\label{sec:scenario5_sinaloa_cartel}

The Sinaloa Cartel, a large, international organized crime syndicate, is known for its illegal drug trafficking and money laundering activities \footnote{\url{https://en.wikipedia.org/wiki/Sinaloa_Cartel}}. Just like many legal businesses, it endeavors to maintain a specific public image and exerts significant influence on its media representation.

\begin{figure}[ht]
  \centering
  \subfloat[Image by Reuters/Fernando Carranza, sourced from \cite{the_mob_museum_-_national_museum_of_organized_crime__law_enforcement_mexican_2020}.]{\includegraphics[height=5cm]{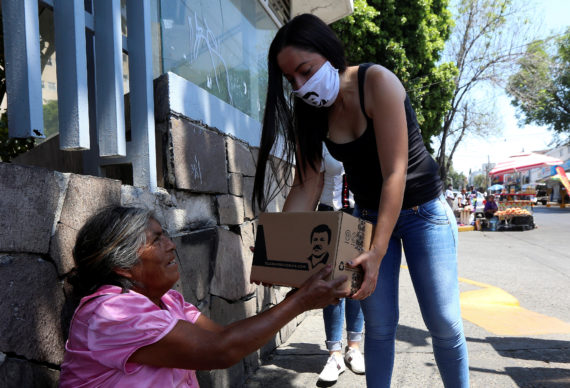}\label{fig:sinaloa_real_a}}
  \quad
  \subfloat[Image by Ulises Ruiz/AFP, sourced from \cite{the_mob_museum_-_national_museum_of_organized_crime__law_enforcement_mexican_2020}.]{\includegraphics[height=5cm]{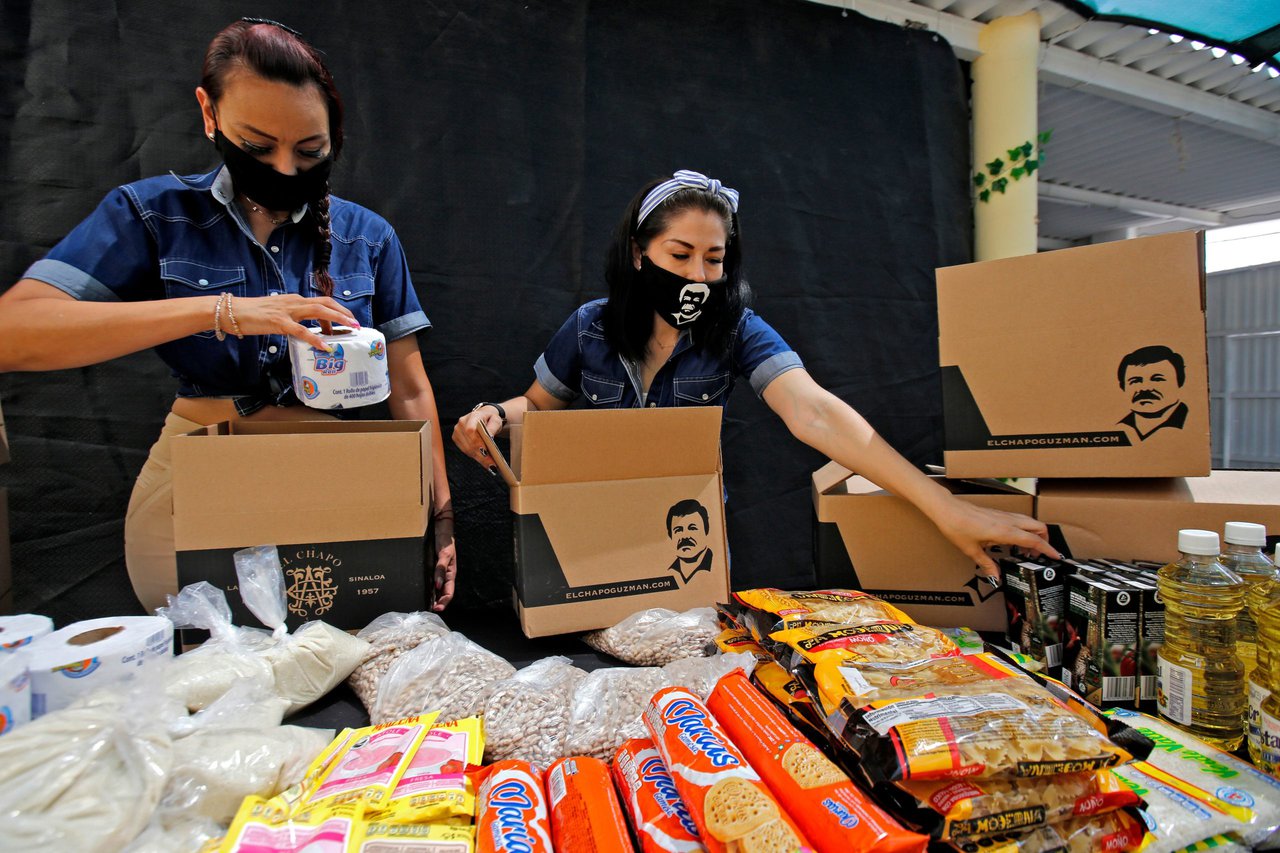}\label{fig:sinaloa_real_b}}
  \quad
  \subfloat[Image by  Francisco Guasco/EPA, sourced from \cite{felbab-brown_covid-19_2020}.]{\includegraphics[height=5cm]{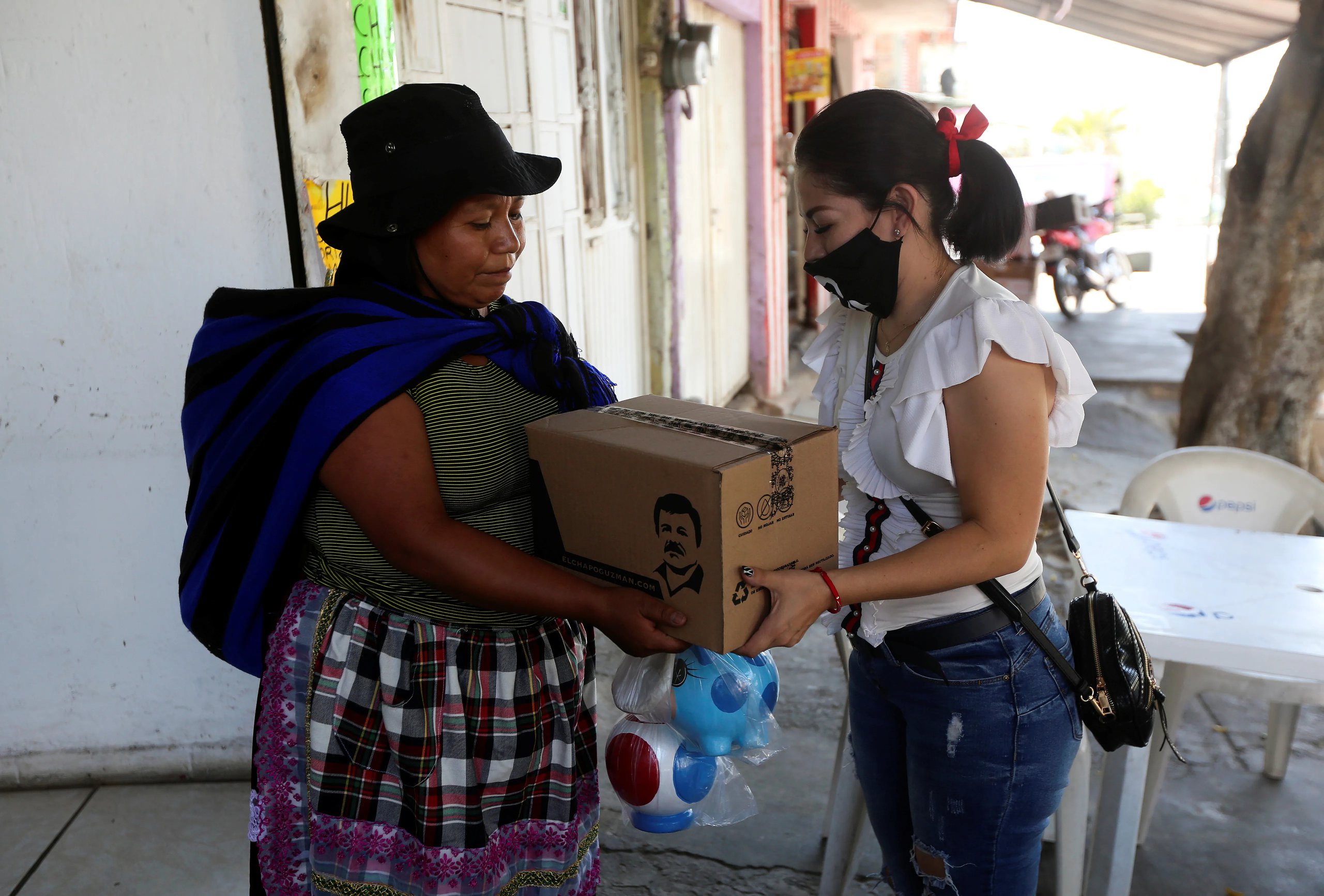}\label{fig:sinaloa_real_c}}
\caption{Employees of a charity foundation, run by the Sinaloa cartel head's daughter, distribute aid packages during the COVID-19 pandemic.
}
\label{fig:images:scenario_5:food_donations:real_examples}
\end{figure}

For this project, we focused on generating two themes of media content associated with the Cartel. Firstly, we simulated scenes similar to the Cartel's humanitarian efforts during the COVID-19 pandemic in 2020 \cite{the_mob_museum_-_national_museum_of_organized_crime__law_enforcement_mexican_2020, felbab-brown_covid-19_2020}. These efforts included the distribution of supplies and food to those in need, as depicted in Figure \ref{fig:images:scenario_5:food_donations:real_examples}. Secondly, we attempted to reproduce images posted on social media highlighting the Cartel members' affluence. In general, these images circulate under hashtags such as \#narcos, \#narcostyle, \#NarcoStyle, and \#NarcoOficia \cite{know_high_2018, baverstock_narcos_2015}.

\begin{figure}[ht]
  \centering
  \subfloat[Image by Instagram/Sinaloa.Guasave, sourced from \cite{know_high_2018}.]{\includegraphics[width=0.23\linewidth]{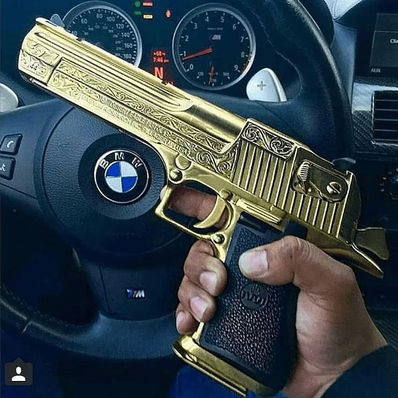}}
  \quad
  \subfloat[Image by Instagram/Narcode Sinola, sourced from \cite{know_high_2018}.]{\includegraphics[width=0.23\linewidth]{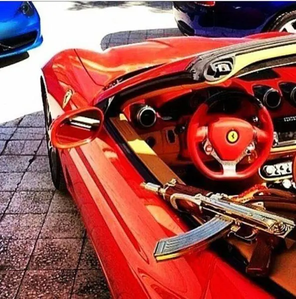}}
  \quad
  \subfloat[Image by Instagram/Sinaloa.Guasave, sourced from \cite{know_high_2018}.]{\includegraphics[width=0.23\linewidth]{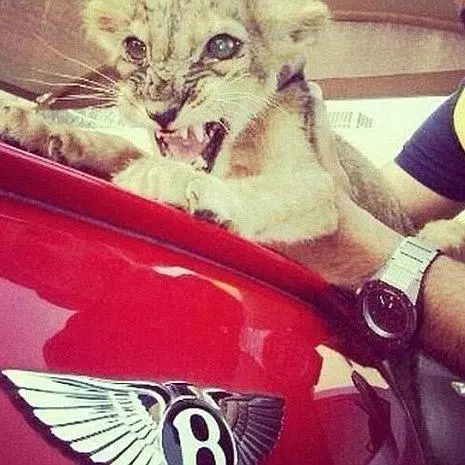}}
  \quad
  \subfloat[Image by Instagram/unidentified, sourced from \cite{baverstock_narcos_2015}.]{\includegraphics[width=0.23\linewidth]{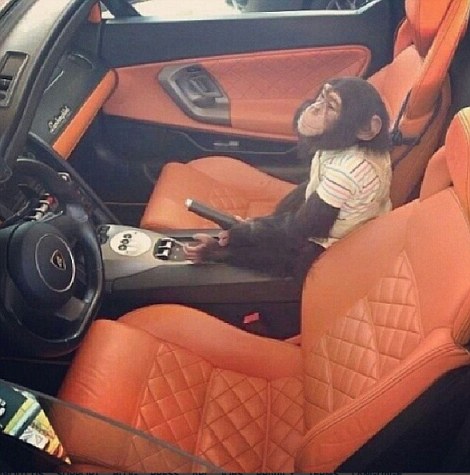}}
\caption{Examples of "Narco Wealth" displayed on social media.}
\label{fig:images:scenario_5:flaunting_wealth:real_examples}
\end{figure}

\begin{figure}[ht!]
  \centering
  \subfloat[]{\includegraphics[width = 0.31\linewidth]{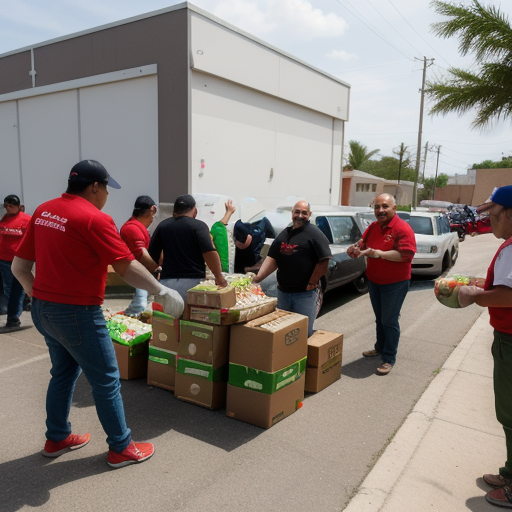}}
    \quad
  \subfloat[]{\includegraphics[width = 0.31\linewidth]{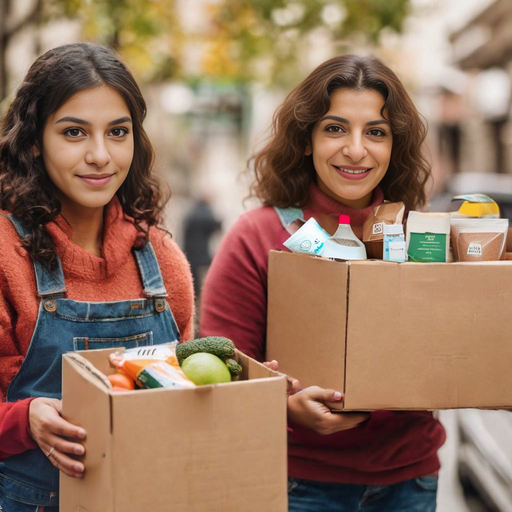}}
      \quad
    \subfloat[]{\includegraphics[width = 0.31\linewidth]{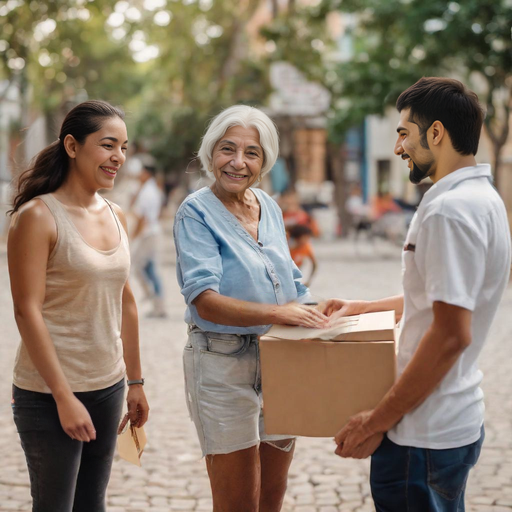}}
\caption{Synthetic images of people distributing food donations.}
\label{fig:images:scenario_5:donations}
\end{figure}

\begin{figure}[ht]
\centering
\subfloat[]{\includegraphics[height = 5.2cm]{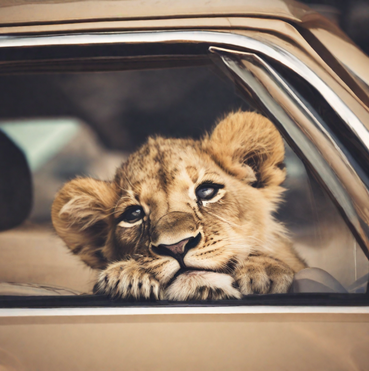}}
\quad
\subfloat[]{\includegraphics[height = 5.2cm]{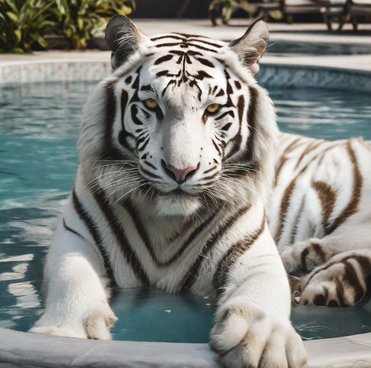}}
\quad
\subfloat[]{\includegraphics[height = 5.2cm]{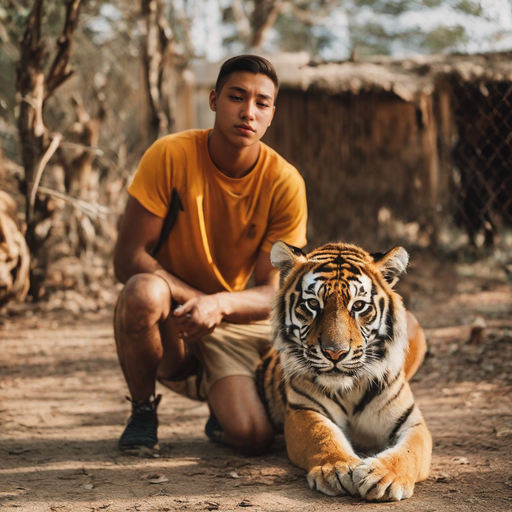}}
\quad
\subfloat[]{\includegraphics[height = 4.2cm]{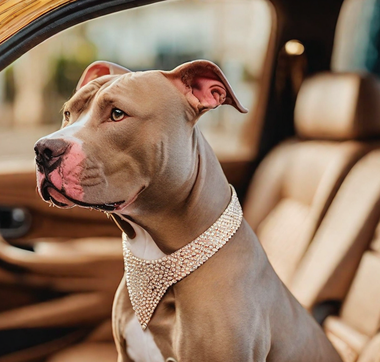}}
\quad
\subfloat[]{\includegraphics[height = 4.2cm]{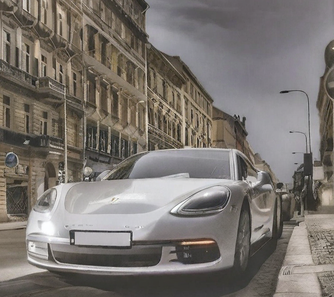}\label{fig:sinaloa_wealth_tropes_f}}
\quad
\subfloat[]{\includegraphics[height = 4.2cm]{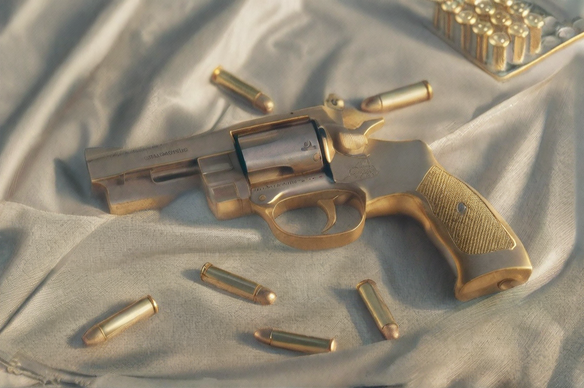}}
\caption{Generated images of \textit{Narco Wealth} tropes: Exotic animals, a luxury car, and a gold-plated gun. Image \protect\subref{fig:sinaloa_wealth_tropes_f} has been generated with a ControlNet based on an existing image (by Tom Def on Unsplash \cite{noauthor_unsplash_nodate}).}
\label{fig:images:scenario_5:flaunting_wealth:animals_guns}
\end{figure}

Taking inspiration from content as shown in Figures \ref{fig:images:scenario_5:flaunting_wealth:real_examples} and \ref{fig:images:scenario_5:flaunting_wealth:animals_guns}, we generated synthetic images of people distributing charity goods in cardboard boxes (Figure \ref{fig:images:scenario_5:donations}), as well as images depicting various luxury items such as exclusive cars, gold-plated guns or exotic animals (Figure \ref{fig:images:scenario_5:flaunting_wealth:animals_guns}).

\FloatBarrier
\clearpage

\subsection{\textit{Environmentalists}}\label{sec:scenario4_environmentalists}

The initial concept of this threat scenario included extreme environmentalists trying to discredit major carbon emitters, possibly inciting consumer boycotts through the distribution of fabricated research or slanderous synthetic imagery of company executives. However, generating such false reports proved to be overly specialized and complex, exceeding our project's scope. Moreover, we believe few companies have executives so well-known and tied to the image of their company that synthetic images would affect the company's reputation significantly.

\begin{figure}[ht!]
  \centering
  \subfloat[
  ]{\includegraphics[width = 0.48 \linewidth]{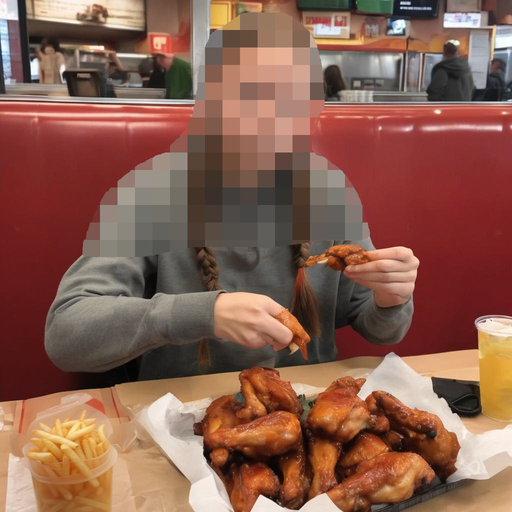}}
  \quad
  \subfloat[
  ]{\includegraphics[width = 0.48 \linewidth]{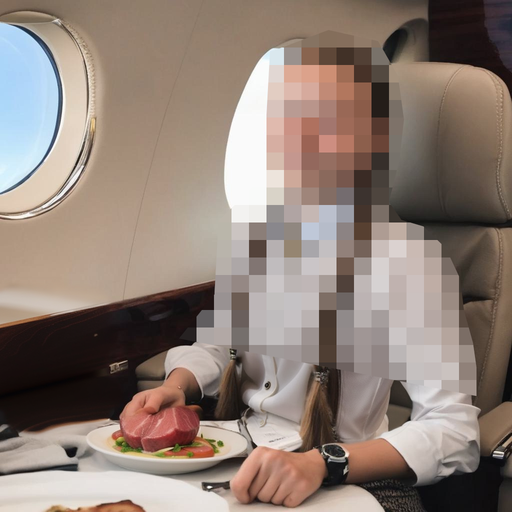}}
\caption{Synthetic images depicting an environmental activist indulging in meat consumption and traveling by private jet, contrary to their public image.
}
\label{fig:images:environmentalists:privatejet}
\end{figure}

Therefore, we shifted our focus to environmental activists becoming victims of disinformation, rather than being perpetrators. The threat actor in this scenario could be a politically or economically motivated individual or group who is/are intent on undermining trust in leading environmental activists. We created synthetic images of environmental activists participating in environmentally harmful behaviors, such as dining on meat or using private jets, to depict them in a hypocritical light (see Figure \ref{fig:images:environmentalists:privatejet}).

\FloatBarrier
\clearpage

\subsection{\textit{Undermining Democracy}}\label{sec:scenario7_undermine_democracy}
In this threat scenario, anti-democratic groups spread conspiracy theories and biased views to influence perceptions of public figures and subvert public discourse. These actors often use humor, satire, and exaggeration, not necessarily aiming for factual accuracy or photorealism in their image content. Instead, their goal is to leave a strong, lasting impression, making memes and cartoons their preferred tools. Moreover, content is frequently tailored towards evoking a strong, typically negative emotional response (e.g. of fear or anger) rendering individuals more susceptible to believe in false information \cite{martel_reliance_2020} and conspiracy theories \cite{molenda_emotion_2023}.

\begin{figure}[H]
  \centering
  \subfloat[]{\includegraphics[width=0.24\linewidth]{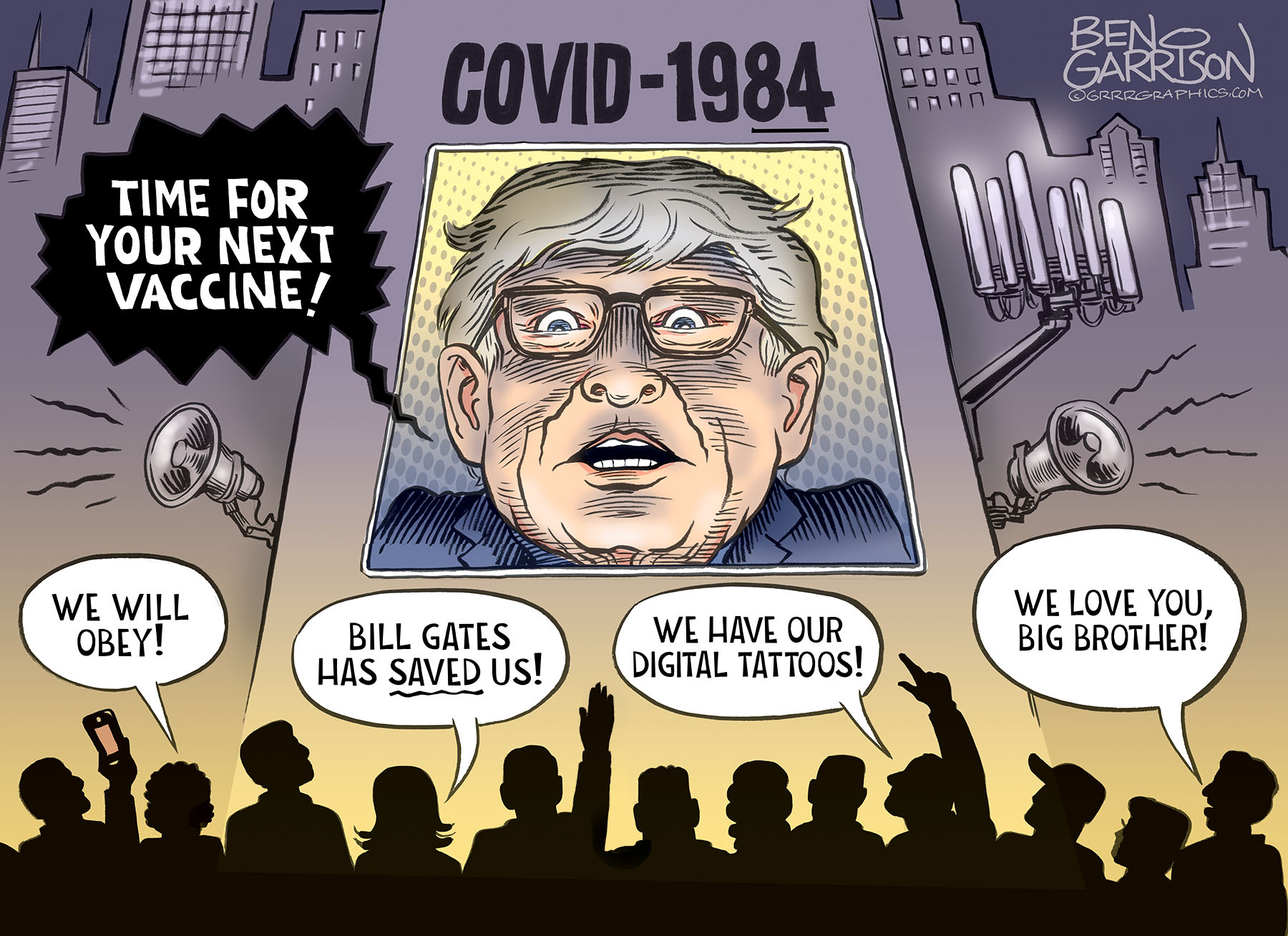}\label{fig:undermine_a}}
  \quad
  \subfloat[]{\includegraphics[width=0.24\linewidth]{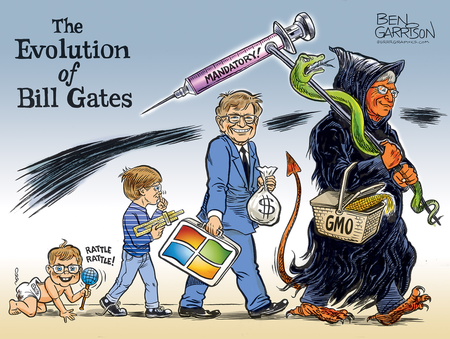}\label{fig:undermine_b}}
  \quad
  \subfloat[]{\includegraphics[width=0.24\linewidth]{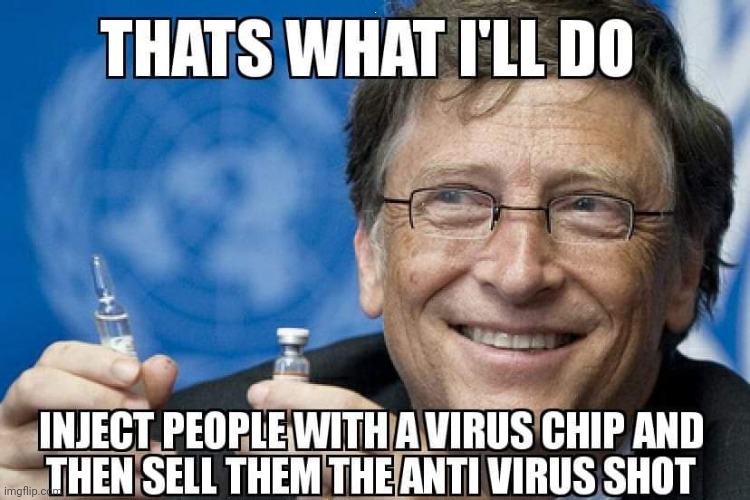}\label{fig:undermine_c}}
  \quad
  \subfloat[]{\includegraphics[width=0.19\linewidth]{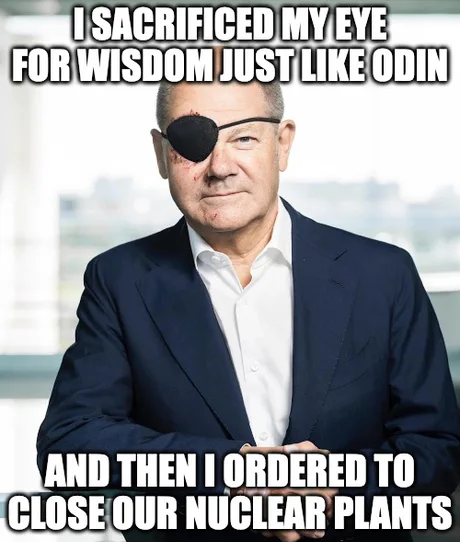}\label{fig:undermine_d}}

\caption{Examples of published content that fits the threat scenario \textit{Undermining Democracy}. Images \protect\subref{fig:undermine_a} (\href{https://grrrgraphics.com/the-evolution-of-bill-gates/}{Source}) and \protect\subref{fig:undermine_b} (\href{https://grrrgraphics.com/big-brother-bill/}{Source}) are cartoons by Ben Garrison \href{https://en.wikipedia.org/wiki/Ben_Garrison}{(Link)} criticizing and devaluing Bill Gates. Images \protect\subref{fig:undermine_c} (one of several memes made with this \href{https://imgflip.com/meme/220417578/Bill-Gates-loves-Vaccines}{template}) and \protect\subref{fig:undermine_d} \href{https://img-9gag-fun.9cache.com/photo/aVb4YXv_460swp.webp}{(Source)} are memes about Bill Gates  and Olaf Scholz, respectively.}
\label{images:undermine:results:non-dl}
\end{figure}

We focus on public figures such as tech mogul Bill Gates, who has been a frequent target of conspiracies\footnote{See for example \href{https://www.xn--verschwrungstheorien-99b.info/enzyklopaedie/bill-gates-verschwoerungstheorien/}{here}.}, and Germany's Chancellor, Olaf Scholz. The cartoons and memes shown in Figure \ref{images:undermine:results:non-dl} are examples of the types of content these groups disseminate and serve as a model for synthetic images we created.
\FloatBarrier

\begin{figure}[ht!]
  \centering
  \subfloat[
  ]{\includegraphics[height=5cm]{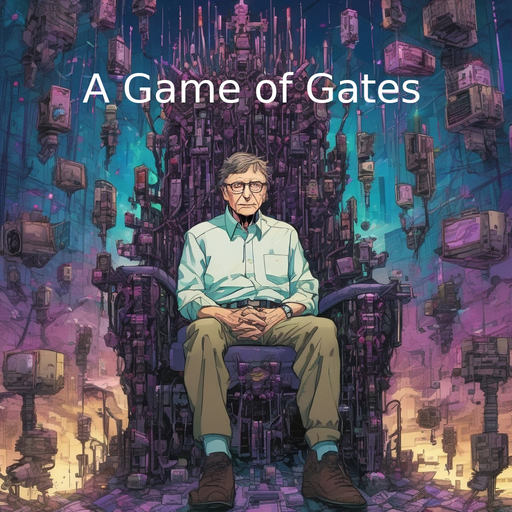}\label{fig:undermine_game_of_gates}}
  \quad
  \subfloat[
  ]{\includegraphics[height=5cm]{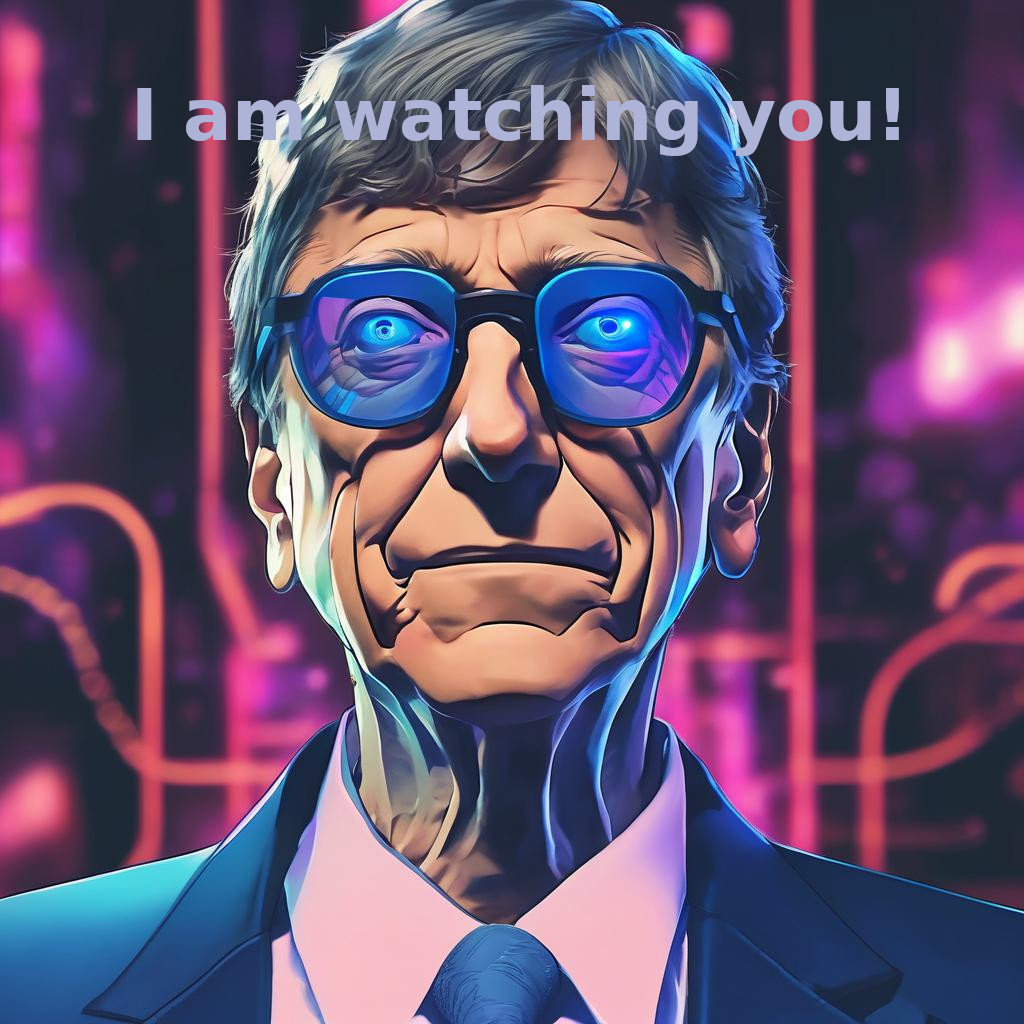}\label{fig:undermine_bill_gates_assimilates}}
\caption{Generated cartoons with manually added text, humorously portraying Bill Gates.}
\label{images:undermine:results:gates_cartoons}
\end{figure}

Figure \ref{images:undermine:results:gates_cartoons} displays cartoon memes of Bill Gates which we generated, combining textual references to pop culture with gloomy, threatening visuals. Internet memes often draw on pop culture to boost their stickiness.

\begin{figure}[th!]
  \centering
  \includegraphics[height=5.5cm]{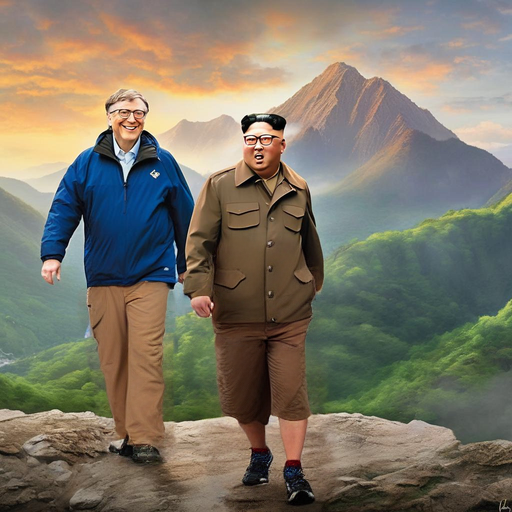}
\caption{A synthetic image representing Bill Gates and North Korean leader Kim Jong Un as close friends, embarking on a hiking trip together.}
\label{images:undermine:results:gates_and_kim}
\end{figure}

\begin{figure}[H]
  \centering
  \subfloat[
  ]{\includegraphics[height=5cm]{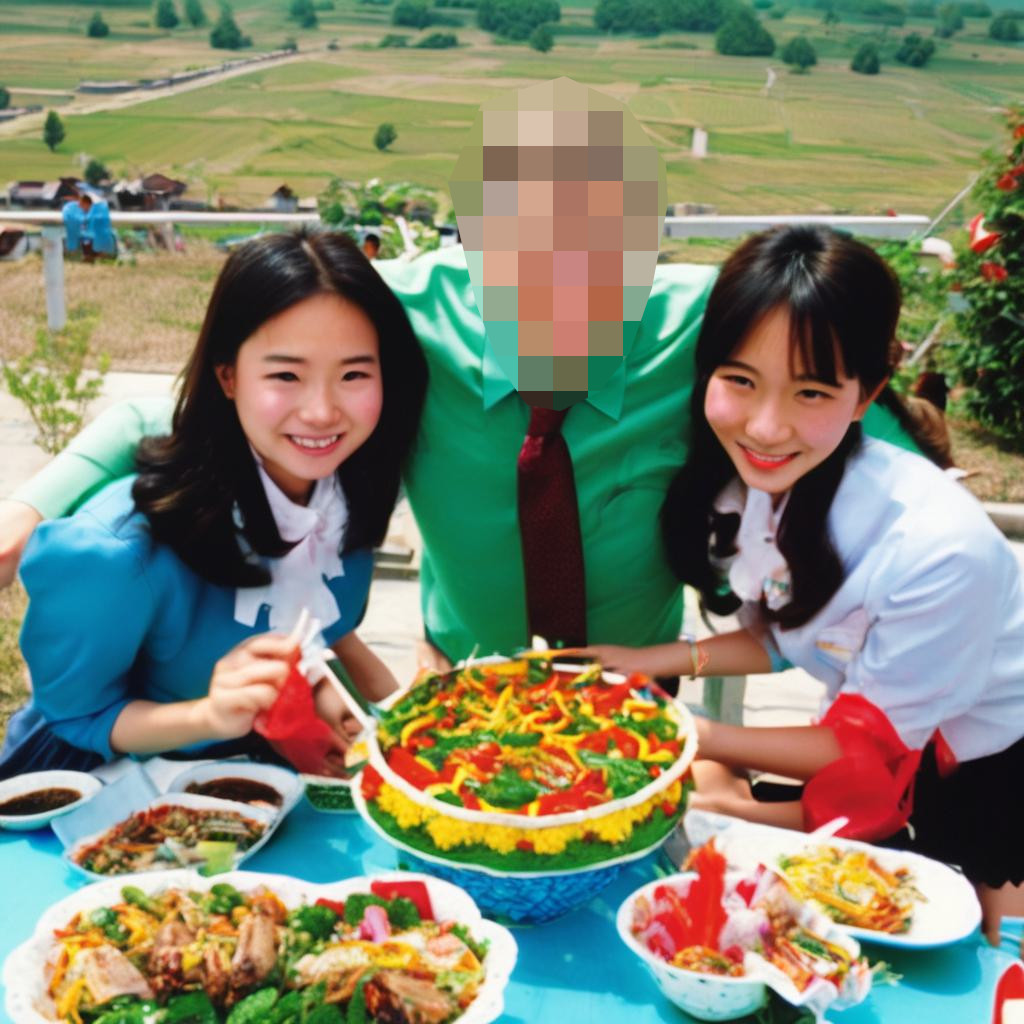}}
  \quad
  \subfloat[
  ]{\includegraphics[height=5cm]{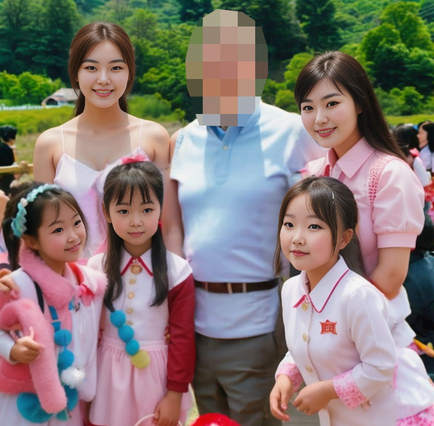}}
    \quad
  \subfloat[
  ]{\includegraphics[height=5cm]{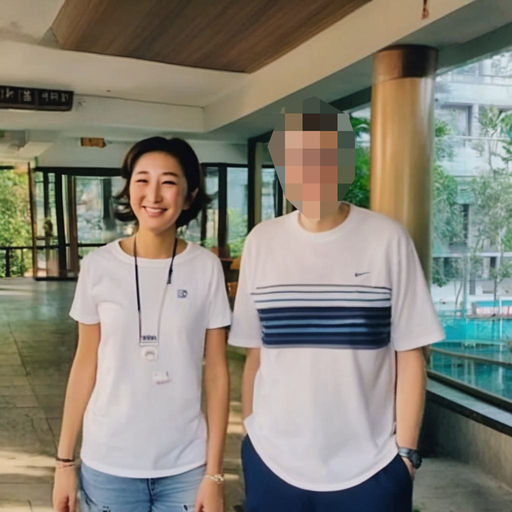}}
\caption{Synthetic images of a specific person surrounded by young North Korean women. A potential background narrative could be that a specific individual maintains amicable relations with the North Korean government and frequently vacations in North Korea, where they particularly enjoy the companionship of young and attractive women. 
}
\label{images:undermine:results:vip_and_girls}
\end{figure}

Additionally, we explore a hypothetical conspiracy theory about Bill Gates' supposed ties with North Korea and Kim Jong Un. Figure \ref{images:undermine:results:gates_and_kim} depicts an image we created with Stable Diffusion XL showing both on a hiking tour in North Korea. We also employed inpainting to smooth out artifacts. Figure \ref{images:undermine:results:vip_and_girls} depicts hypothetical \enquote{leaked} photos of a prominent person meeting with young North Korean women.

\begin{figure}[H]
\centering
\subfloat[]{\includegraphics[height=5cm]{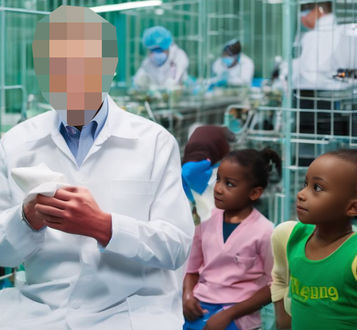}\label{fig:guinea_pig_1}}
\quad
\subfloat[]{\includegraphics[height=5cm]{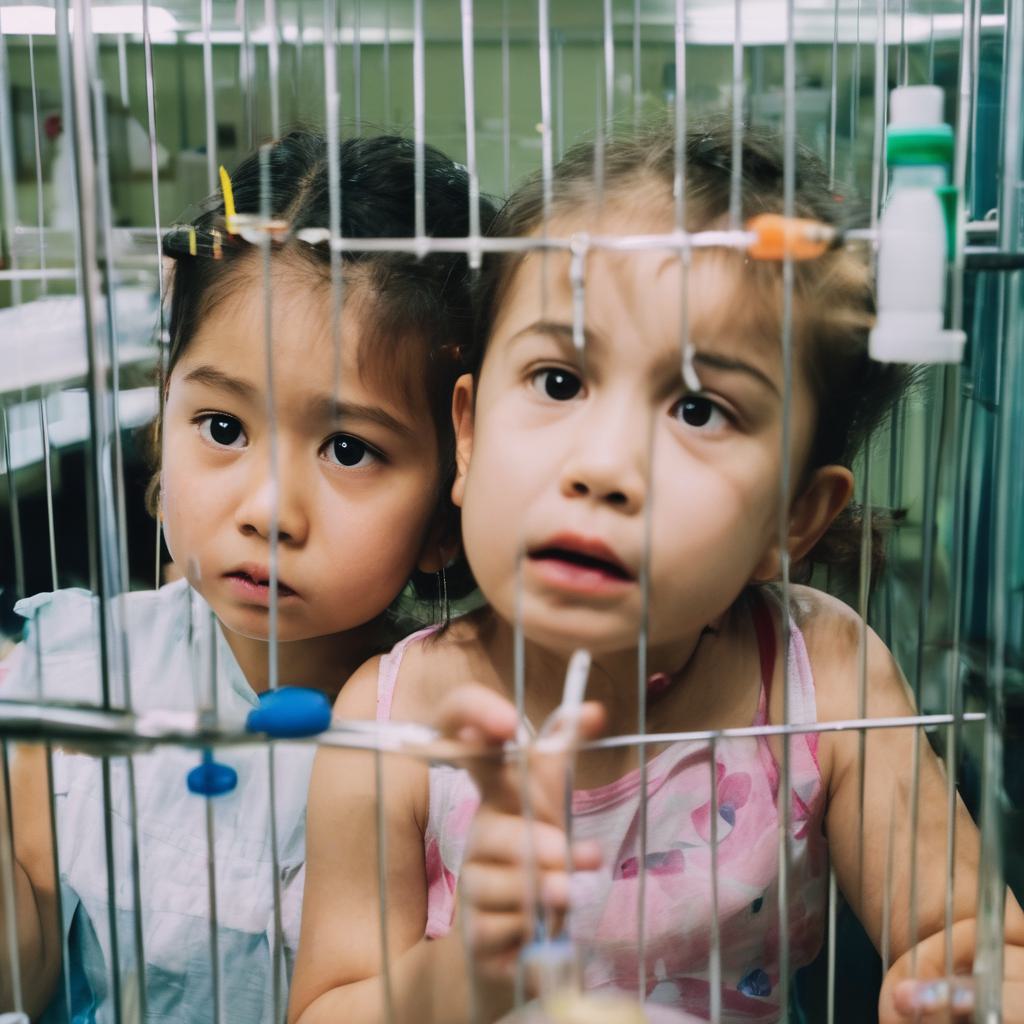}\label{fig:guinea_pig_2}}
\caption{Generated images portraying a specific individual in a laboratory with young children \protect\subref{fig:guinea_pig_1} and two very young, fearful children in a laboratory cage \protect\subref{fig:guinea_pig_2}. These could suggest a false narrative that this individual subjects children to vaccine trials.}
\label{fig:images:undermine:guinea_pigs}
\end{figure}

Lastly, Figure \ref{fig:images:undermine:guinea_pigs} visualizes claims that a specific person\footnote{This could be any person a threat actor may consider useful in a specific cyber influence operation.} was deeply involved in COVID-19 vaccine development, allegedly using children as test subjects in laboratories.

\clearpage


\clearpage

\begin{figure}[H]
  \centering
  \subfloat[]{\includegraphics[width=0.31\linewidth]{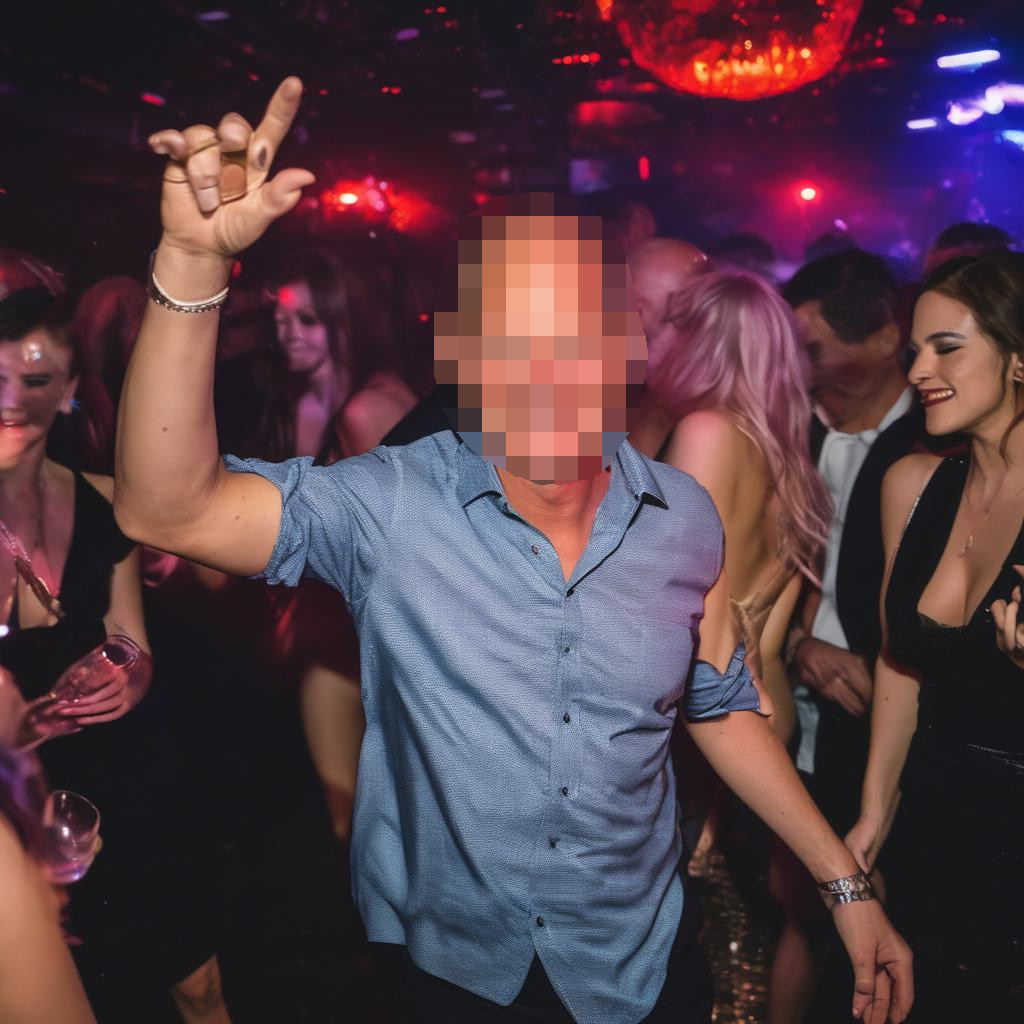}\label{fig:vip2_club}}
  \quad
  \subfloat[]{\includegraphics[width=0.31\linewidth]{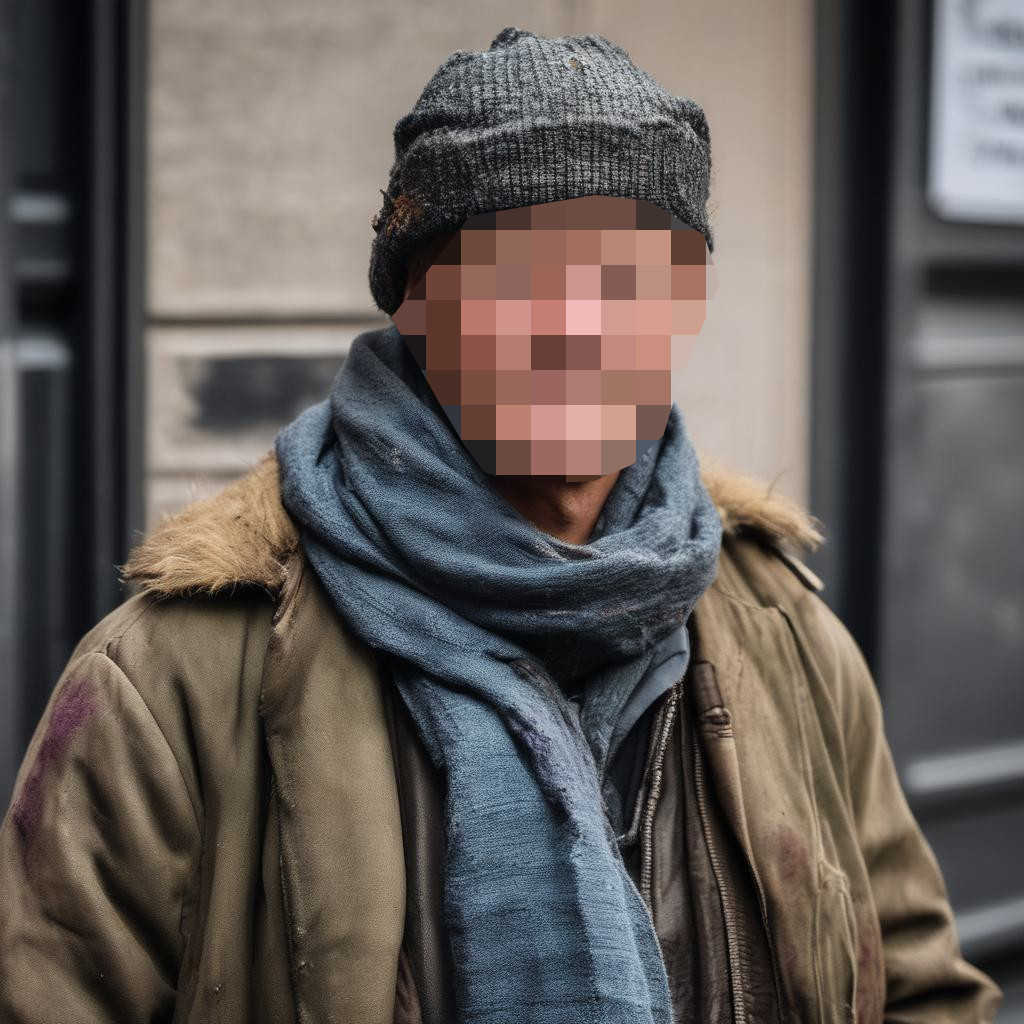}\label{fig:vip2_homeless}}
  \quad
  \subfloat[]{\includegraphics[width=0.31\linewidth]{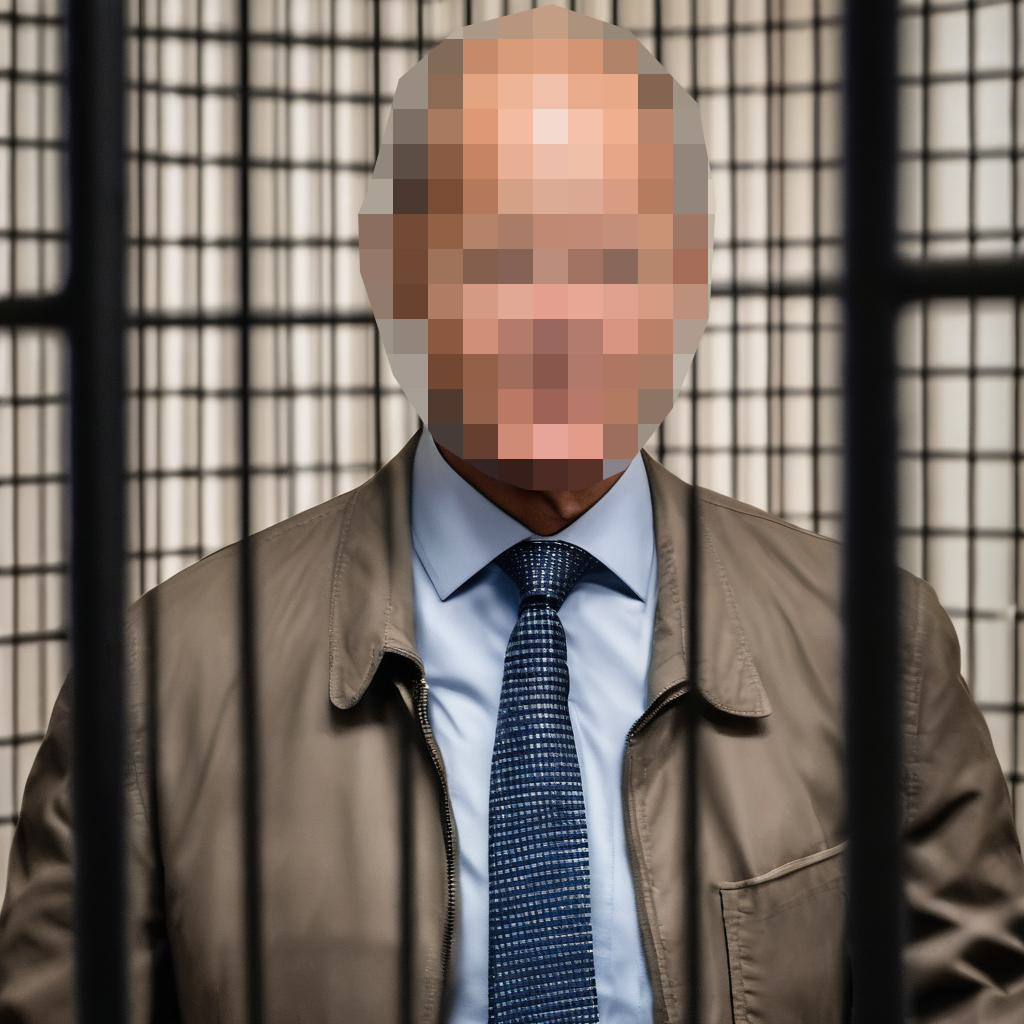}\label{fig:vip2_prison}}
\caption{Synthetic images of a specific individual in various scenarios: \protect\subref{fig:vip2_club} shows them partying in a nightclub, \protect\subref{fig:vip2_homeless} depicts them as homeless, and \protect\subref{fig:vip2_prison} portrays them in prison.
}
\label{fig:images:undermine:olaf}
\end{figure}

Finally, synthetic images in Figure \ref{fig:images:undermine:olaf} show a specific person in situations that seem at odds with their public image.

\FloatBarrier
\clearpage

\section{Potential of Generative Models}

\subsection{Out-of-the-box Capabilities}

Pre-trained generative deep learning models, such as the Stable Diffusion family of models, offer great versatility. These advanced models are ready to produce a diverse array of visual content out-of-the-box, effectively circumventing the need for fine-tuning which requires specialized know-how and resources. This flexibility not only accelerates the content creation process but also opens up a wealth of possibilities for creators to experiment with different styles and themes. By leveraging such models, one can effortlessly generate high-quality images, from realistic photographs to various artistic renditions, all while significantly reducing the time and resources traditionally required for such tasks. This accessibility to a broad spectrum of visual outputs makes such models a potentially useful tool for cyber influence operations.

\subsection{Illustrations and Cartoons}\label{illustrations_and_cartoons}

\begin{figure}[ht!]
\centering
\subfloat[]
{\includegraphics[width=0.31\linewidth]{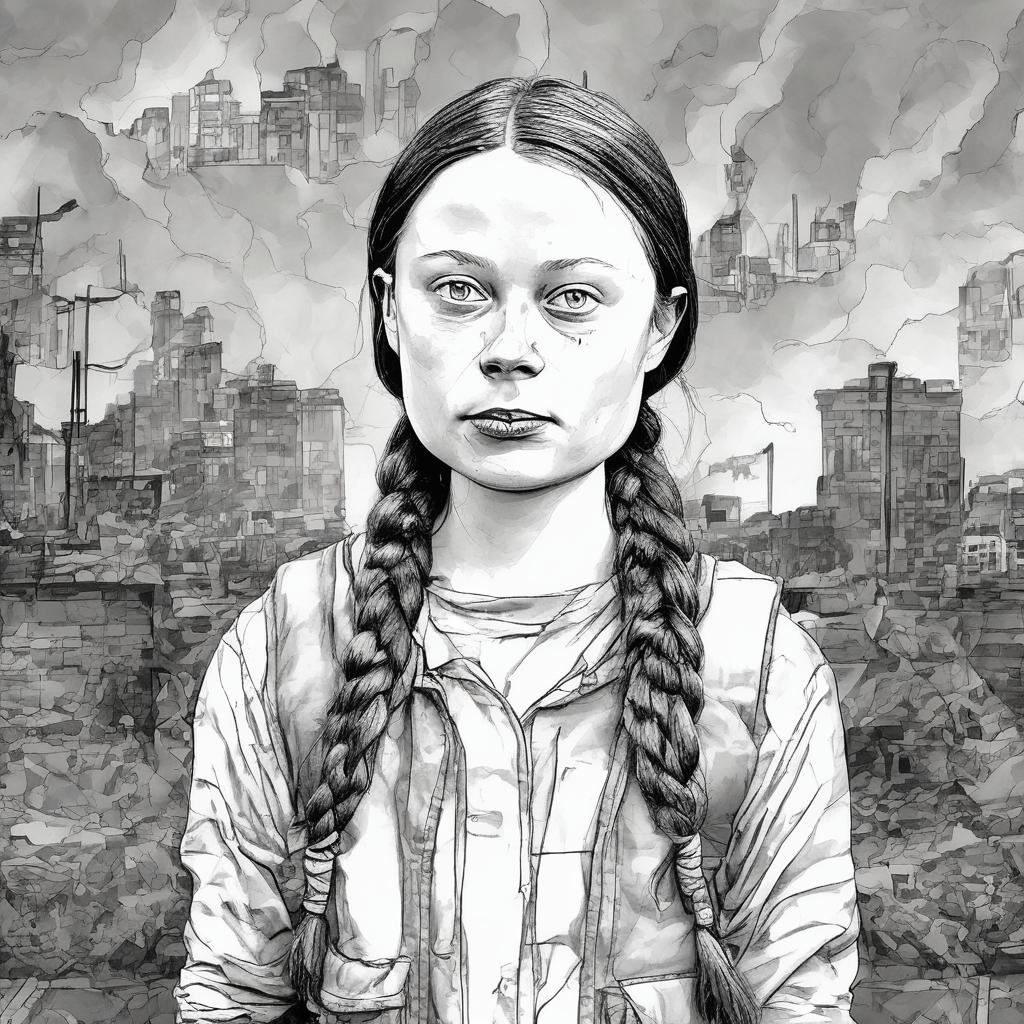}}
\quad
\subfloat[]
{\includegraphics[width=0.31\linewidth]{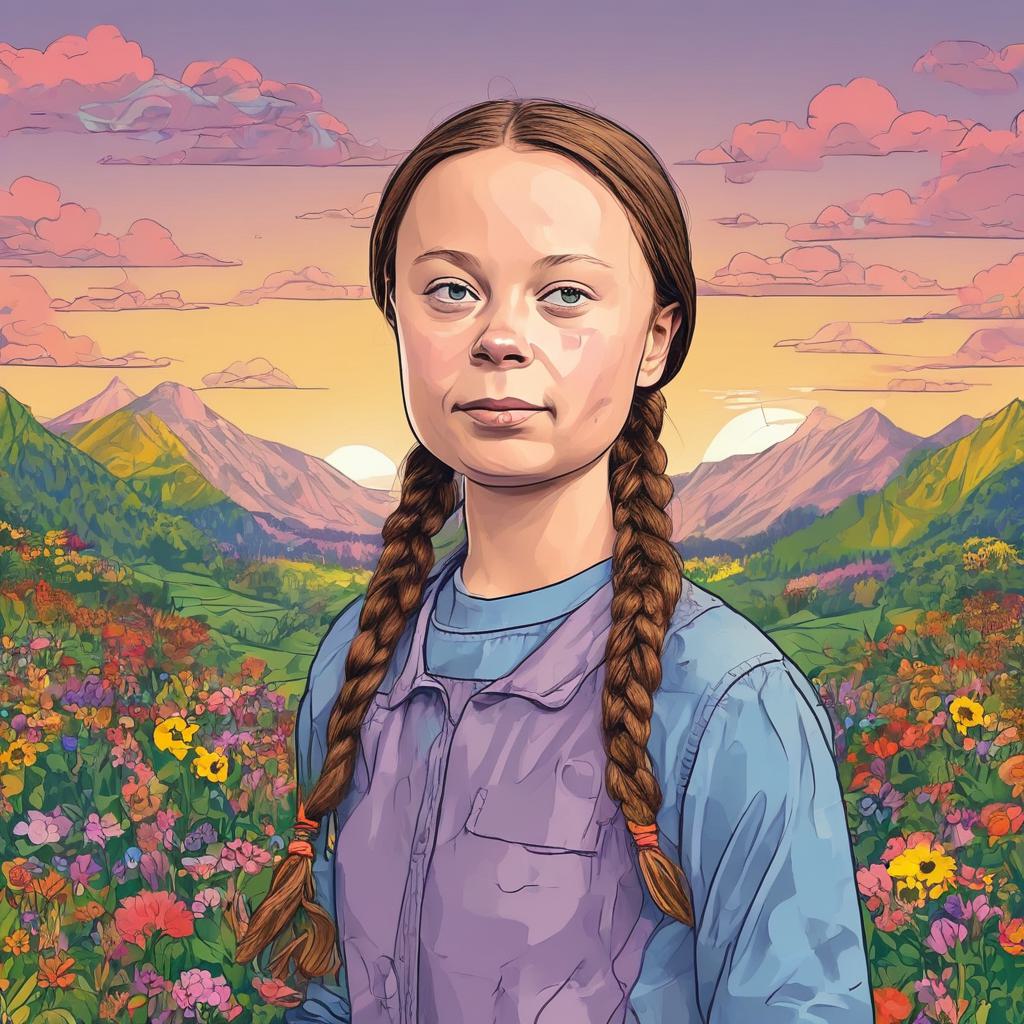}}
\quad
\subfloat[]
{\includegraphics[width=0.31\linewidth]{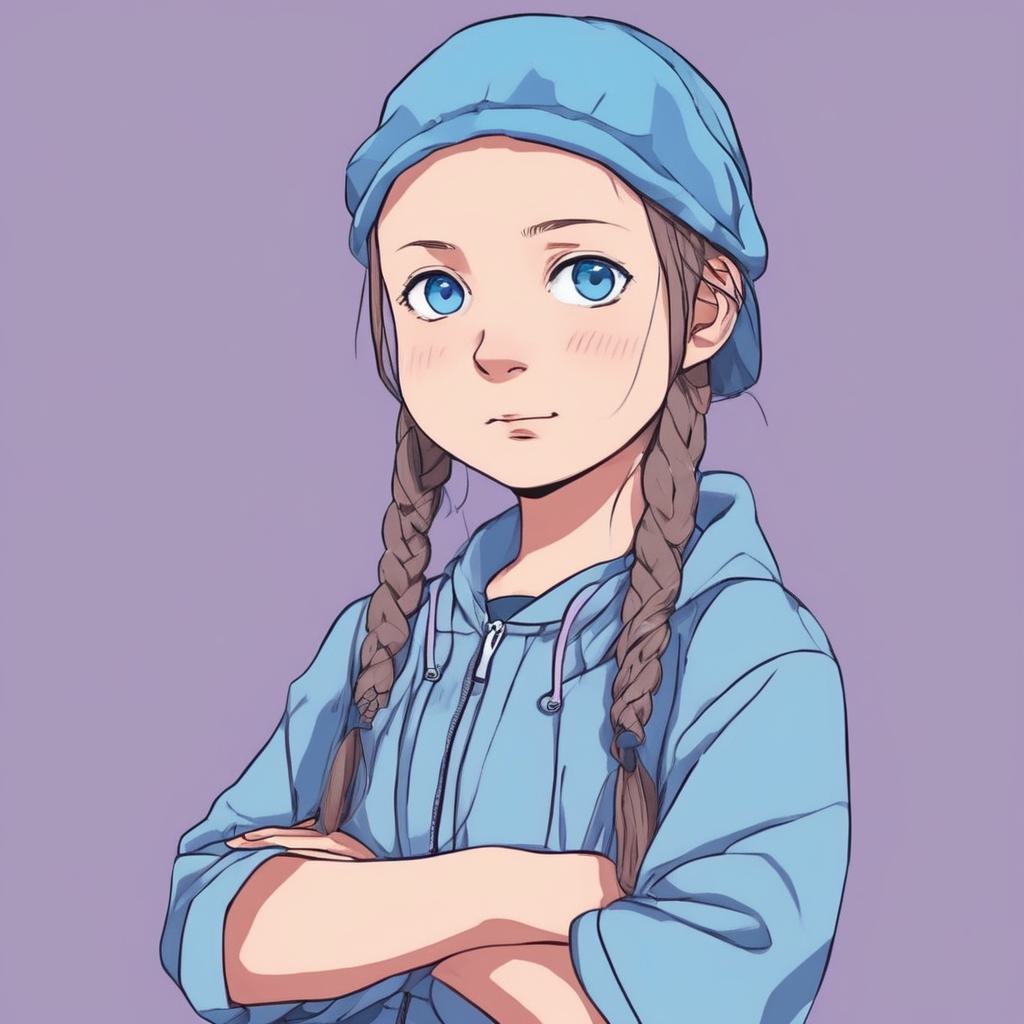}}
\quad
\subfloat[]
{\includegraphics[width=0.31\linewidth]{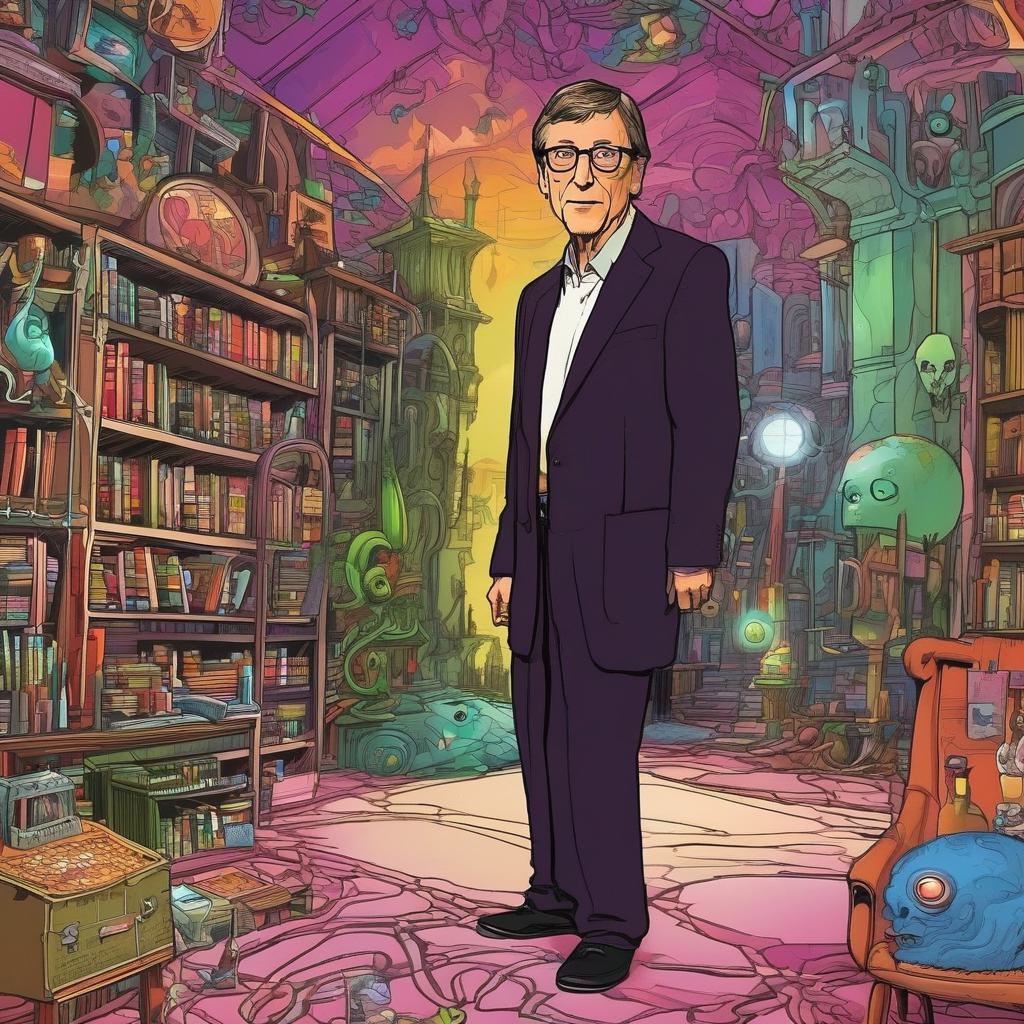}}
\quad
\subfloat[]
{\includegraphics[width=0.31\linewidth]{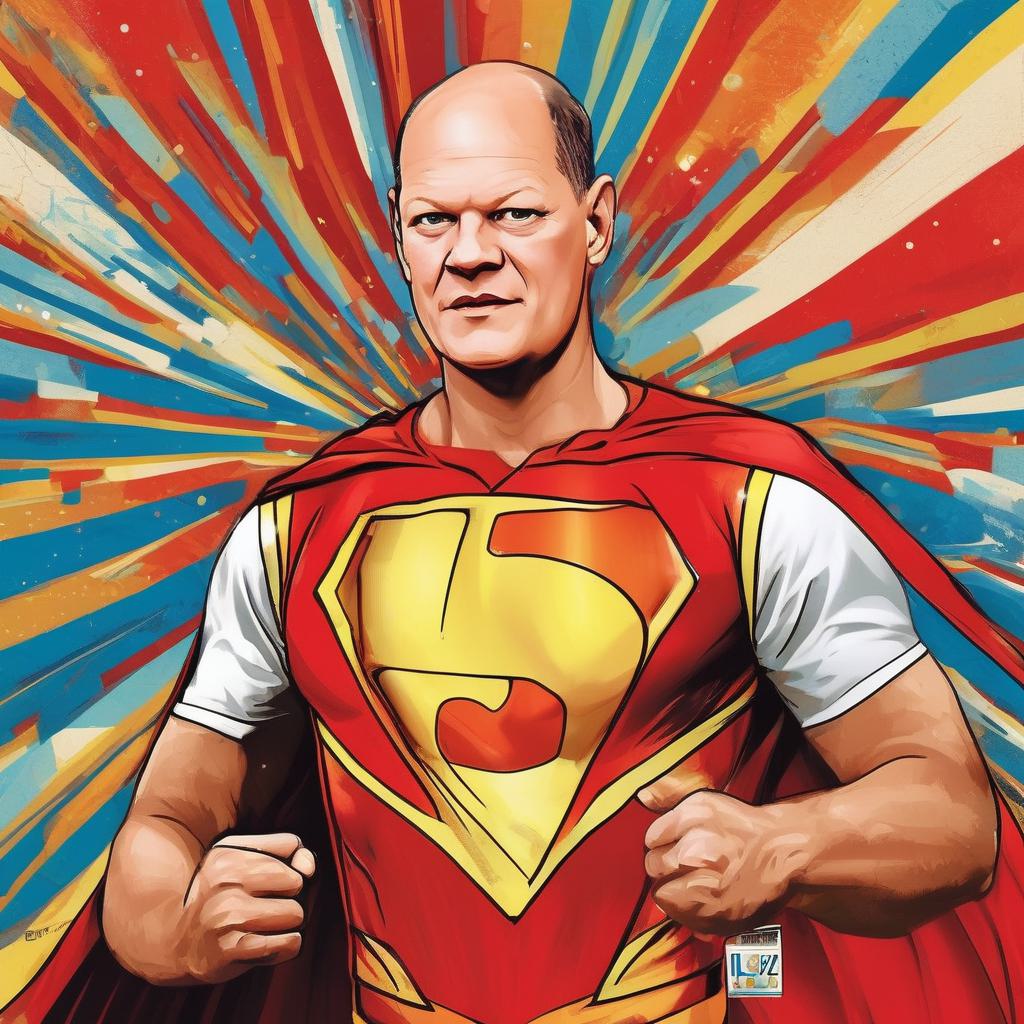}}
\caption{Illustrations of Greta Thunberg, Bill Gates, and Olaf Scholz rendered in various artistic styles. Generated with Stable Diffusion XL.}
\label{image:potential:artistic}
\end{figure}

Deep generative models, such as the Stable Diffusion models, offer vast opportunities for creative and stylistic exploration, as exemplified in Figure \ref{image:potential:artistic}. These models excel in generating humorous and novel content, such as cartoons and memes, which can significantly boost cyber influence operations where authenticity is less critical than originality. They enable the creation of content with enhanced symbolism, exaggeration, and stereotypes beyond what is possible with photographs.

\FloatBarrier

\subsection{Emotional expressiveness in human faces}

\begin{figure}[ht!]
  \centering
  \subfloat[
]{\includegraphics[height=3.2cm]{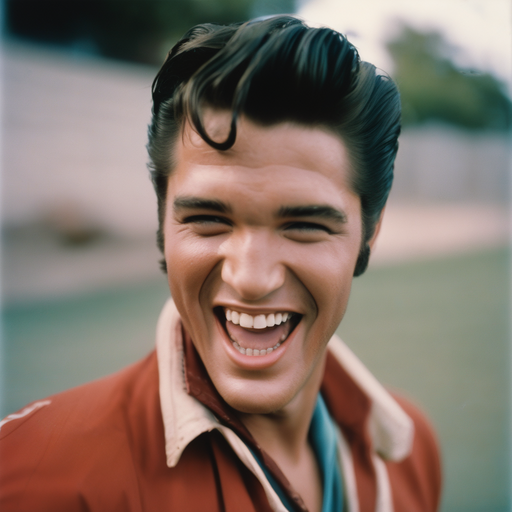}}
  \quad
  \subfloat[
  ]{\includegraphics[height=3.2cm]{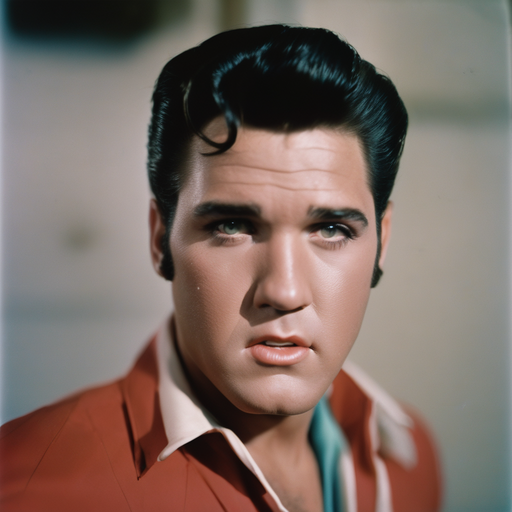}}
  \quad
  \subfloat[
  ]{\includegraphics[height=3.2cm]{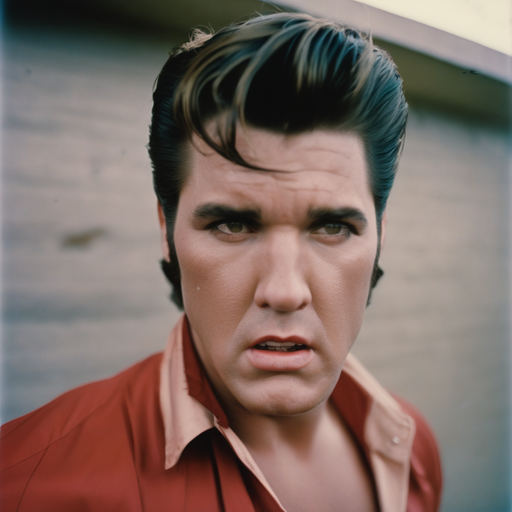}}
  \quad
  \subfloat[
  ]{\includegraphics[height=3.2cm]{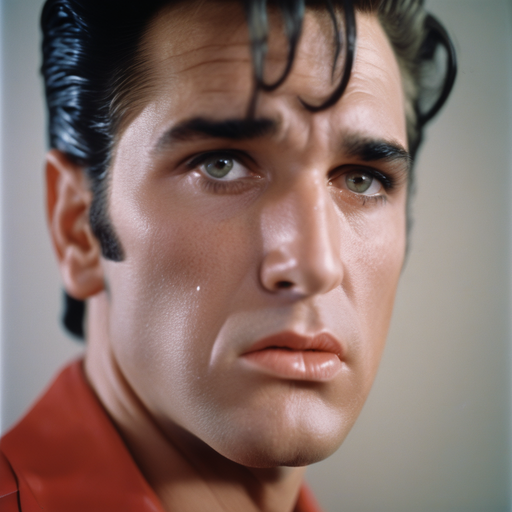}}
\caption{Elvis Presley with various emotional expressions. }
\label{fig:images:potential:emo_vip}
\end{figure}

\begin{figure}[ht!]
\centering
\subfloat[]{\includegraphics[height=3.2cm]{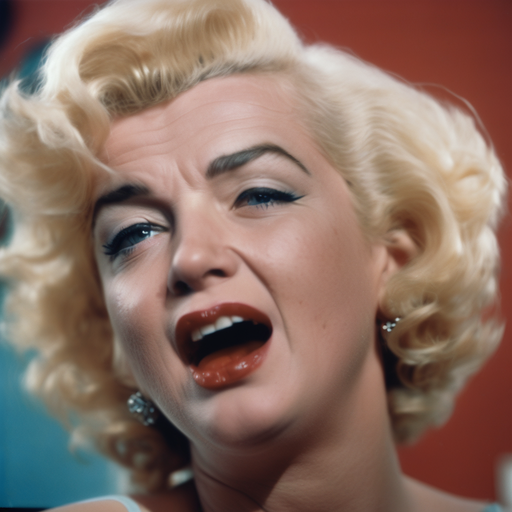}}
\quad
\subfloat[]{\includegraphics[height=3.2cm]{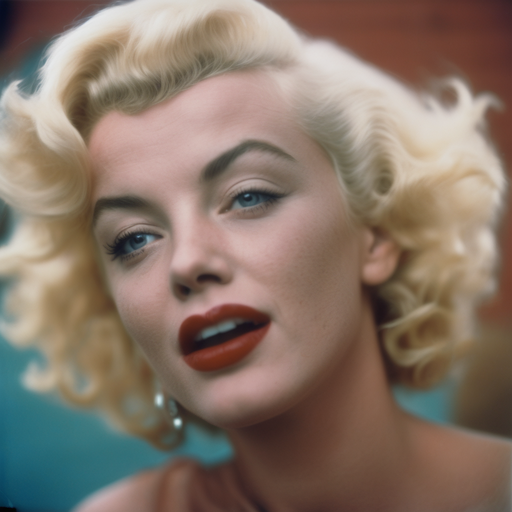}}
\quad
\subfloat[]{\includegraphics[height=3.2cm]{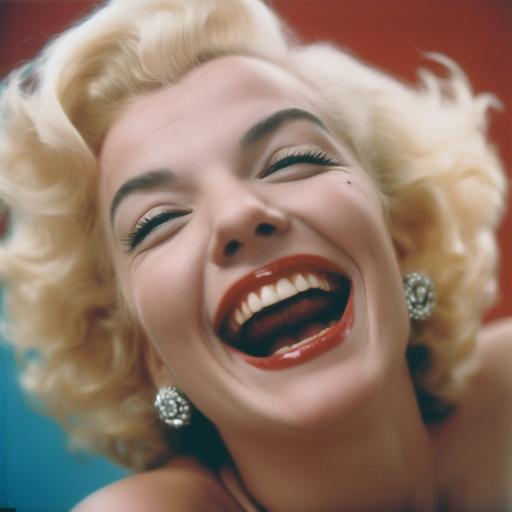}}
\quad
\subfloat[]{\includegraphics[height=3.2cm]{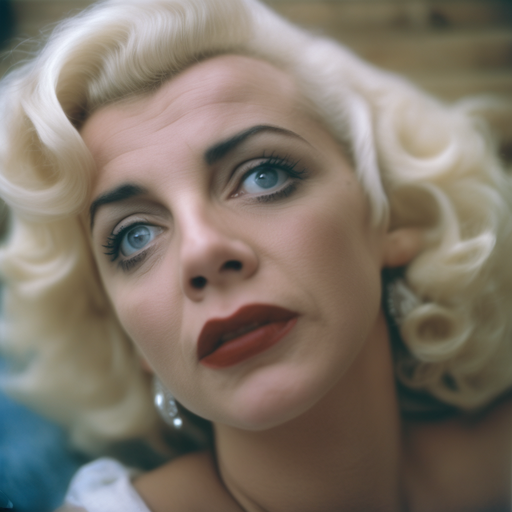}}
\caption{Marylin Monroe with various emotional expressions. }
\label{fig:images:potential:emo_marilyn}
\end{figure}

Stable Diffusion XL excels in capturing and intensifying emotional expressions, enabling it to generate synthetic images with exaggerated emotions for effects like attracting attention or conveying humor. This is evident in the overemphasized facial expressions of Elvis Presley and Marylin Monroe in Figures \ref{fig:images:potential:emo_vip} and \ref{fig:images:potential:emo_marilyn}. Similar to illustrations (see Section \ref{illustrations_and_cartoons}), this feature is particularly beneficial for cyber influence operations that rely on humor and hyperbole to influence audiences.

\FloatBarrier
\subsection{Content with Shock Value}

\begin{figure}[ht!]
\centering
\subfloat[]{\includegraphics[height=4.8cm]{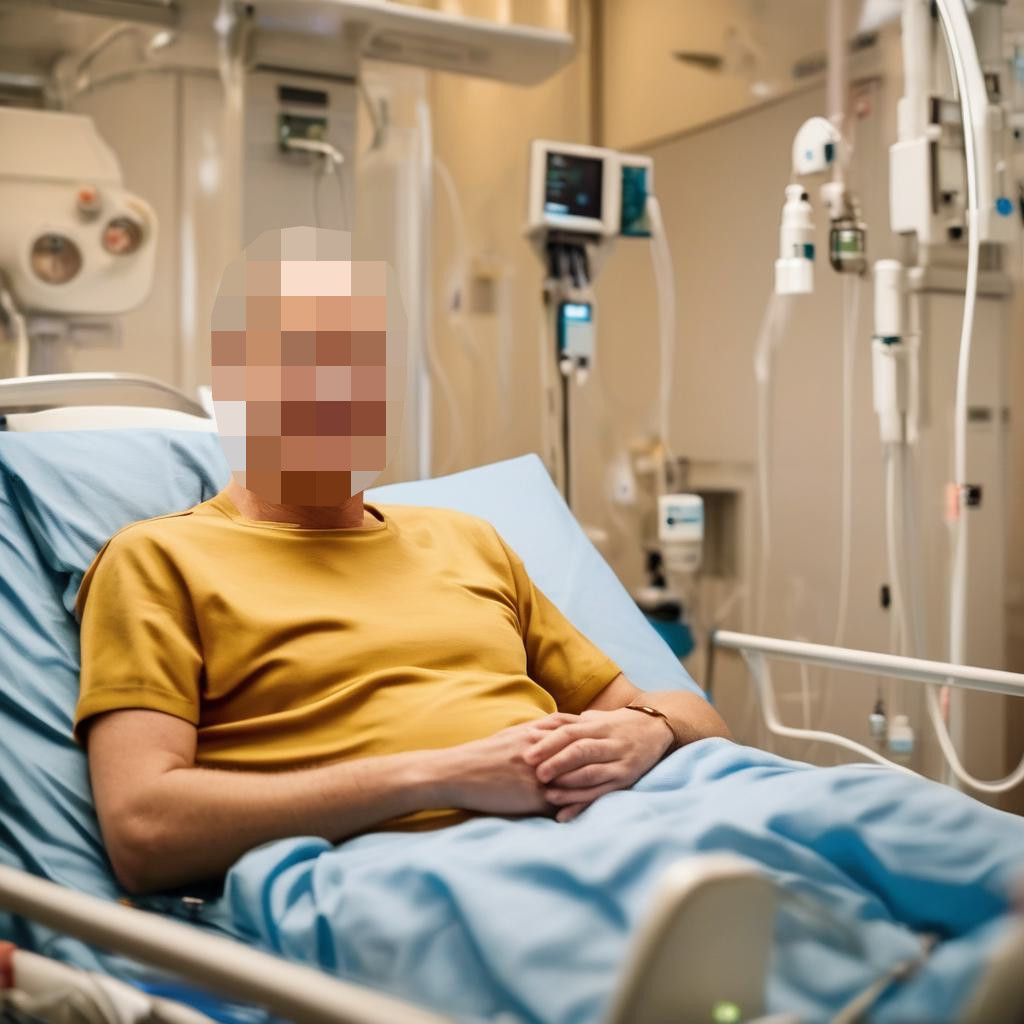}}
\quad
\subfloat[]{\includegraphics[height=4.8cm]{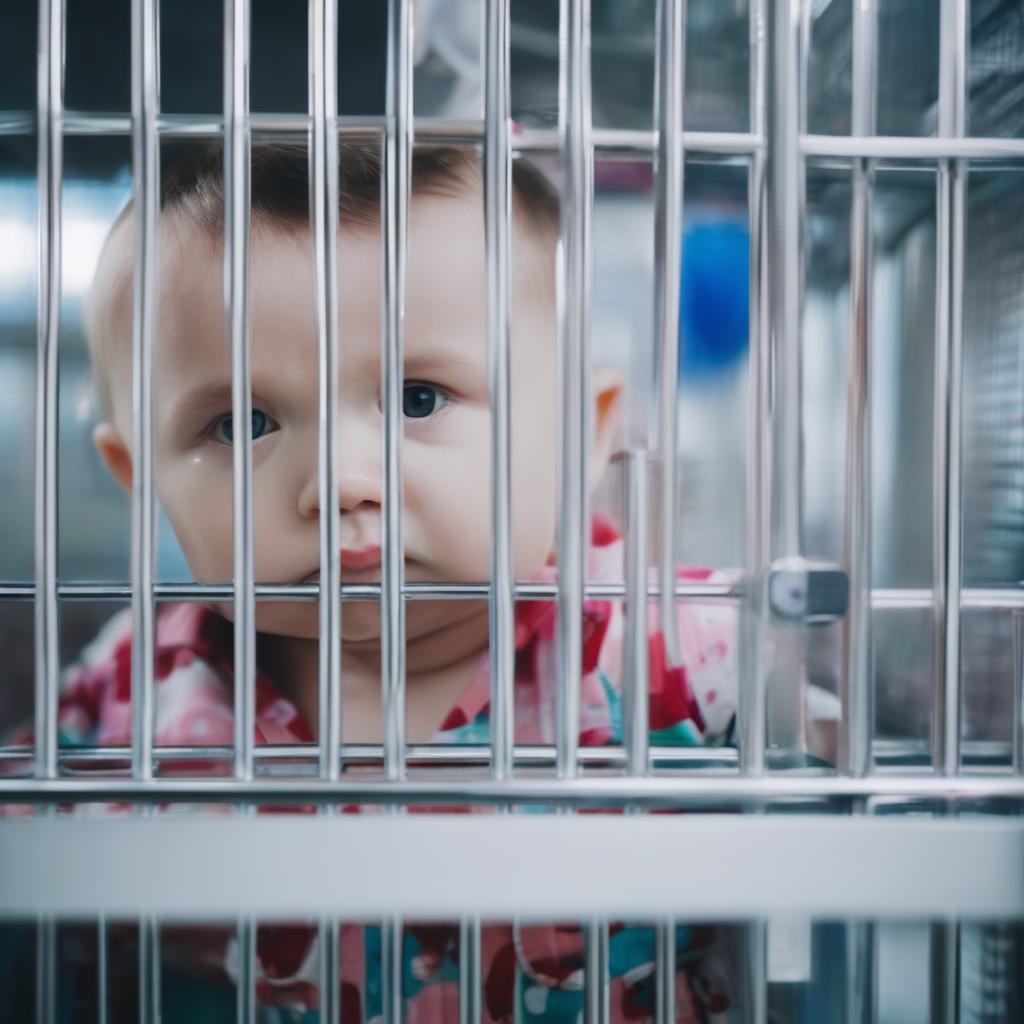}}
\quad
\subfloat[]{\includegraphics[height=4.8cm]{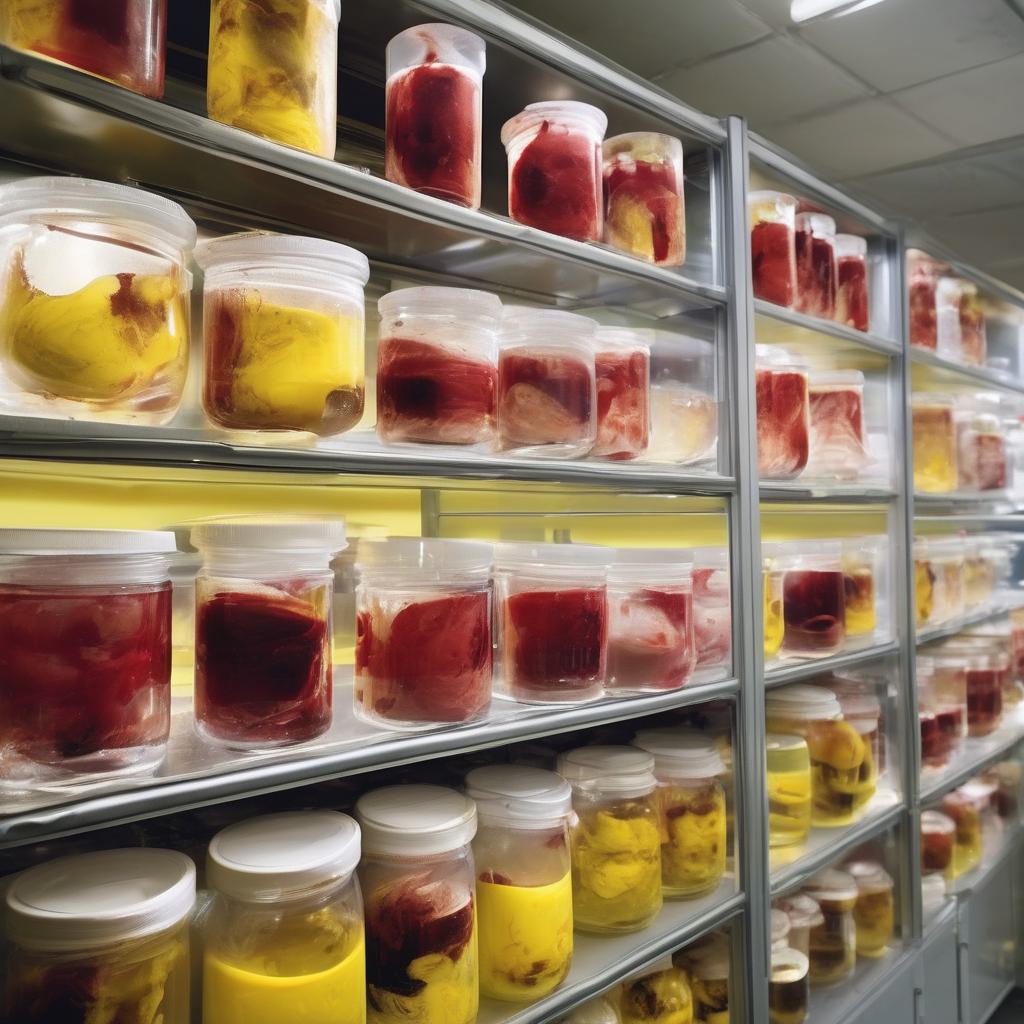}}
\caption{Scenes that might incite an immediate, potent emotional response (like pity or disgust) upon first glance, even if the viewer eventually identifies the image as synthesized.}
\label{images:shocking}
\end{figure}

Synthetic images are useful in cyber influence operations that use shock value to evoke strong emotional reactions like disgust, revulsion, or pity as a strategic element (see Section \ref{sec:scenario7_undermine_democracy}). Fig. \ref{images:shocking} show possible examples of this type of content.

\FloatBarrier
\clearpage

\section{Limitations of Generative Models}

\subsection{Artifacts}

Artifacts in image generation are unintended errors or inconsistencies that deviate from the desired outcome. This section will detail common artifact occurrences while some sections in  \ref{section:learnings}  will detail strategies to mitigate them.

\label{artefacts_human_anatomy}

\begin{figure}[ht]
  \centering
  \subfloat[Face and hand artifacts.]{\includegraphics[width=0.31\linewidth]{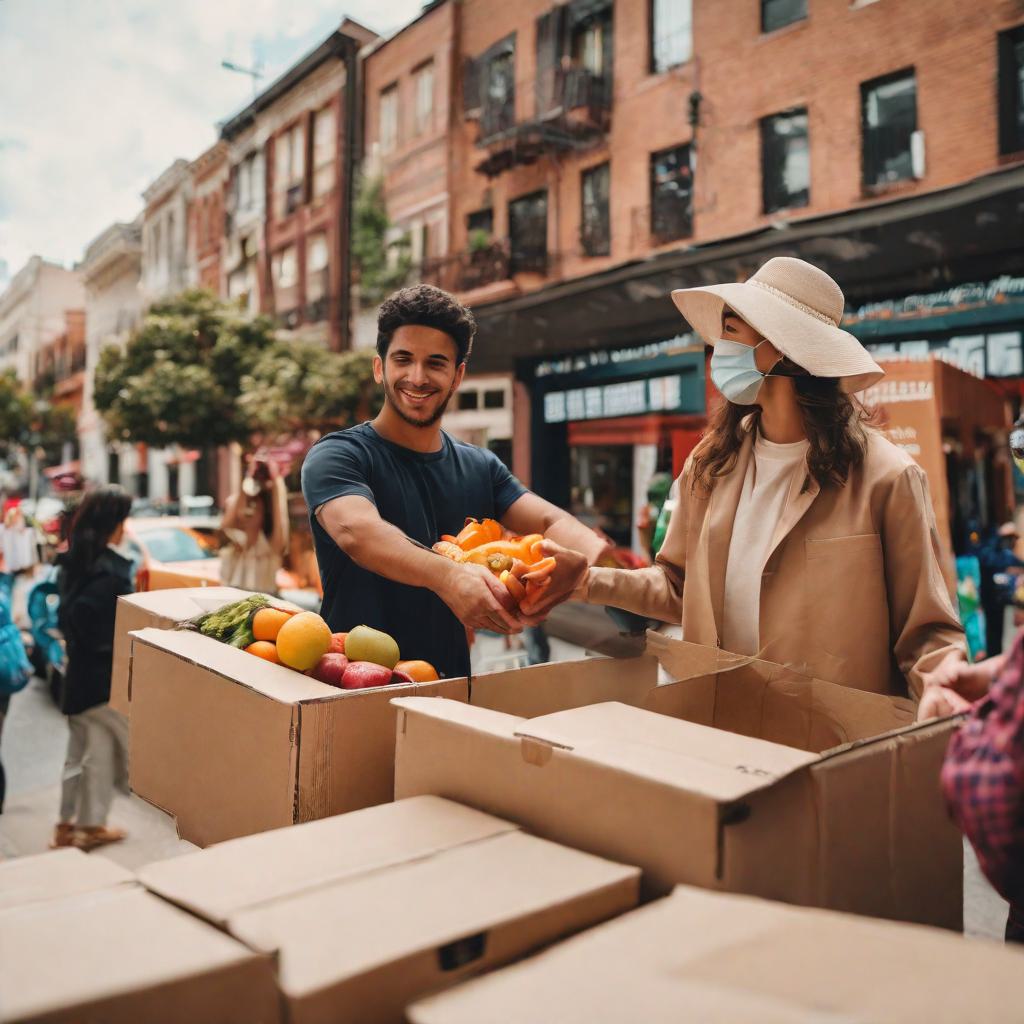}}
  \quad
  \subfloat[Face and hand artifacts, surplus hands and arms.]{\includegraphics[width=0.31\linewidth]{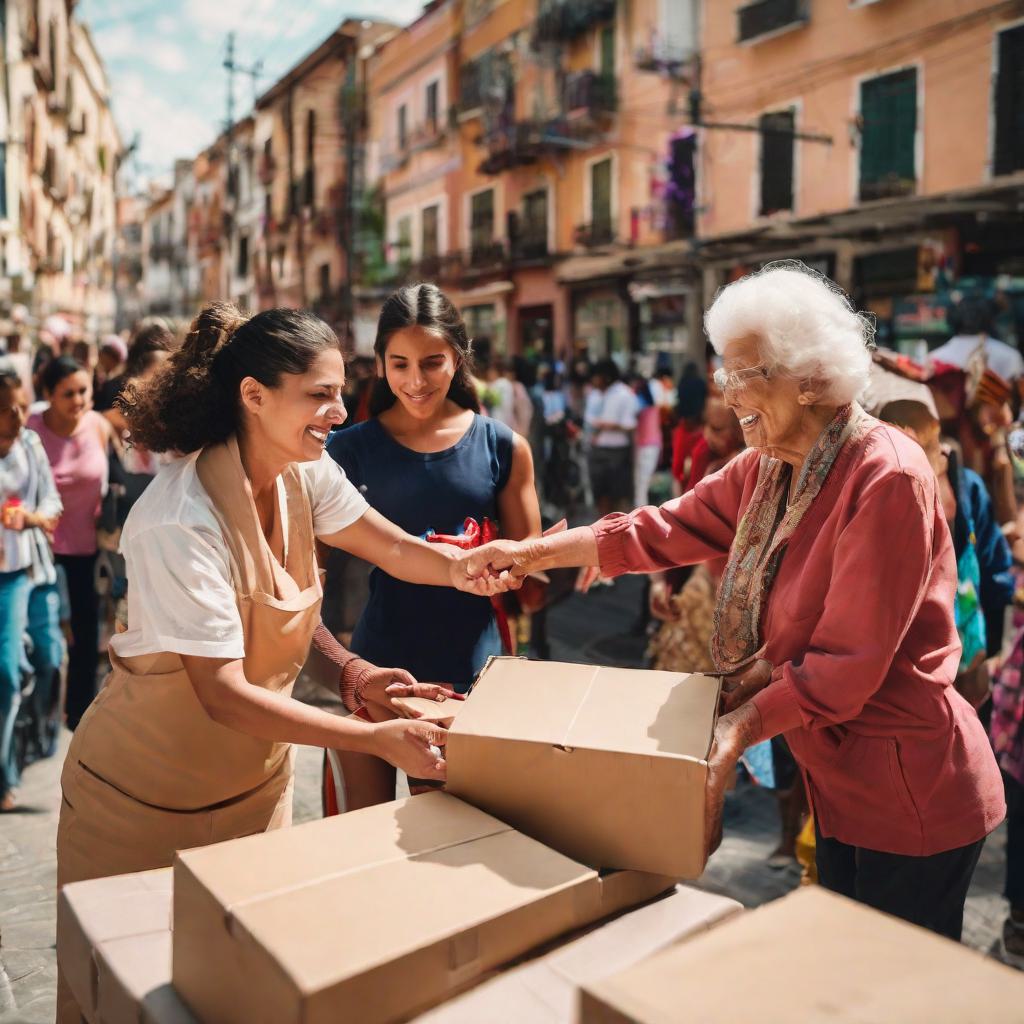}}
  \quad
  \subfloat[Face and hand artifacts, unnatural arm positions.]{\includegraphics[width=0.31\linewidth]{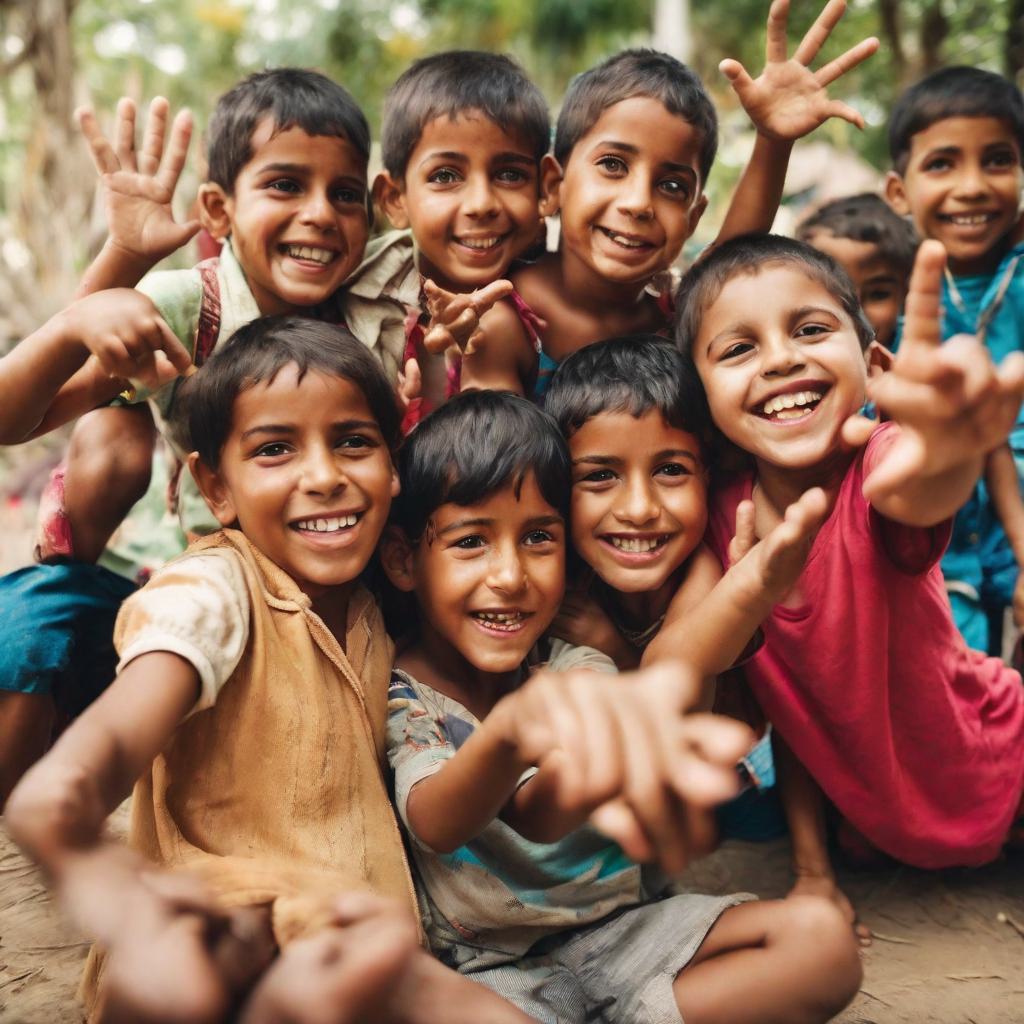}}
  
  \subfloat[Face and hand artifacts.]{\includegraphics[width=0.31\linewidth]{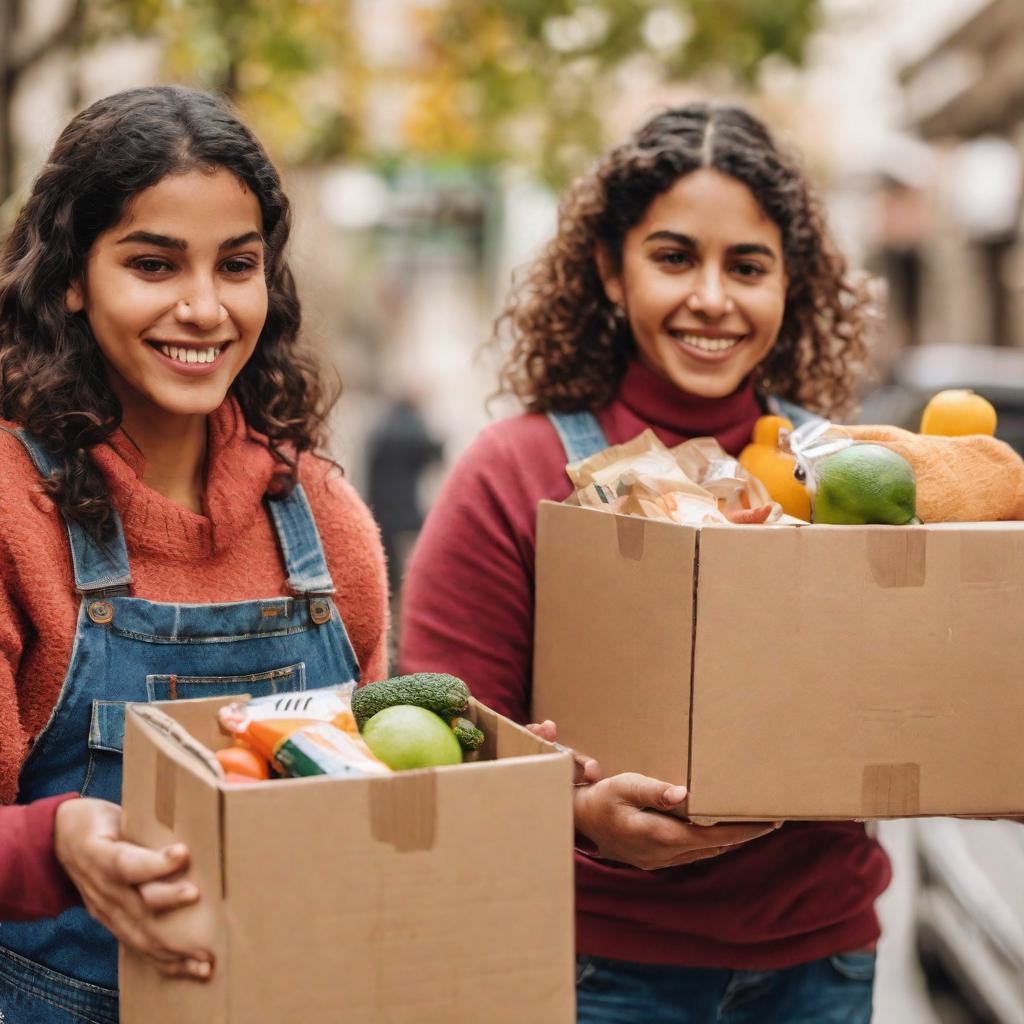}}
  \quad
  \subfloat[Face and hand artifacts, unnatural arm  and arm positions.]{\includegraphics[width=0.3\linewidth]{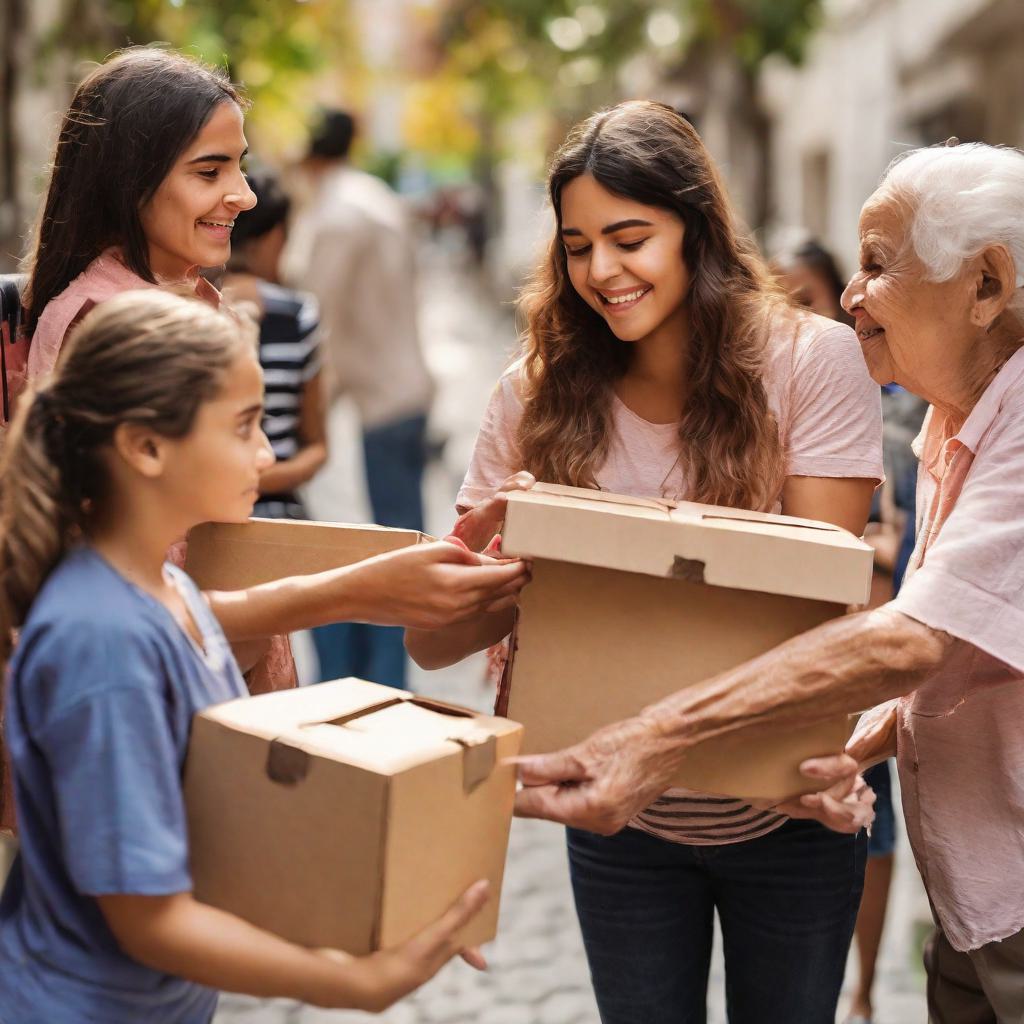}}
  \quad
  \subfloat[Face and hand artifacts.]{\includegraphics[width=0.31\linewidth]{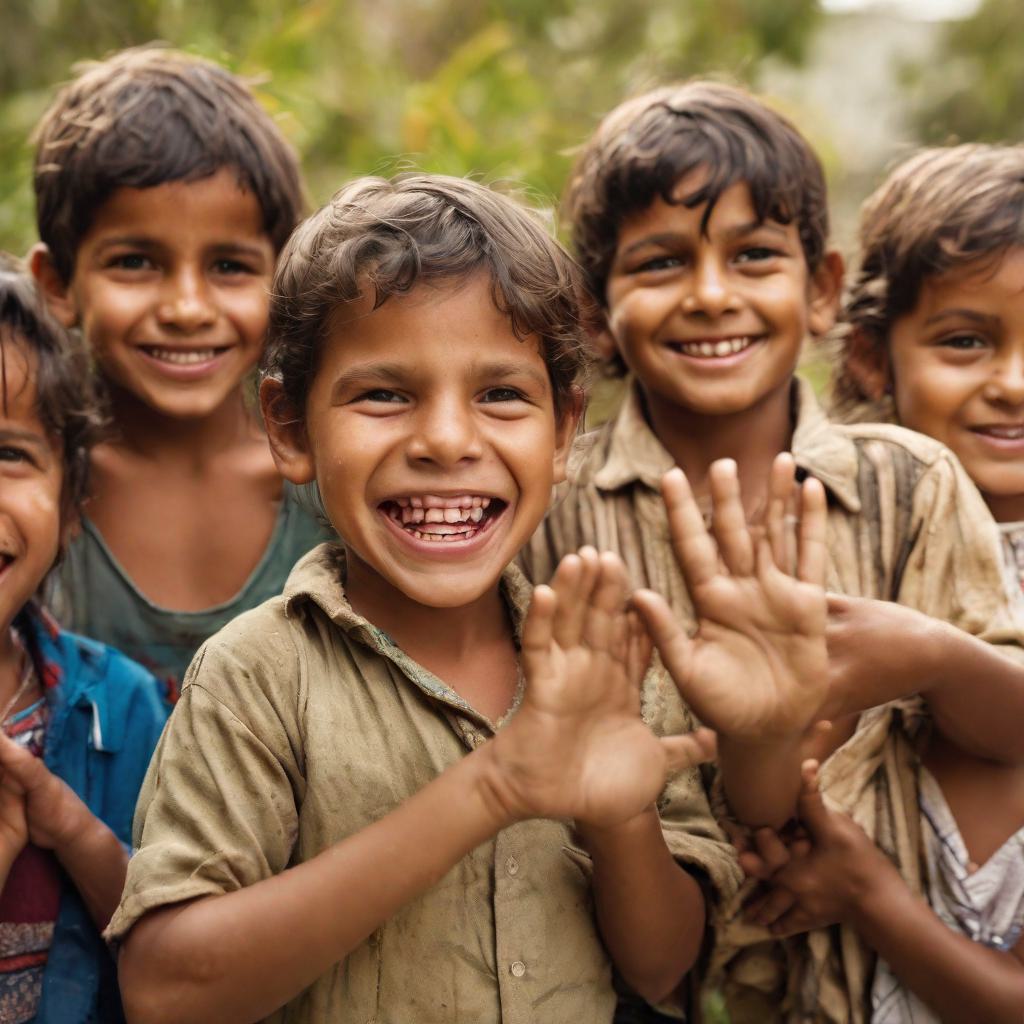}}
\caption{Example of artifacts concerning human anatomy, from images generated with Stable Diffusion XL.
}
\label{images:artefacts:human_anatomy:complete_examples}
\end{figure}

\begin{figure}[ht]
  \centering
      \subfloat[]
      {\includegraphics[height = 4.5cm]{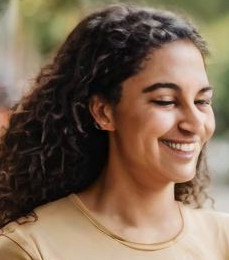}}
      \quad
      \subfloat[]
      {\includegraphics[height = 4.5cm]{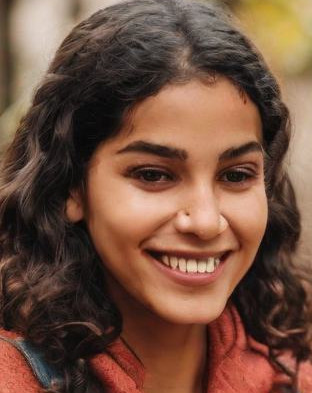}}
      \quad
      \subfloat[]
      {\includegraphics[height = 4.5cm]{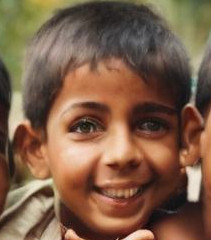}}
      \quad
      \subfloat[]
      {\includegraphics[height = 4.5cm]{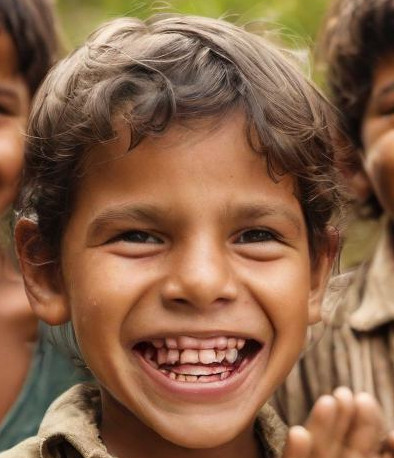}}

\caption{Close-up of artifacts in faces from images in Figure \ref{images:artefacts:human_anatomy:complete_examples}: Dark eye sockets, disproportionate or protruding teeth, multiple rows of teeth, split lips, various unnatural asymmetries.}
\label{images:artefacts:human_anatomy:face_examples}
\end{figure}

\begin{figure}[ht]
  \centering
  \subfloat[]
  {\includegraphics[height = 2.8cm]{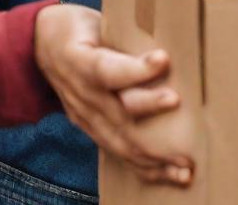}}
  \quad
  \subfloat[]
  {\includegraphics[height = 2.8cm]{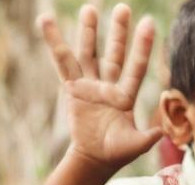}}
  \quad
  \subfloat[]
  {\includegraphics[height = 2.8cm]{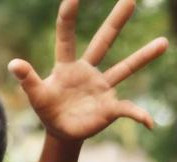}}
  \quad
  \subfloat[]
  {\includegraphics[height = 2.8cm]{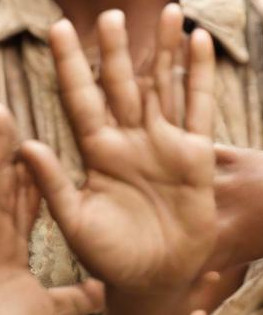}}
  \quad
  \subfloat[]
  {\includegraphics[height = 2.8cm]{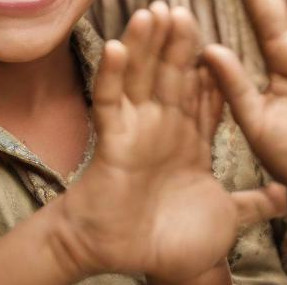}}
  
\caption{Close-up of hand artifacts from images in Figure \ref{images:artefacts:human_anatomy:complete_examples}: Too many fingers, too few fingers, missing thumbs, fingers of unnatural lengths, too much space between fingers, bent fingers or hands where there wouldn't be any joints, split or "broken"-looking hands and fingers, hands that look like multiple hands grown together.}
\label{images:artefacts:human_anatomy:hande_examples}
\end{figure}

\begin{figure}[H]
  \centering
  	\subfloat[]{\includegraphics[height=5cm]{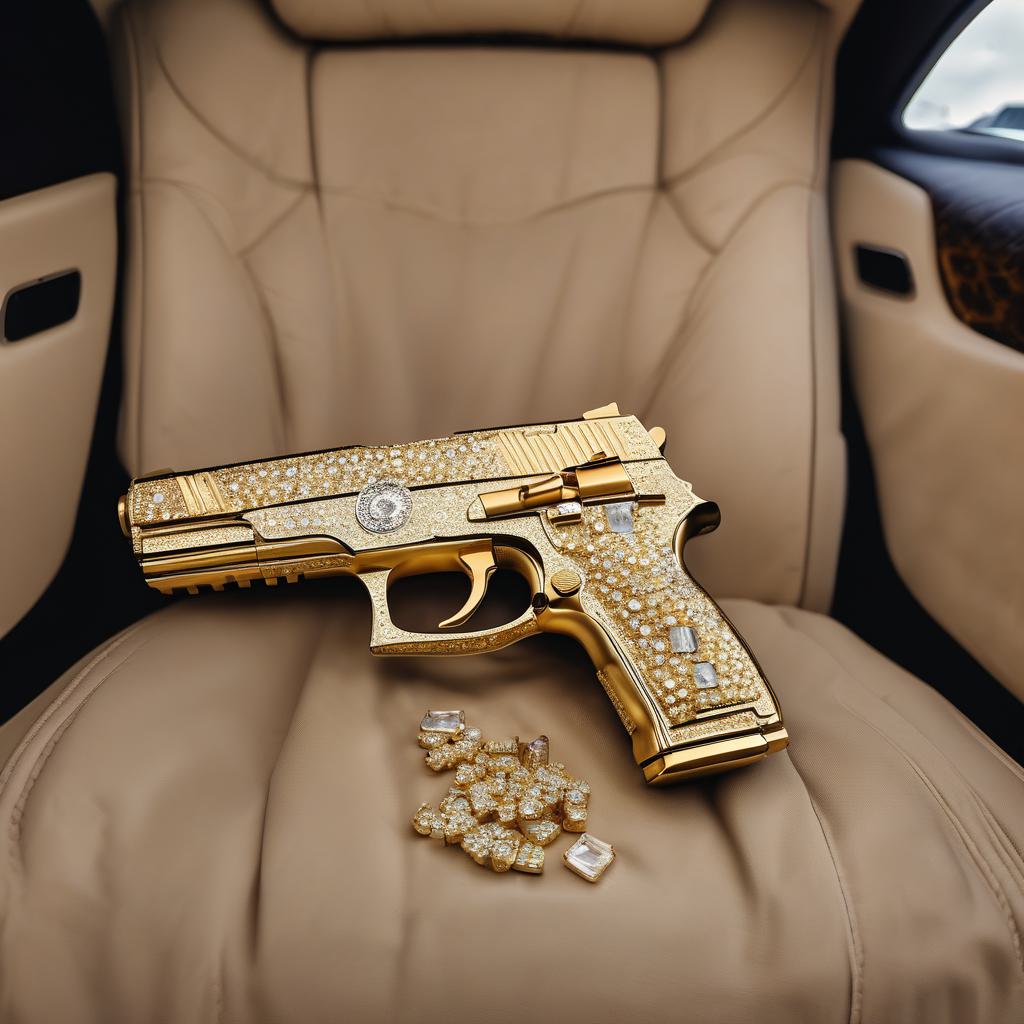}\label{images:artefacts:other:proportions:wobbly_gun1}}
	\quad
  	\subfloat[]{\includegraphics[height=5cm]{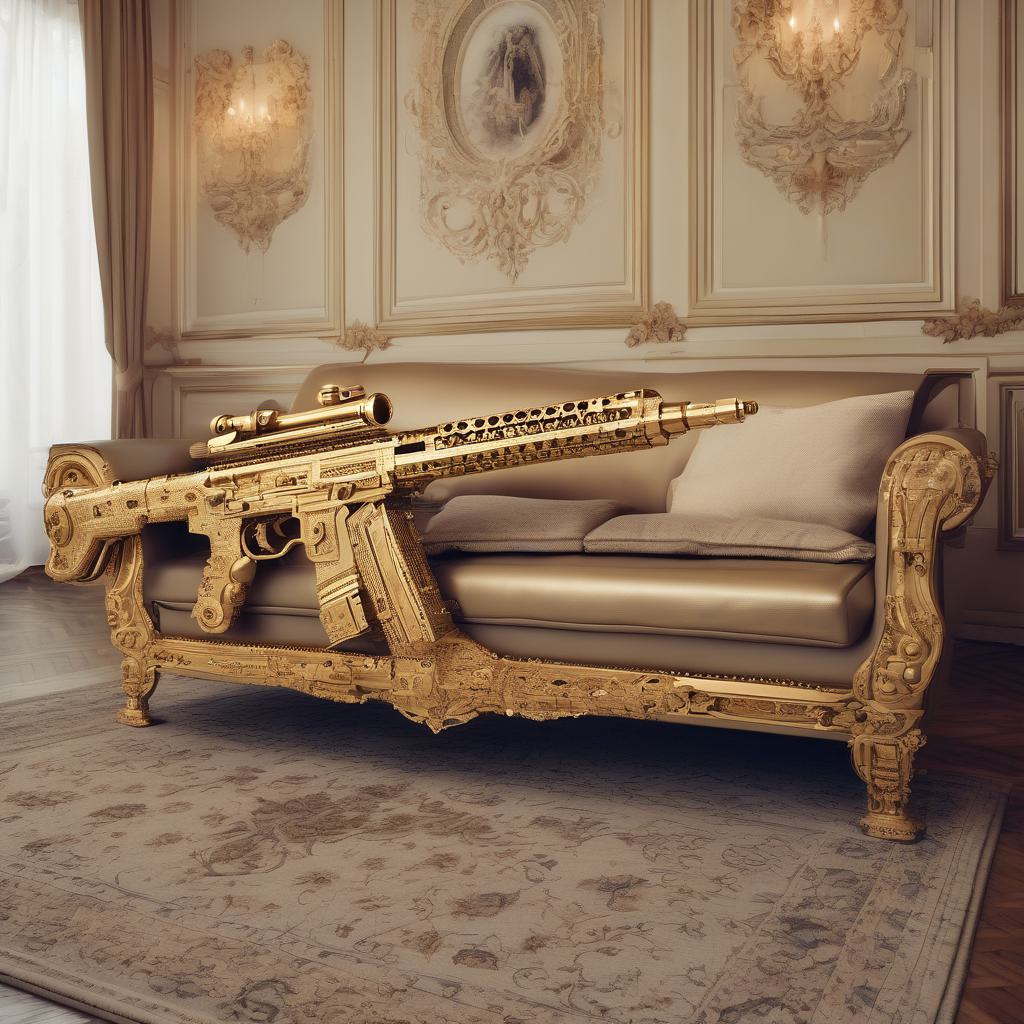}\label{images:artefacts:other:proportions:wobbly_gun2}}
	\quad
	\subfloat[]{\includegraphics[height=5cm]{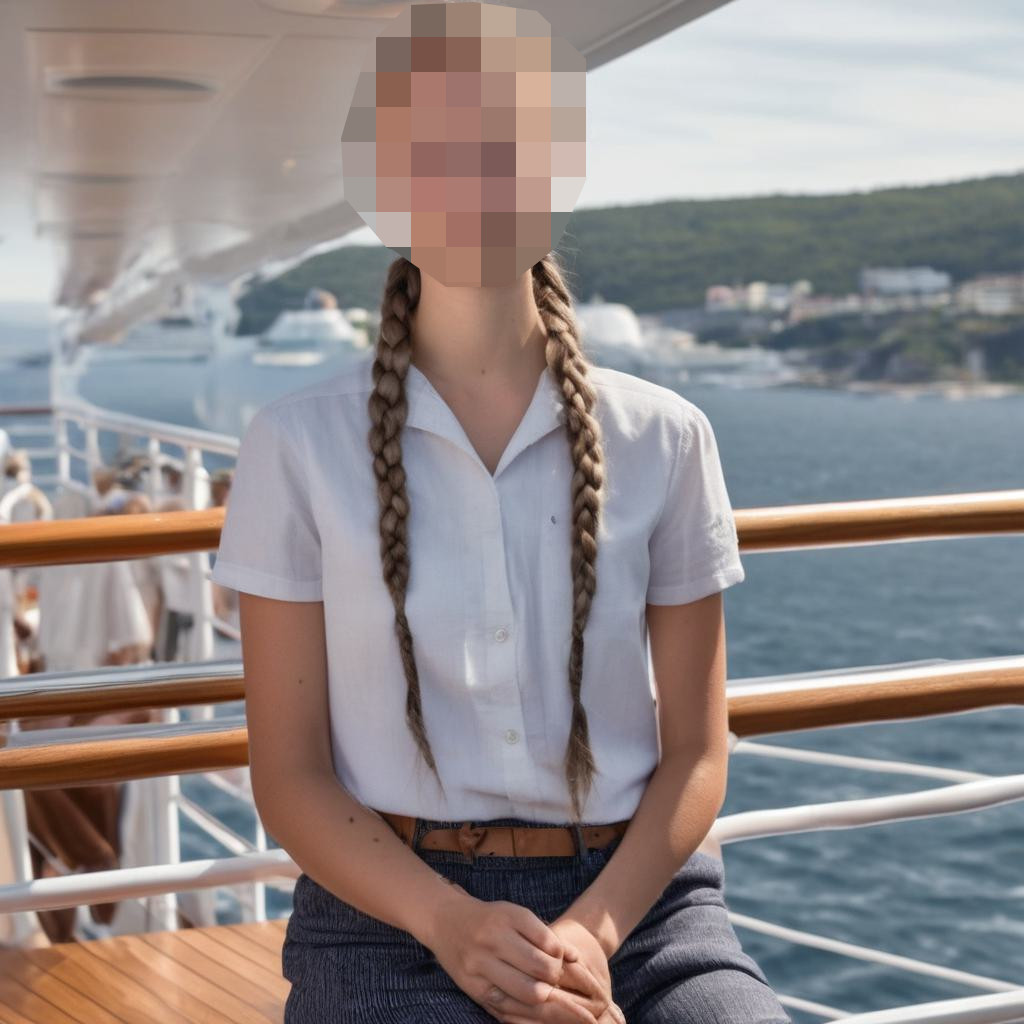}\label{images:artefacts:other:proportions:person}}
\caption{Unnatural or impossible proportions: The two guns in \protect\subref{images:artefacts:other:proportions:wobbly_gun1} and \protect\subref{images:artefacts:other:proportions:wobbly_gun2} are improbably large compared to their environment. The person's neck in \protect\subref{images:artefacts:other:proportions:person} is unnaturally long. The two gun images also display concept leakage artifacts (see Section \protect\ref{leakage_and_duplication}), and the railings at the sides of \protect\subref{images:artefacts:other:proportions:person} are mismatched and not straight.}
\label{images:artefacts:other:proportions}
\end{figure}

Achieving realistic human likenesses is a complex task for generative models such as the Stable Diffusion family of models. Figure \ref{images:artefacts:human_anatomy:complete_examples} demonstrates various instances of such artifacts, and Figures \ref{images:artefacts:human_anatomy:face_examples} and \ref{images:artefacts:human_anatomy:hande_examples} zoom in on facial and hand regions with noticeable artifacts, whereas Figure \ref{images:artefacts:other:proportions:person} highlights an instance of distorted proportions, showcasing an overly elongated neck of a person.

\FloatBarrier
\label{artefacts_proportion}

\begin{figure}[H]
\centering
\subfloat[]
{\includegraphics[height = 5cm]{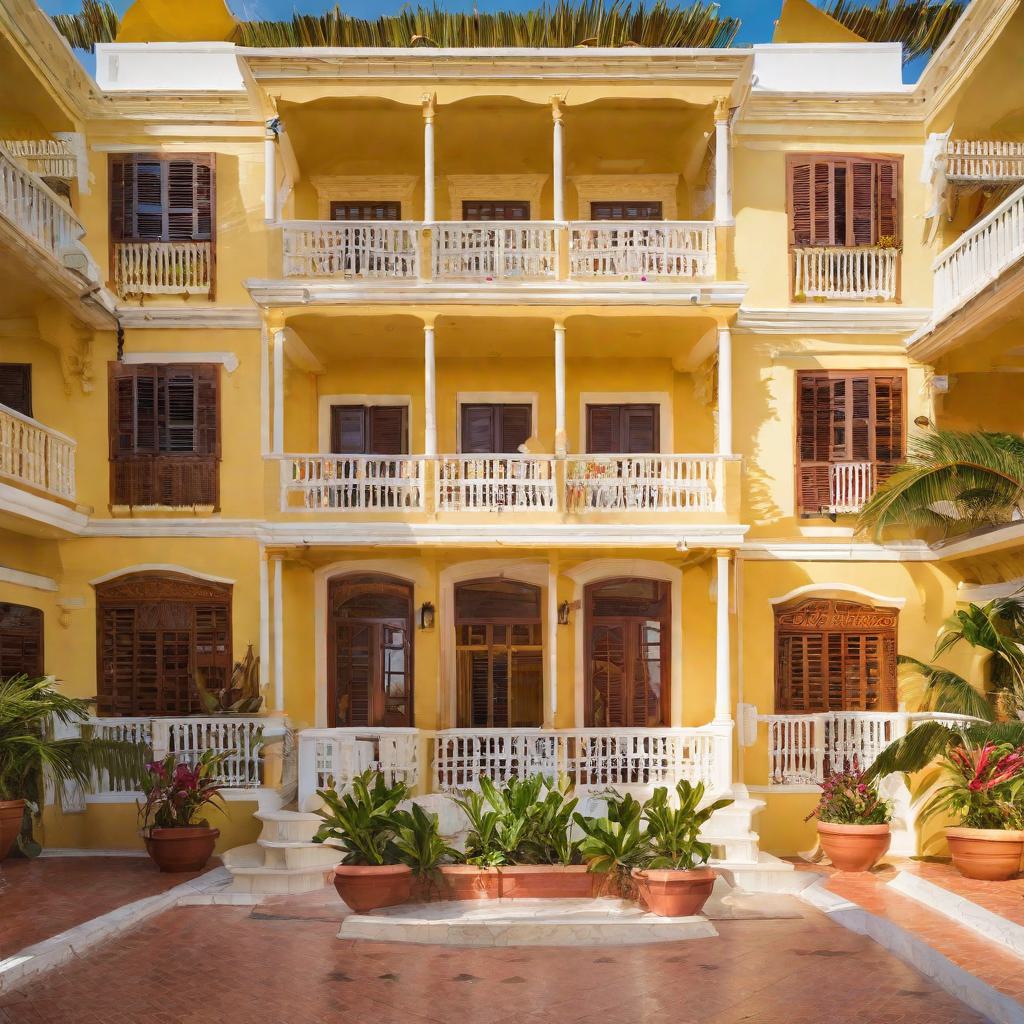}}
\quad
\subfloat[]
{\includegraphics[height = 5cm]{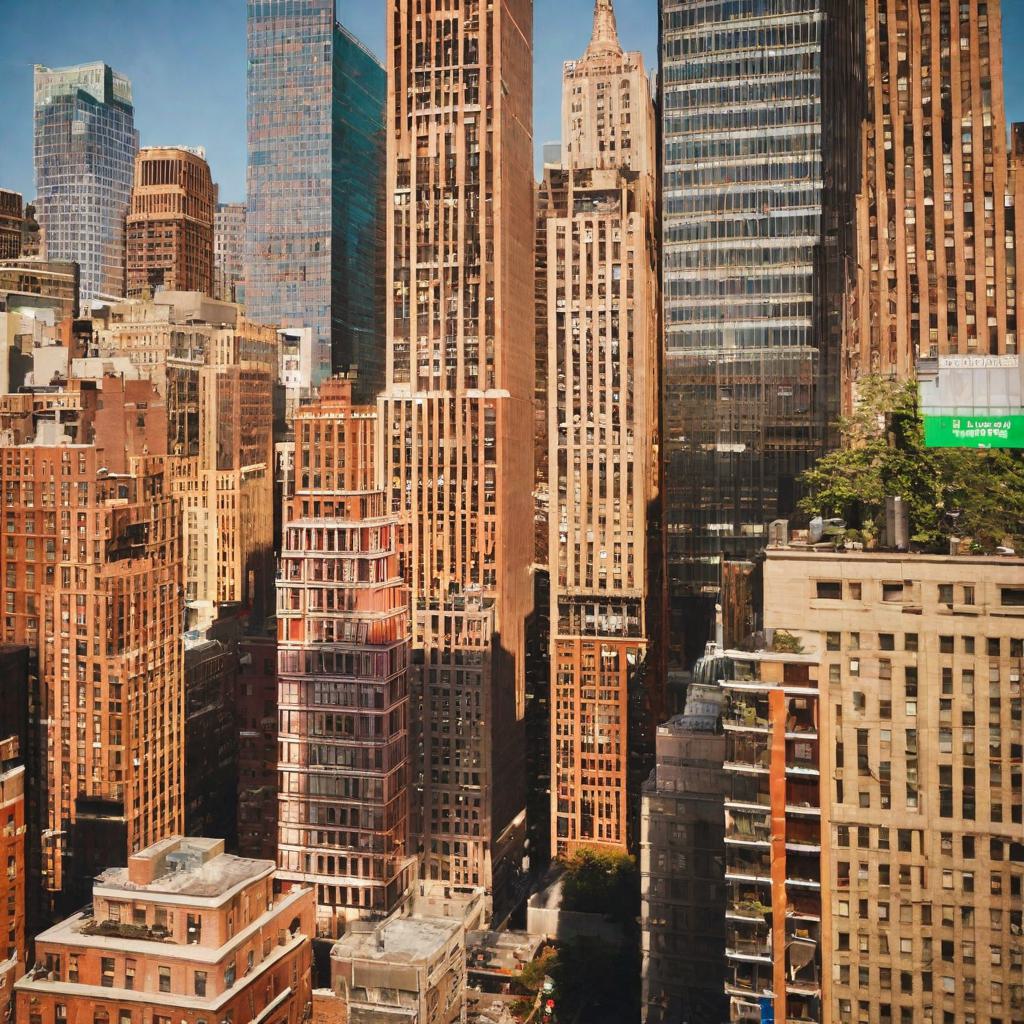}}
\quad
\subfloat[]
{\includegraphics[height = 5cm]{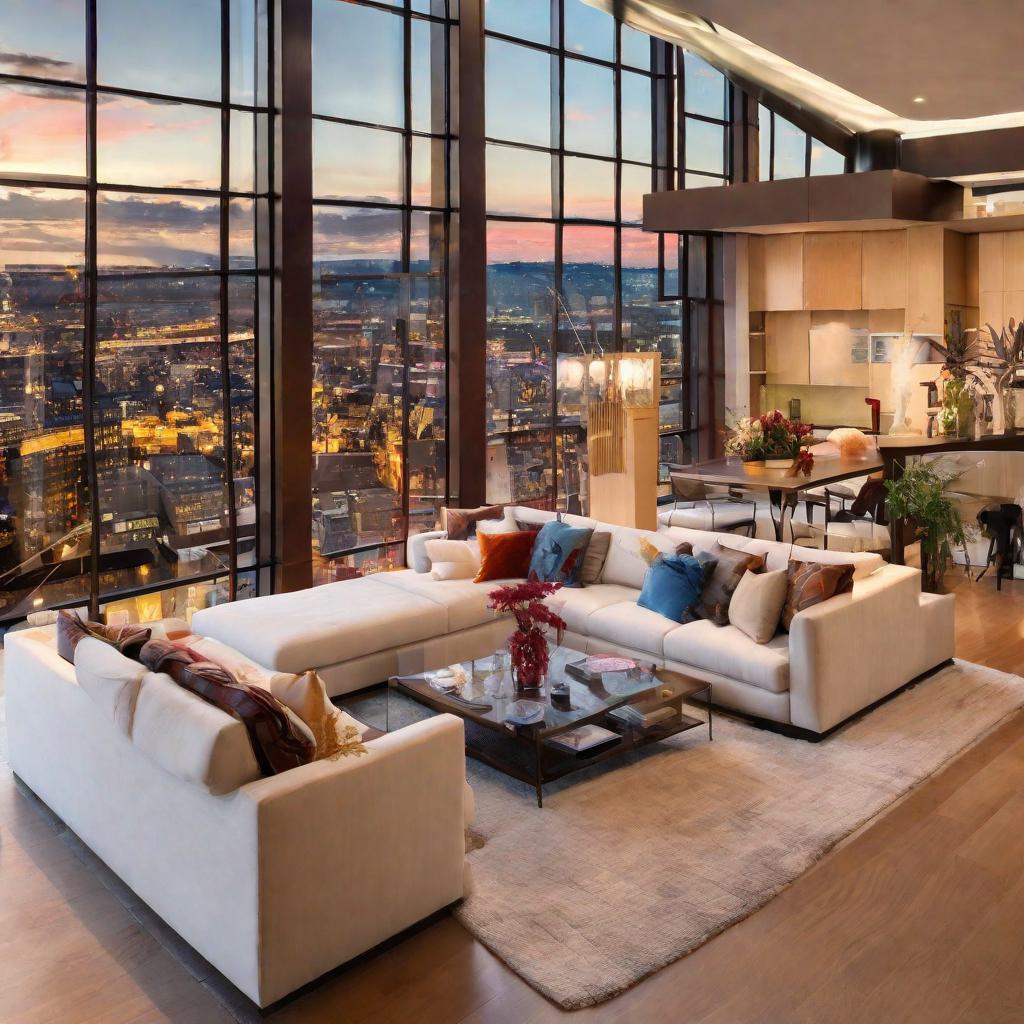}}

\subfloat[]
{\includegraphics[height = 2.75cm]{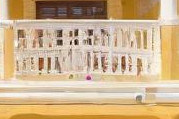}}
\quad
\subfloat[]
{\includegraphics[height = 2.75cm]{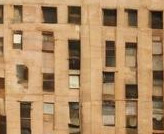}}
\quad
\subfloat[]
{\includegraphics[height = 2.75cm]{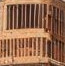}}
\quad
\subfloat[]
{\includegraphics[height = 2.75cm]{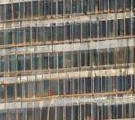}}
\quad
\subfloat[]
{\includegraphics[height = 2.5cm]{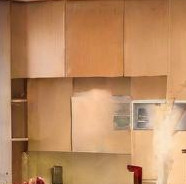}}
\quad
\subfloat[]
{\includegraphics[height = 2.5cm]{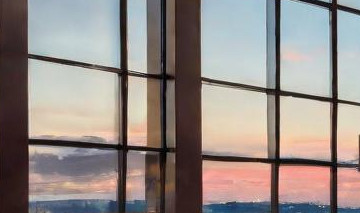}}
\quad
\subfloat[]
{\includegraphics[height = 2.5cm]{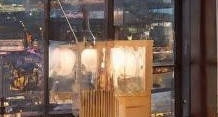}}
\quad
\subfloat[]
{\includegraphics[height = 2.5cm]{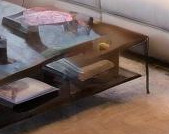}}
\caption{Irregular, interrupted, blurred, entangled or \enquote{wobbly} lines and patterns, lack of straight lines and orthogonal angles.}
\label{images:artefacts:other:lines_and_patterns}
\end{figure}

\begin{figure}[ht!]
\centering
    \subfloat[]
    {\includegraphics[height = 2.75cm]{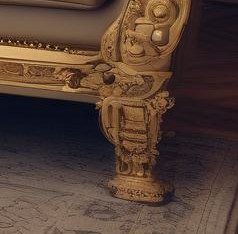}}
    \quad
    \subfloat[]
    {\includegraphics[height = 2.75cm]{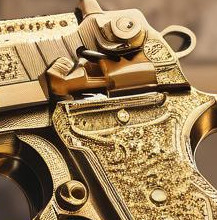}}
    \quad
    \subfloat[]
    {\includegraphics[height = 2.75cm]{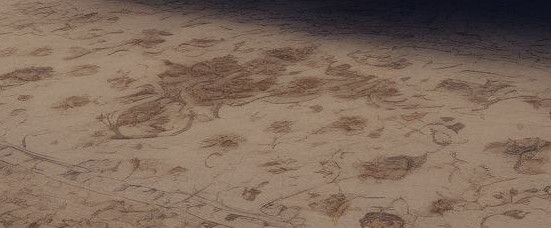}}
    \quad
    \subfloat[]
    {\includegraphics[height = 2.5cm]{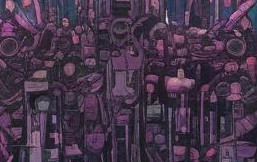}}
    \quad
    \subfloat[]
    {\includegraphics[height = 2.5cm]{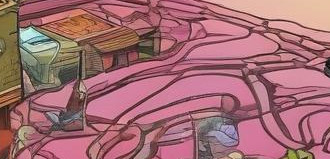}}
    \quad
    \subfloat[]
    {\includegraphics[height = 2.5cm]{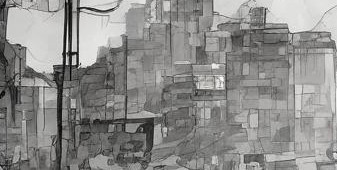}}
\caption{Close-up views of characteristic patterns that may not initially stand out as apparent artifacts, but can be a potential indicator that the image is generated with a model such as Stable Diffusion.}
\label{subtle_patterns}
\end{figure}

Images generated with Stable Diffusion models frequently have artifacts characterized by distorted, \enquote{wobbly}, or tangled lines and patterns, particularly in places where uniformity is expected, such as windows on a building. These artifacts are evident in examples shown in Figure \ref{images:artefacts:other:lines_and_patterns}. There are also more subtle patterns which may not seem especially disturbing at first glance, but on closer examination appear unusual if not unlikely. Figure \ref{subtle_patterns} shows some examples of these patterns.

\begin{figure}[H]
\centering
\subfloat[]
    {\includegraphics[height=5cm]{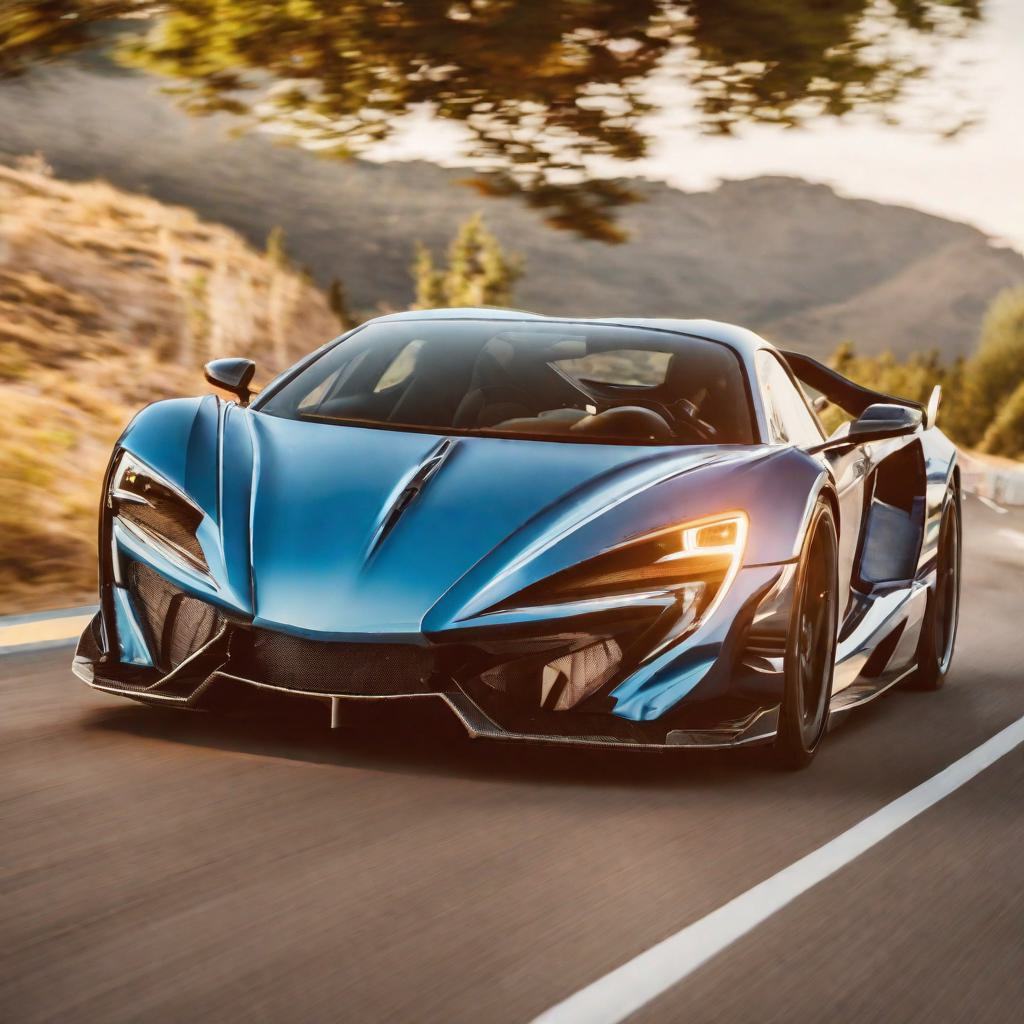}}
    \quad
    \subfloat[]
    {\includegraphics[height=5cm]{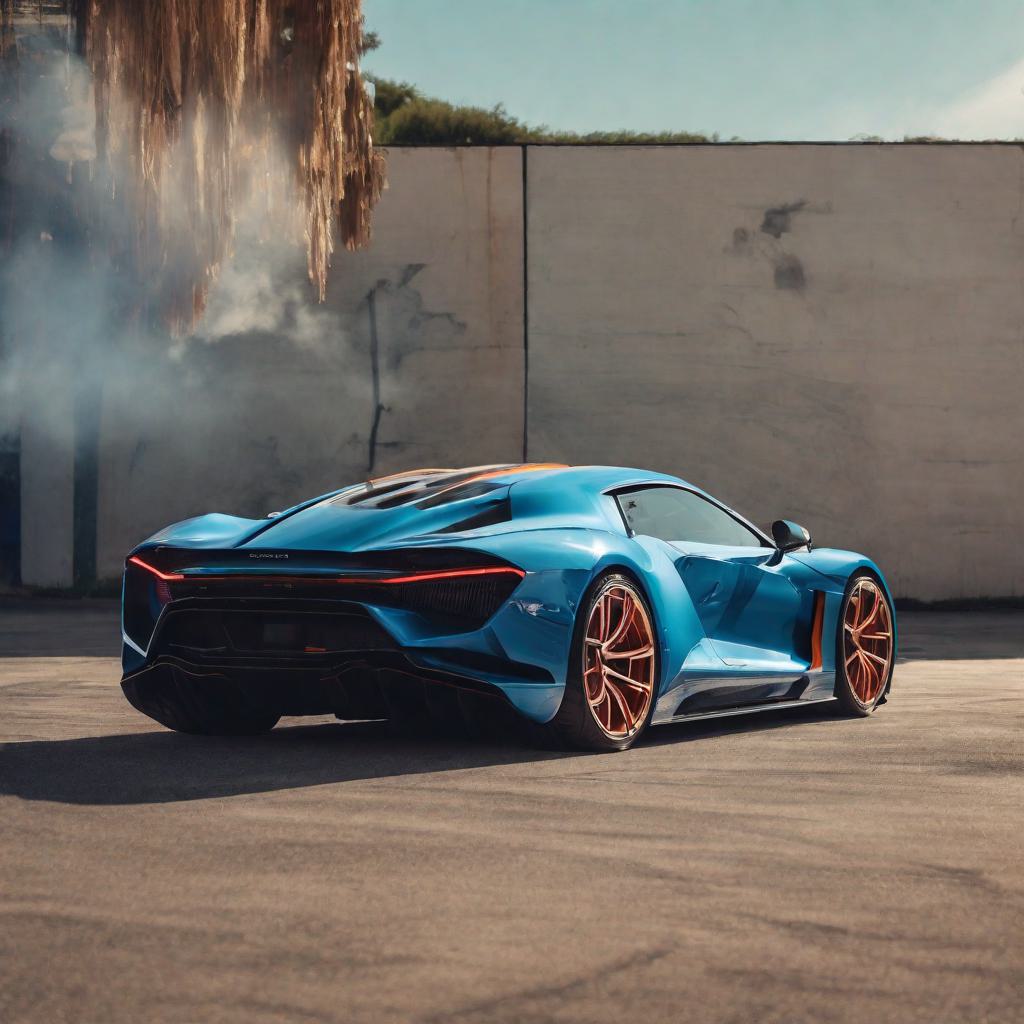}}
    \quad
    \subfloat[]
    {\includegraphics[height=5cm]{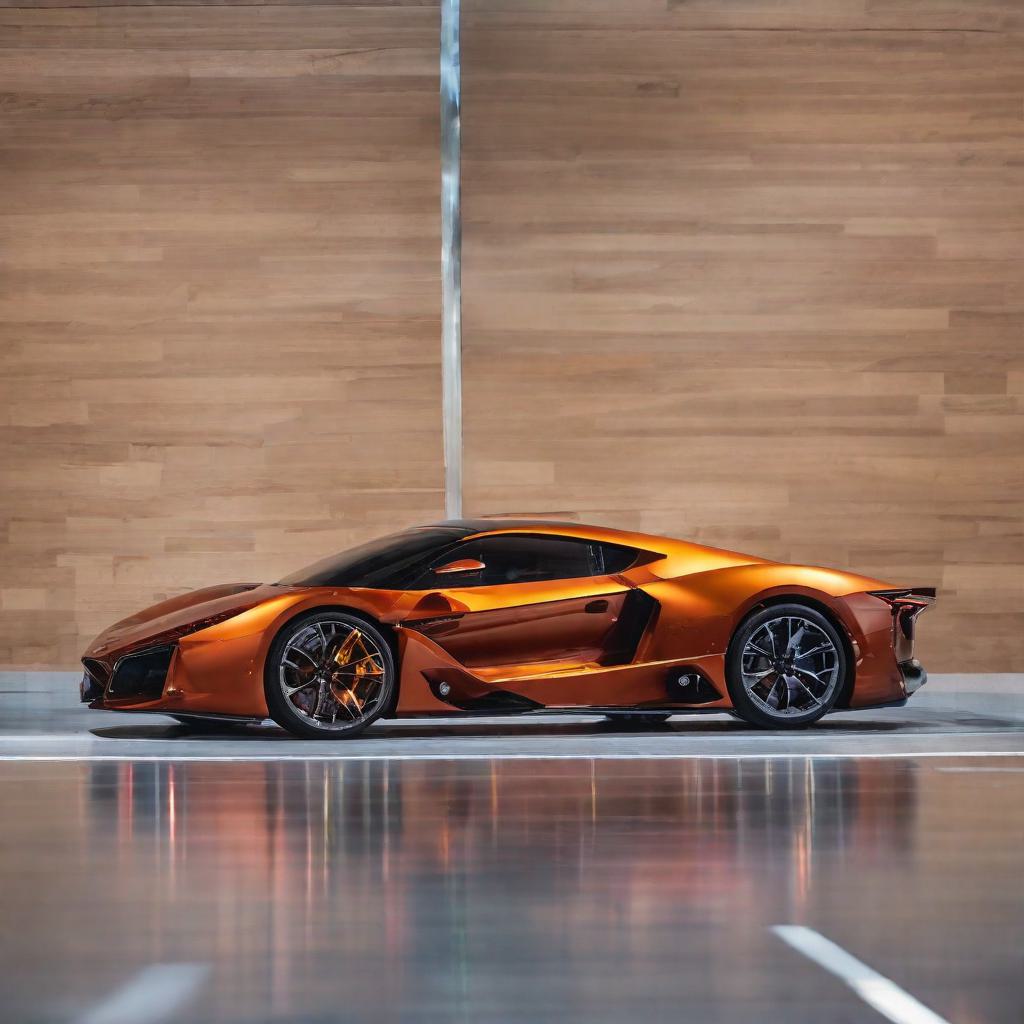}}
\caption{Lack of symmetry and undulating lines and shapes as artifacts in generated images of cars.}
\label{images:artefacts:other:cars}
\end{figure}

\begin{figure}[H]
\centering
    \subfloat[]
    {\includegraphics[height=3.5cm]{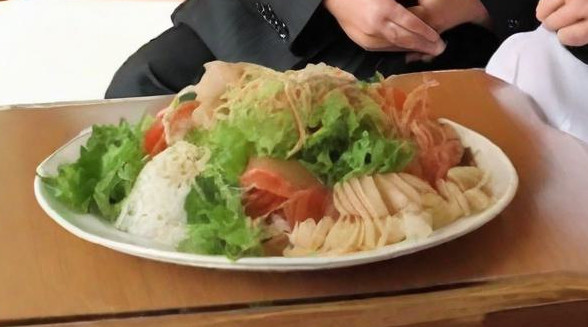}}
    \quad
    \subfloat[]
    {\includegraphics[height=3.5cm]{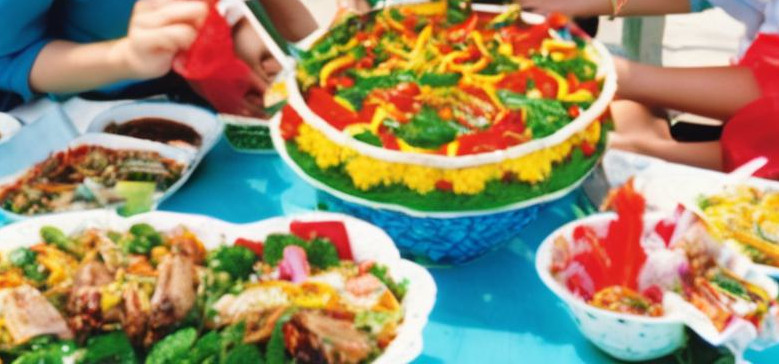}}
    \quad
    \subfloat[]
    {\includegraphics[height=3.5cm]{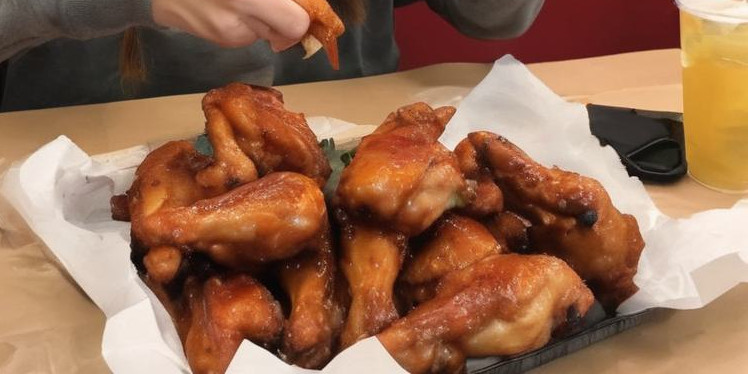}}
\caption{Examples of "artificial" or unprocessed looking food and unnaturally large food portions in generated images.}
\label{images:artefacts:other:proportions_food}
\end{figure}

Generated images also often exhibit noticeable asymmetries and inconsistencies, especially where symmetry is expected. Such artifacts are especially apparent in images of vehicles, as seen in Figure \ref{images:artefacts:other:cars}. Similarly, objects, individuals, or body parts may appear in disproportionate scales in these images, as shown in Figure \ref{images:artefacts:other:proportions}. The issue extends to food representations, which can appear unnaturally unprocessed or feature artificial textures and colors, as depicted in Figure \ref{images:artefacts:other:proportions_food}.

\begin{figure}[ht!]
  \centering
      \subfloat[
      ]{\includegraphics[height=5cm]{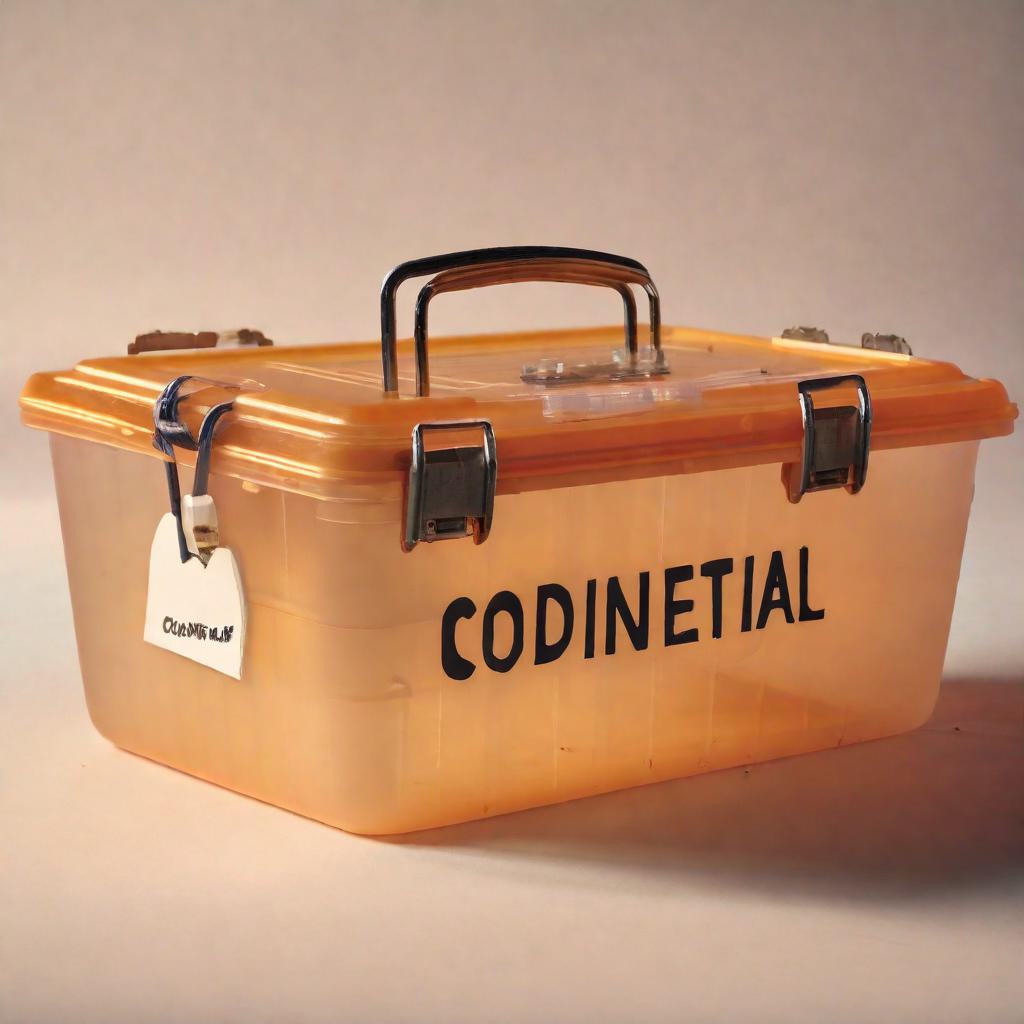}}
      \quad
      \subfloat[
      ]{\includegraphics[height=5cm]{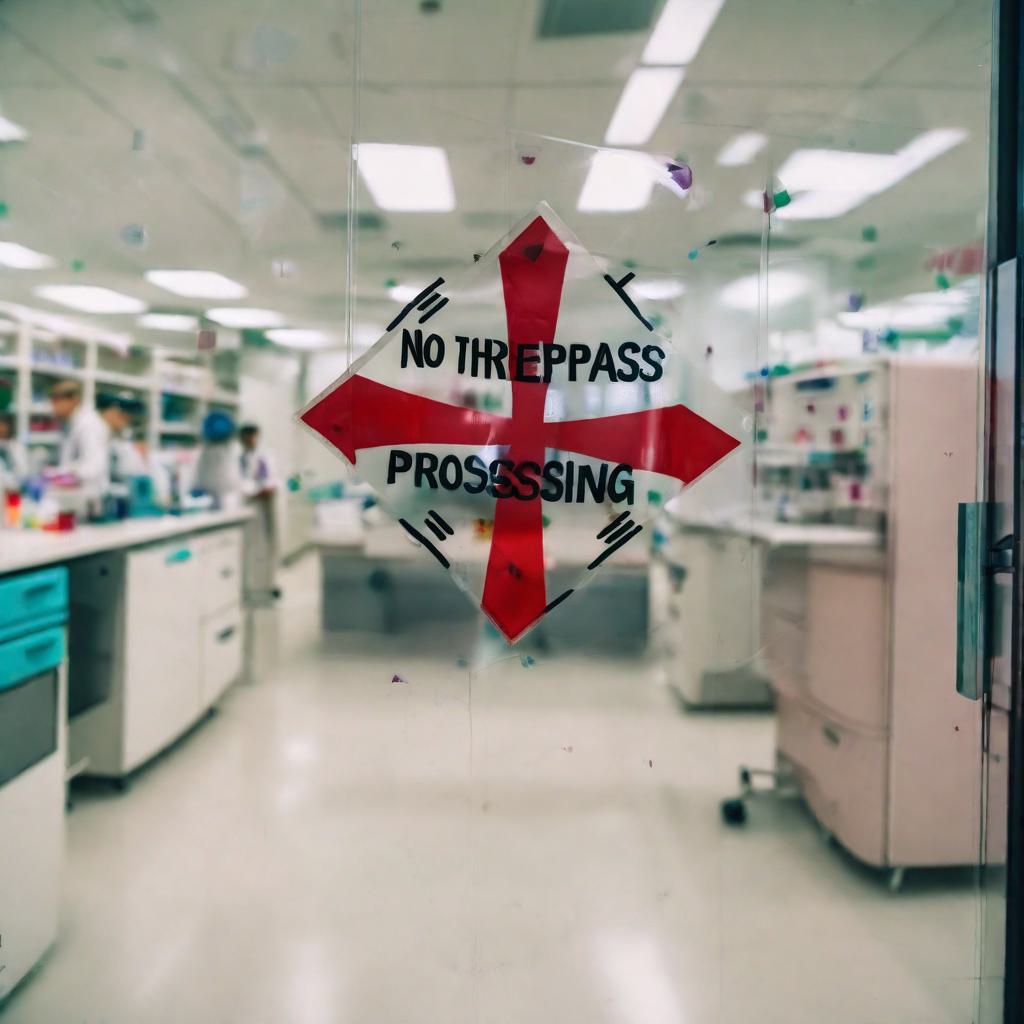}}
      \quad
      \subfloat[
      ]{\includegraphics[height=5cm]{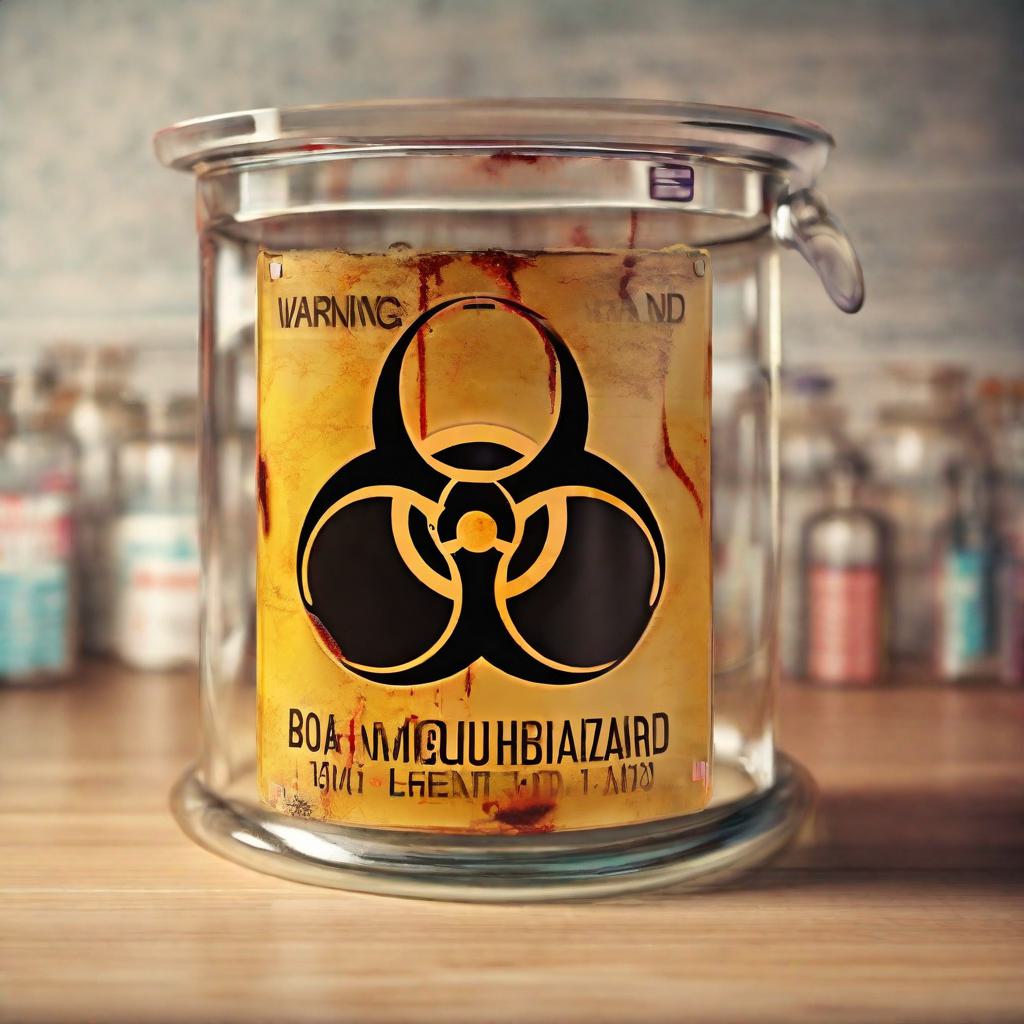}}
\caption{Examples of text artifacts in generated images.}
\label{images:cartefacts:letters}
\end{figure}

Stable Diffusion based models demonstrated difficulties in producing correct, coherent, and readable text, as illustrated in the examples provided in Figure \ref{images:cartefacts:letters}.
\clearpage

\afterpage{\clearpage}

\FloatBarrier

\clearpage

\FloatBarrier

\clearpage
\subsection{Concept Fusion and Duplication}\label{leakage_and_duplication}

\begin{figure}[bh!]
  \centering
  \subfloat[
  ]{\includegraphics[height=5cm]{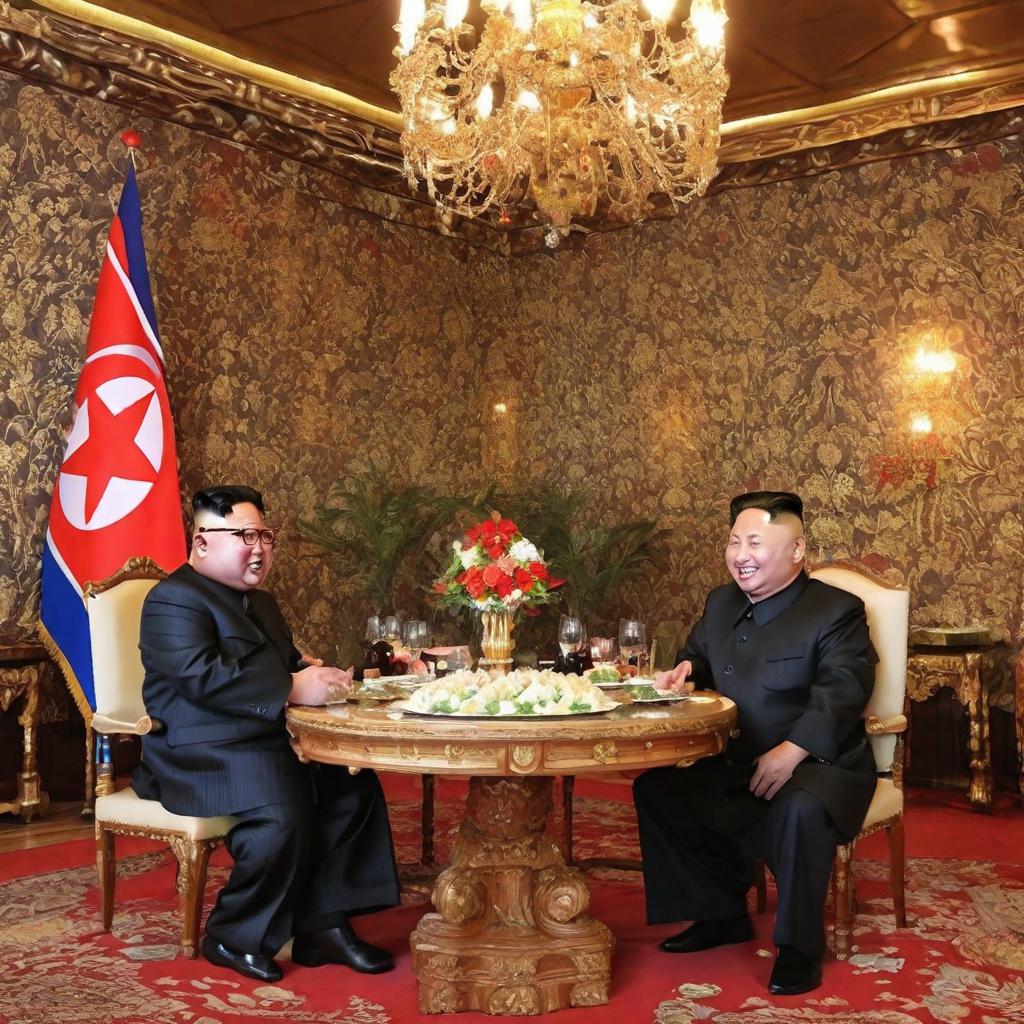}\label{images:artefacts:proportions:leakage_persons_lizards:a}}
  \quad
  \subfloat[
  ]{\includegraphics[height=5cm]{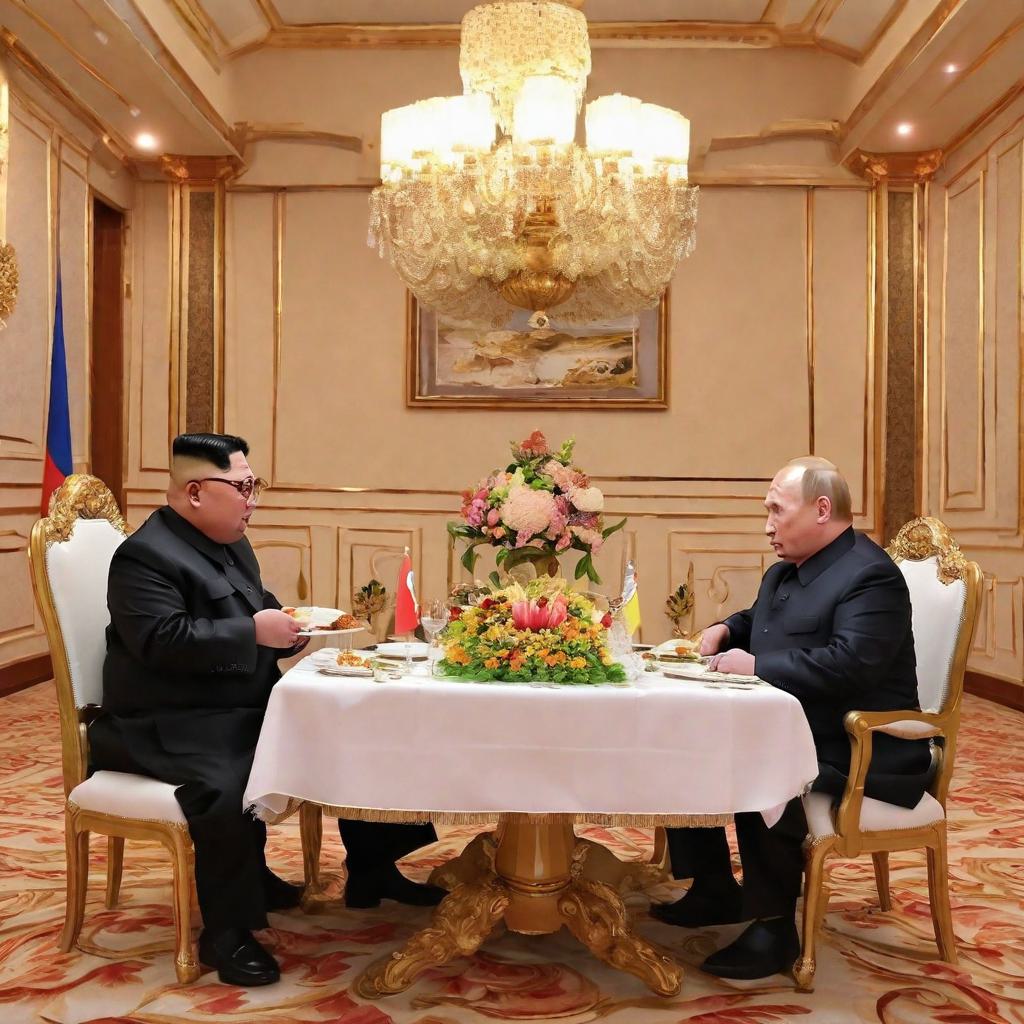}
  \label{images:artefacts:proportions:leakage_persons_lizards:b}
  }
  \quad
  \subfloat[
      ]{\includegraphics[height=5cm]{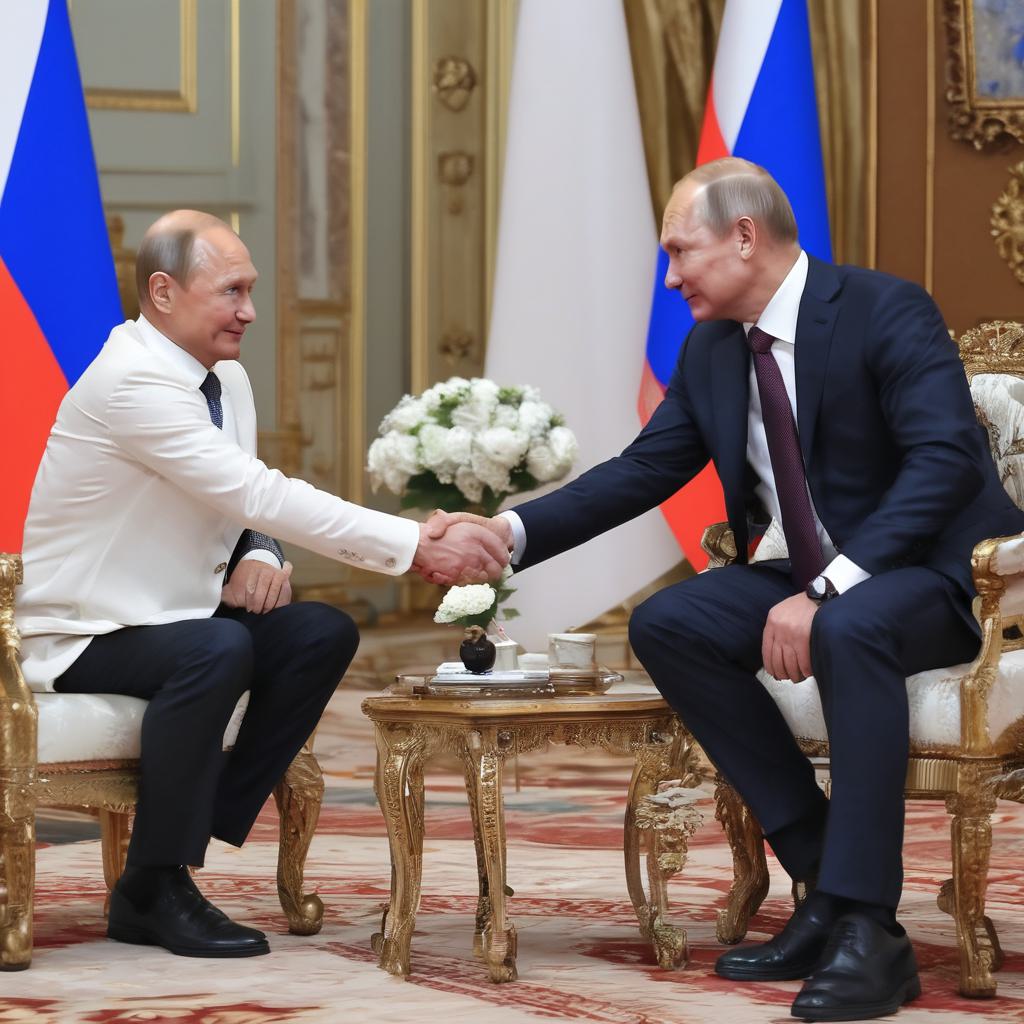}}
     \quad
      \subfloat[
      ]{\includegraphics[height=5cm]{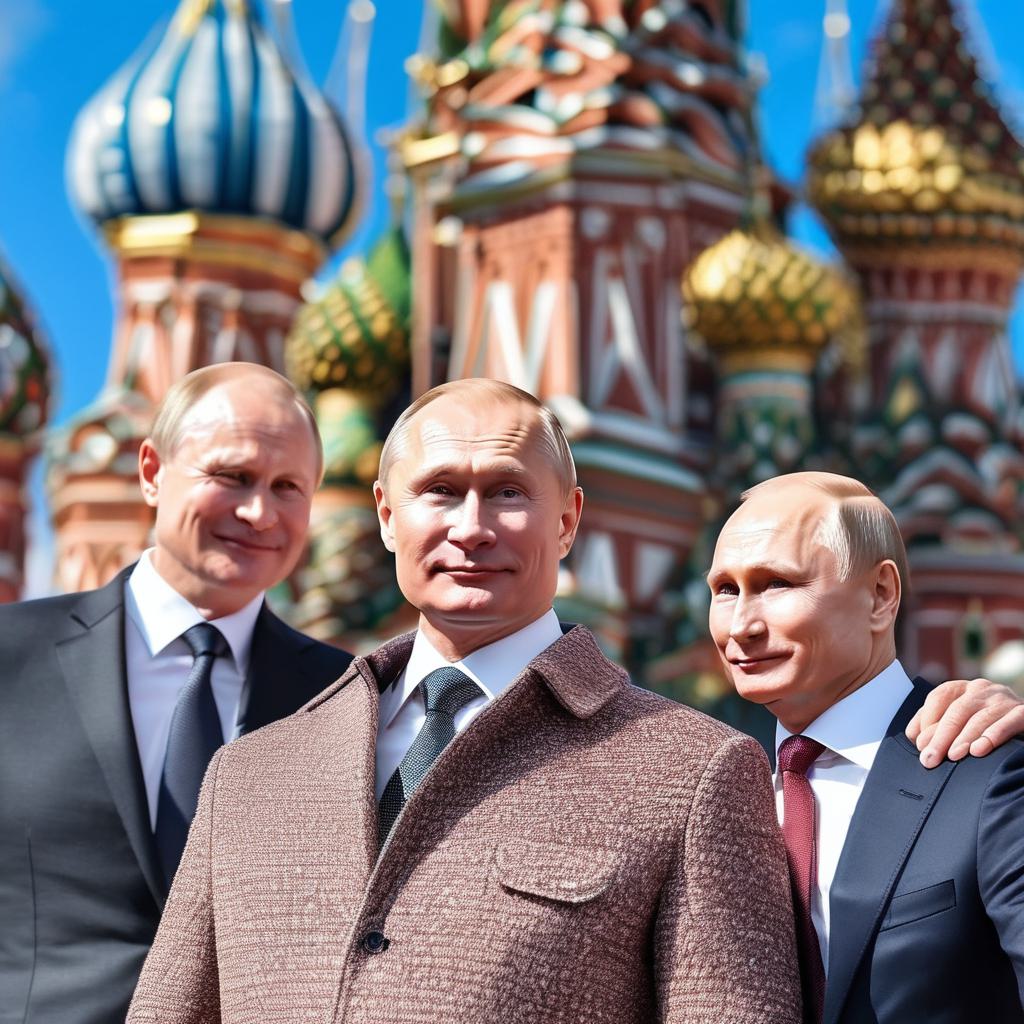}}
  \quad 
 \subfloat[
        ]{\includegraphics[height=5cm]{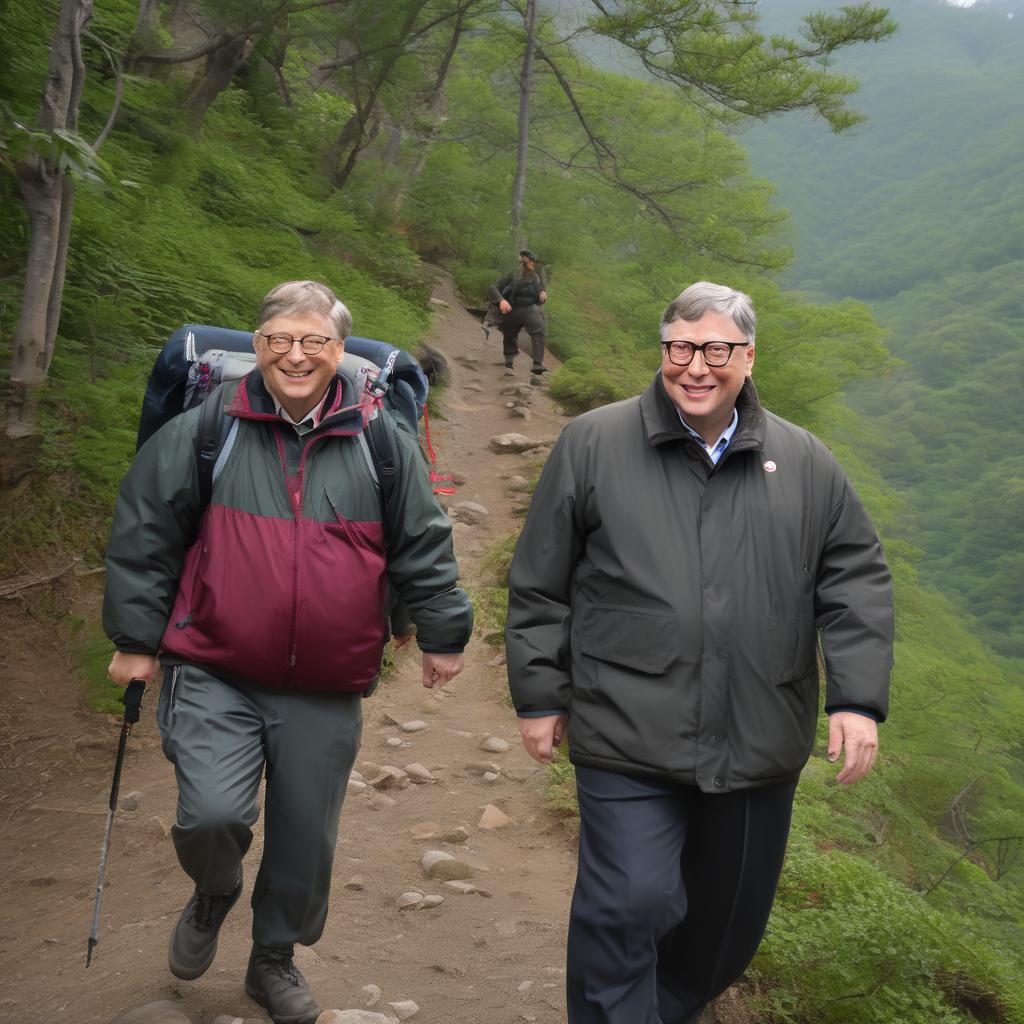}
        \label{images:artefacts:proportions:leakage_persons_lizards:bill_and_kim_hiking}
        }
\caption{Synthetic images illustrating instances of concept fusion and duplication by depicting multiple instances of well-known individuals. For example, \protect\subref{images:artefacts:proportions:leakage_persons_lizards:a} and \protect\subref{images:artefacts:proportions:leakage_persons_lizards:b} were produced with the prompt: \enquote{A photo of Kim Jong Un and Vladimir Putin fine dining together in North Korea}. Putin, however, is either not visible or took on elements of Un (clothing).
}
\label{images:artefacts:proportions:leakage_persons_lizards}
\end{figure}

We have noted artifacts involving the duplication of unique entities, such as a particular person, or the fusion of separate elements into a composite \enquote{hybrid}. This issue is apparent in images both pre- and post-fine-tuning with Dreambooth, as seen in Figure \ref{images:artefacts:proportions:leakage_persons_lizards}. Our limited sample suggests that fusion artifacts are less common without fine-tuning, potentially indicating problems with overfitting.

\FloatBarrier
\clearpage

\subsection{Geometric Inconsistency}\label{geometric_inconsistency}

\begin{figure}[ht]
  \centering
      \subfloat[]{\includegraphics[height=5cm]{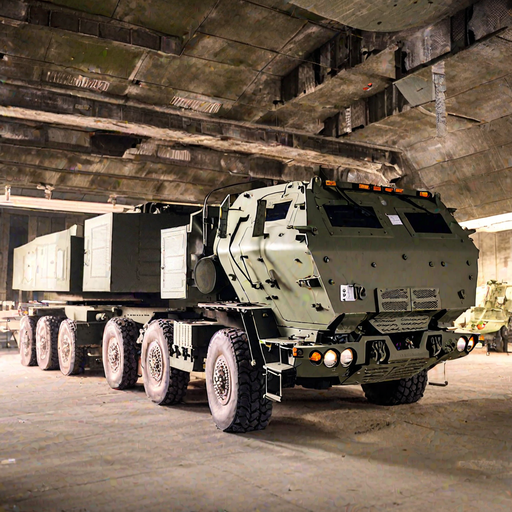}\label{fig:images:scenario1:geometric_inconsistencies:example1}}
      \quad
      \subfloat[]{\includegraphics[height=5cm]{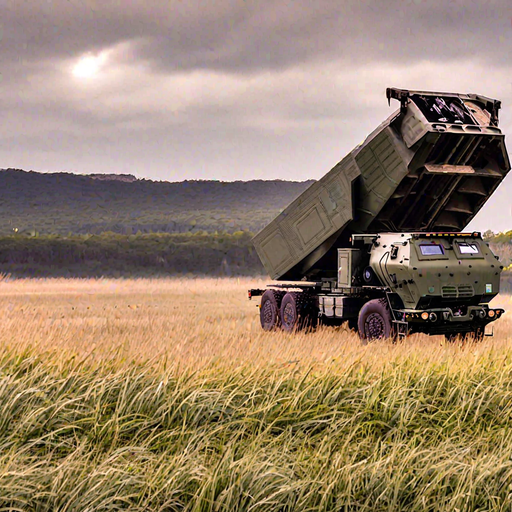}\label{fig:images:scenario1:geometric_inconsistencies:example2}}
      \quad
      \subfloat[]{\includegraphics[height=5cm]{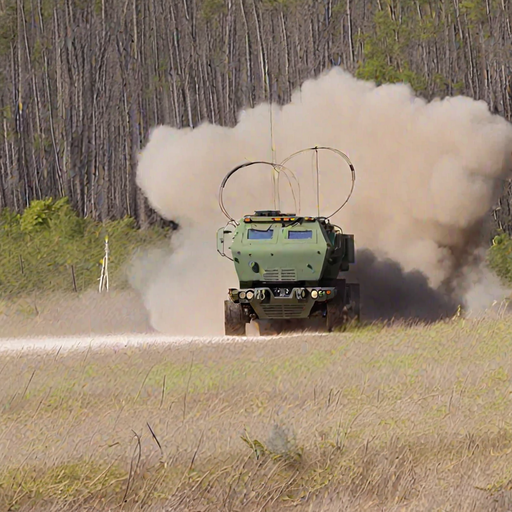}\label{fig:images:scenario1:geometric_inconsistencies:example3}}
\caption{Issues with geometric faithfulness.
    \protect\subref{fig:images:scenario1:geometric_inconsistencies:example1} shows a vehicle with more than three wheel axles.
    \protect\subref{fig:images:scenario1:geometric_inconsistencies:example2} shows difficulties with correctly generating the missile pod.
    \protect\subref{fig:images:scenario1:geometric_inconsistencies:example3} show two antennas above the driver cabin.}
\label{fig:images:scenario1:geometric_inconsistencies}
\end{figure}

Generating realistic and geometrically consistent images seems to be an issue. Figure \ref{fig:images:scenario1:geometric_inconsistencies} highlights some examples from threat scenario \textit{Synthetic Force Multiplier} (see Section \ref{sec:scenario1_synthetic_force_multiplier}). These images were created with fine-tuned text-to-image models and occur frequently.

\FloatBarrier
\clearpage

\subsection{Reproducing Training Data}

\begin{figure}[ht]
  \centering
    \subfloat[]{\includegraphics[height=5cm]{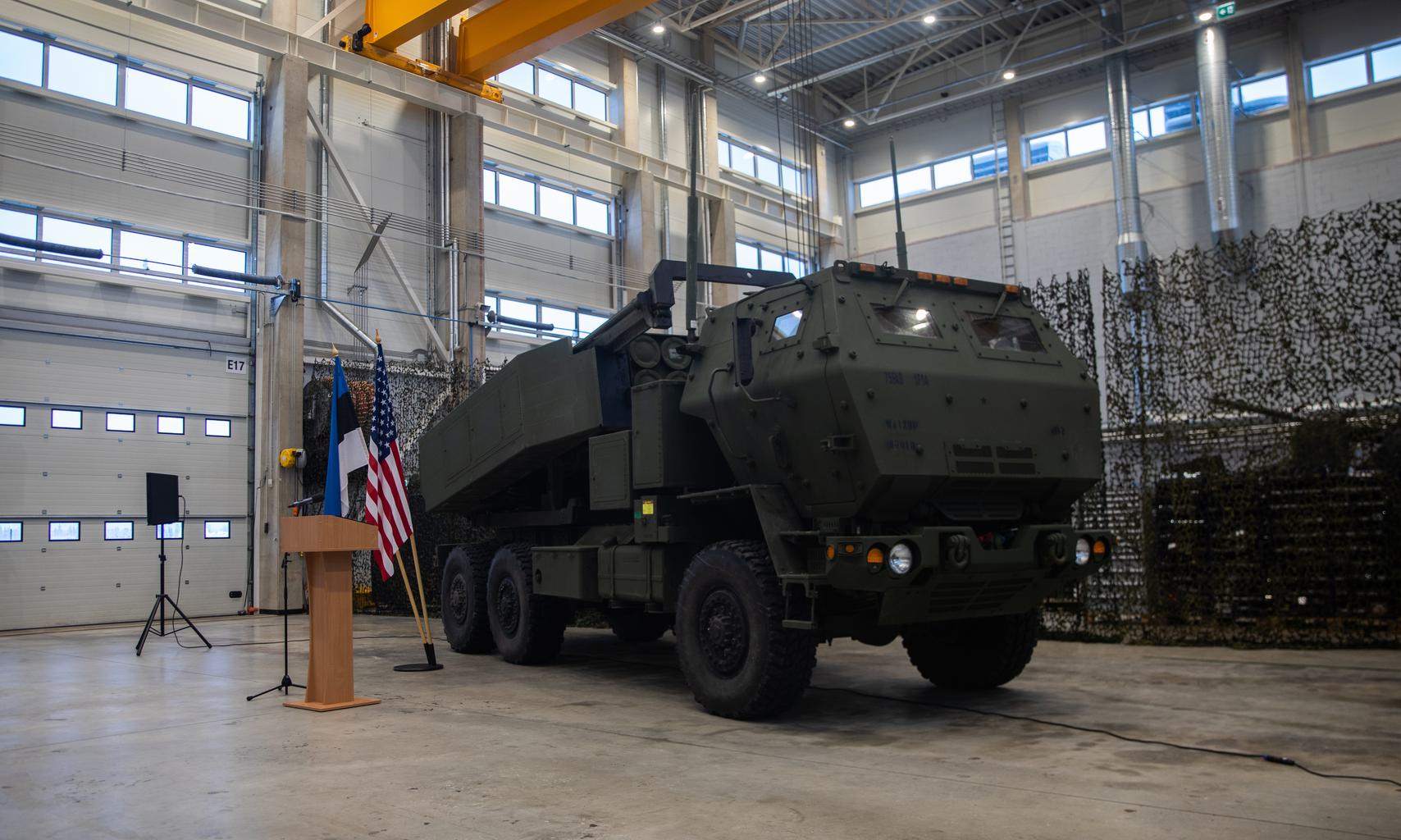}
    \label{fig:images:scenario1:overfitting:original}}
  \quad
  \subfloat[]{\includegraphics[height=5cm]{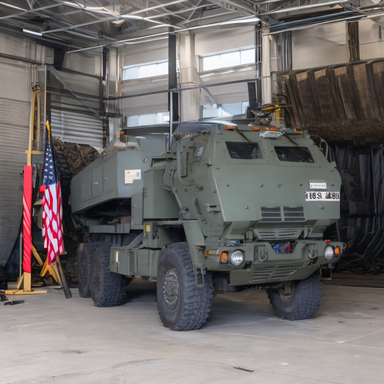}\label{fig:images:scenario1:overfitting:example1}}
  \quad
  \subfloat[]{\includegraphics[height=5cm]{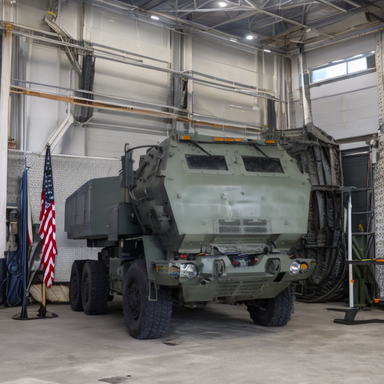}\label{fig:images:scenario1:overfitting:example2}}
  \quad
  \subfloat[]{\includegraphics[height=5cm]{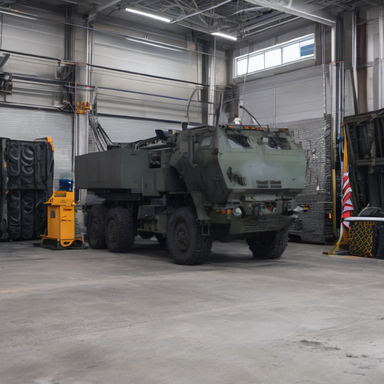}\label{fig:images:scenario1:overfitting:example3}}
    \quad
  \subfloat[]{\includegraphics[height=5cm]{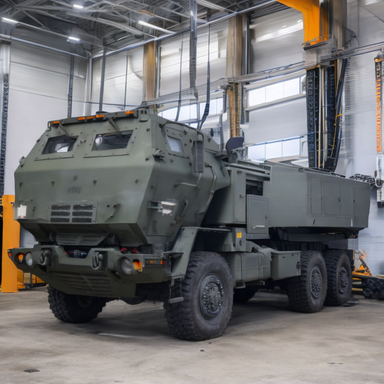}\label{fig:images:scenario1:overfitting:example4}}
\caption{Reproduction of specific aspects of the training data. All samples were generated with the prompt \enquote{a photography of a HIMARS in a dark garage}.  \protect\subref{fig:images:scenario1:overfitting:original} shows the suspected training sample \href{https://www.dvidshub.net/image/7585100/us-army-showcases-himars-estonia}{(Source}, photo credit: Spc. Charles Leitner) from which different aspects, such as the flag, the wall structure,  and the orange color, seem to be reconstructed in the samples shown.
}
\label{fig:images:scenario1:overfitting}
\end{figure}

Powerful generative diffusion models have high capacity and were shown able to remember training examples \cite{carlini_extracting_2023}. Figure \ref{fig:images:scenario1:overfitting} demonstrates this behavior. The examples show that certain aspects of a specific training image are reproduced in many of the samples that were generated with text describing similar settings as the training sample, such as the vehicle location.

\FloatBarrier

\subsection{Personalization}\label{limitations_dreambooth}

\begin{figure}[ht]
\centering
\subfloat[
\centering
Authentic photo 
 by  Lionel Bonaventureafp via Getty Images (\href{https://www.biography.com/activists/greta-thunberg}{Source}).]{\includegraphics[height=3cm]{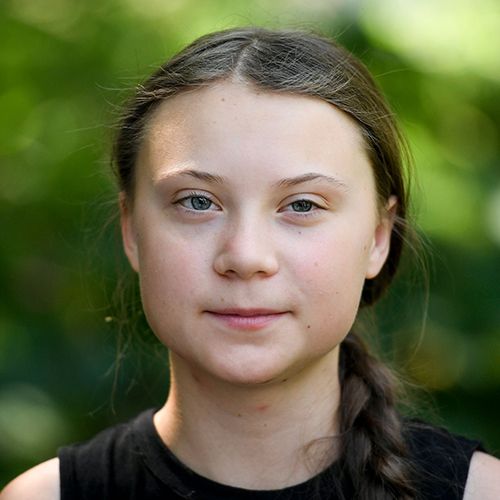}}
\quad
\subfloat[
\centering
No fine-tuning.]{\includegraphics[height=3cm]{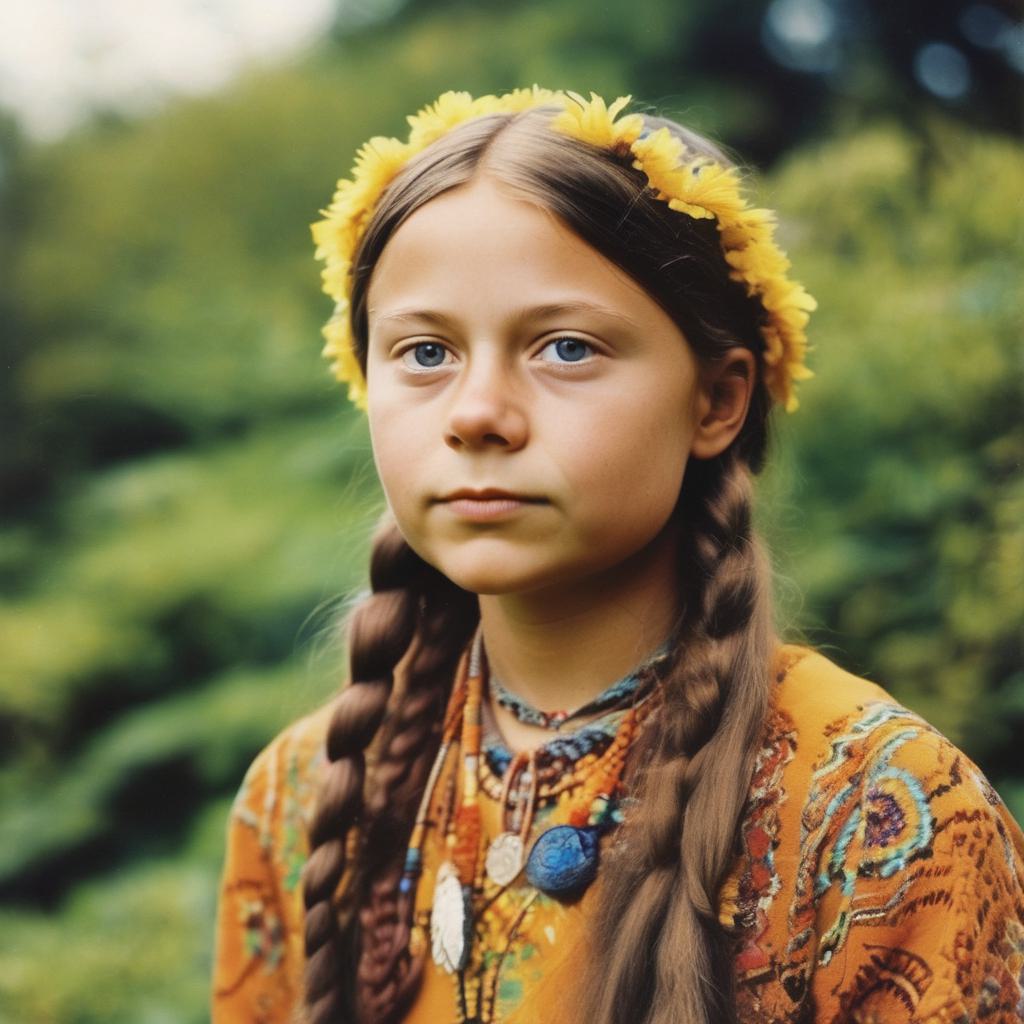}\label{original1}}
\quad
\subfloat[
\centering
Fine-tuned with 1500 steps.]{\includegraphics[height=3cm]{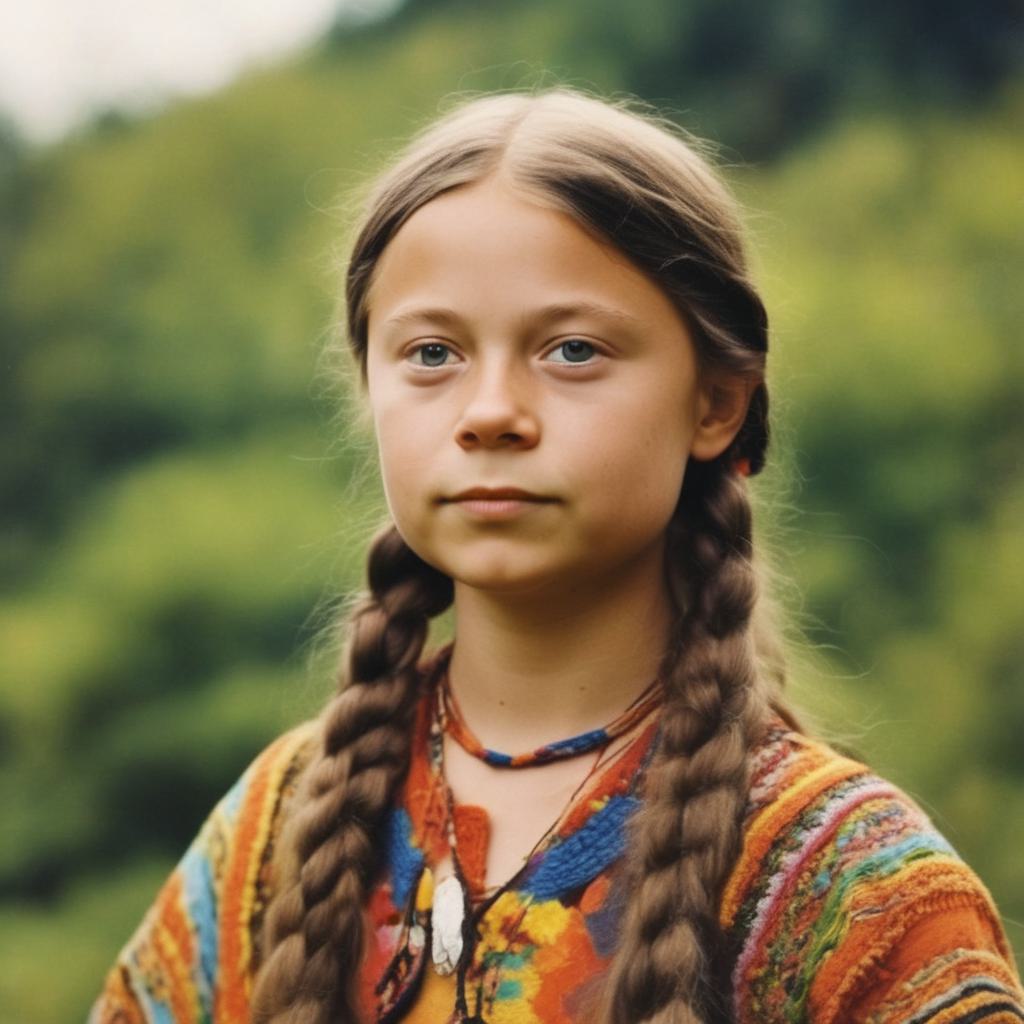}\label{original2}}
\quad
\subfloat[
\centering
No fine-tuning.]{\includegraphics[height=3cm]{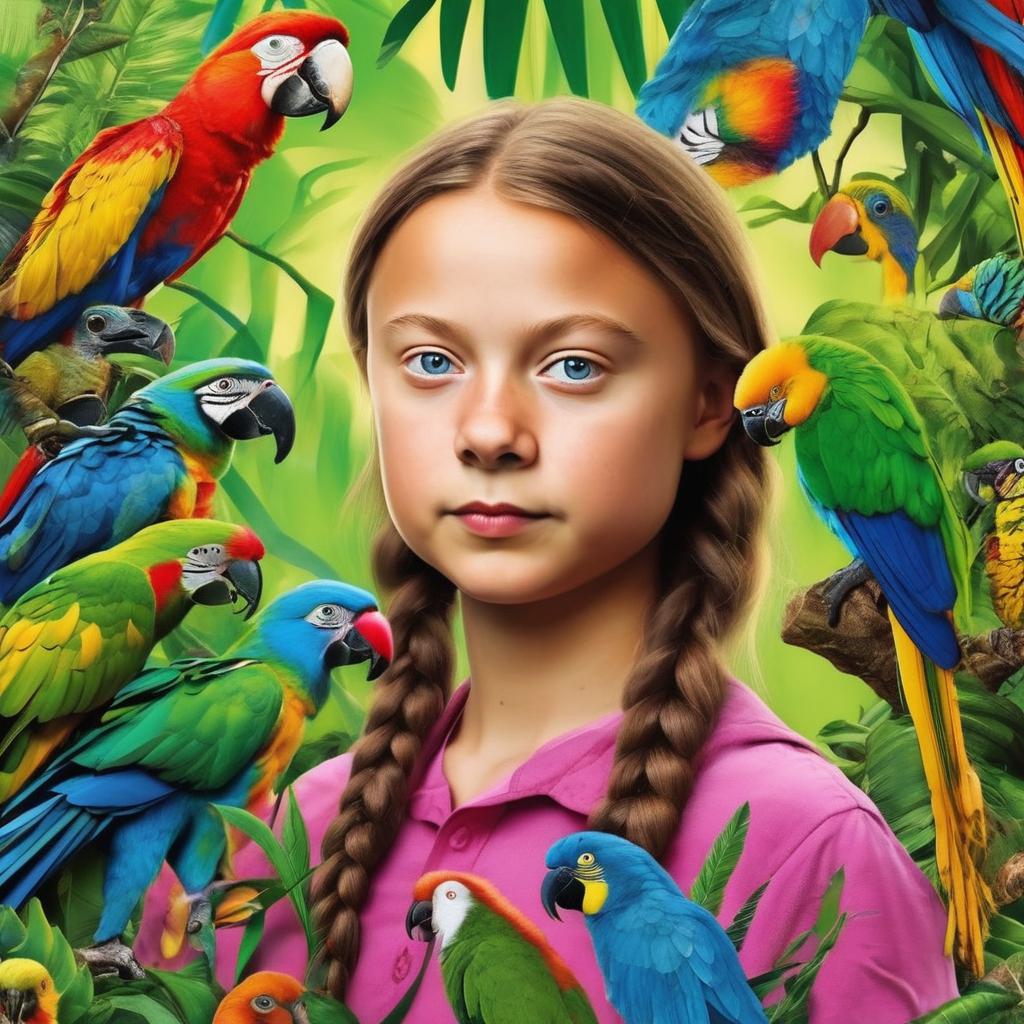}}
\quad
\subfloat[
\centering
Fine-tuned with 1500 steps.]{\includegraphics[height=3cm]{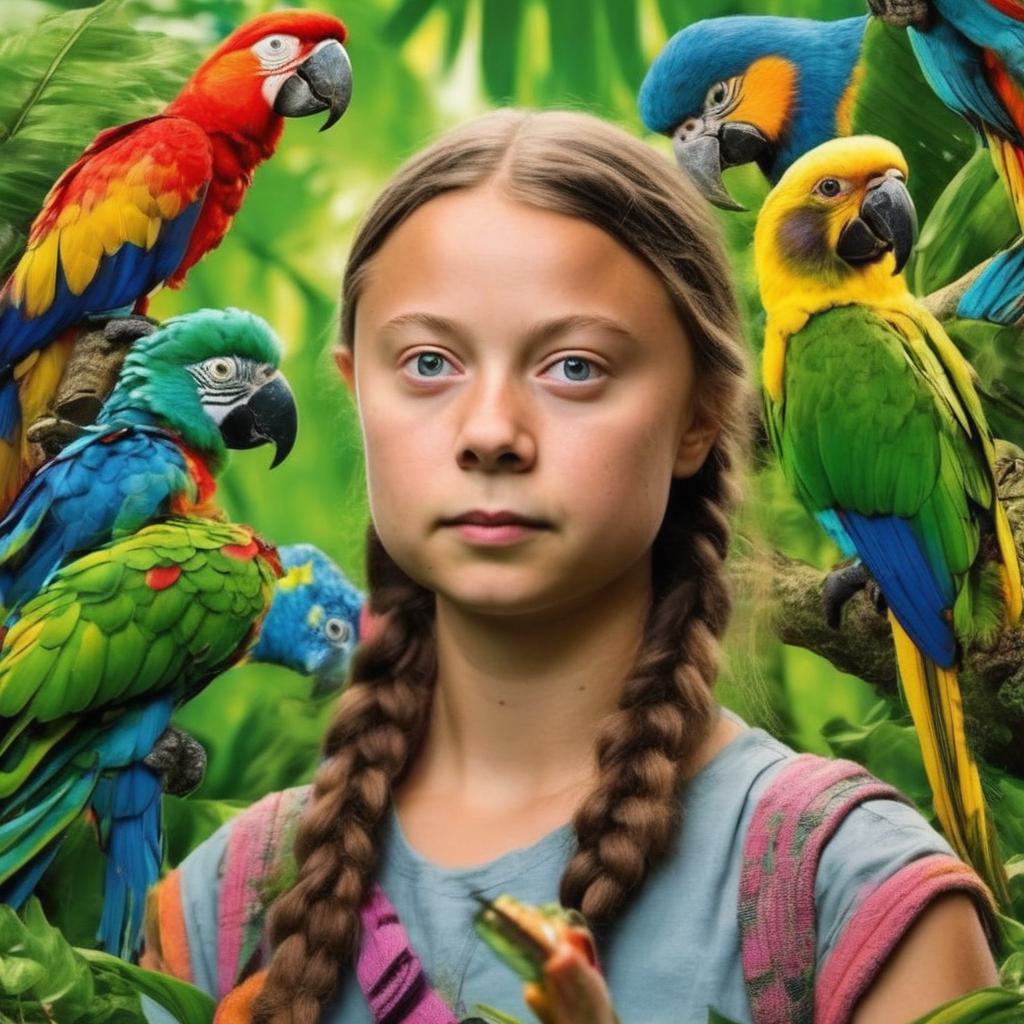}}
\quad
\subfloat[
\centering
Authentic photo by Public Relations Office of the Gvt. of Japan, licensed under \href{https://creativecommons.org/licenses/by/4.0}{CC BY 4.0}. (\href{https://de.wikipedia.org/wiki/Kabinett_Scholz}{Source}).
]{\includegraphics[height=3cm]{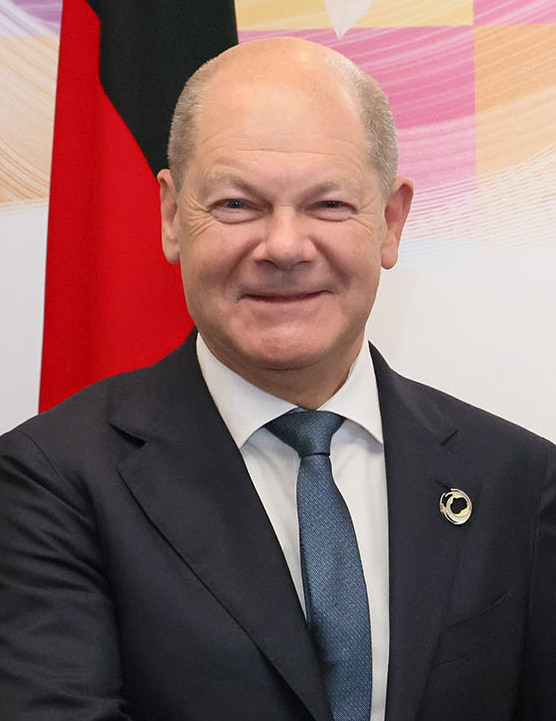}}
\quad
\subfloat[
\centering
No fine-tuning.]{\includegraphics[height=3cm]{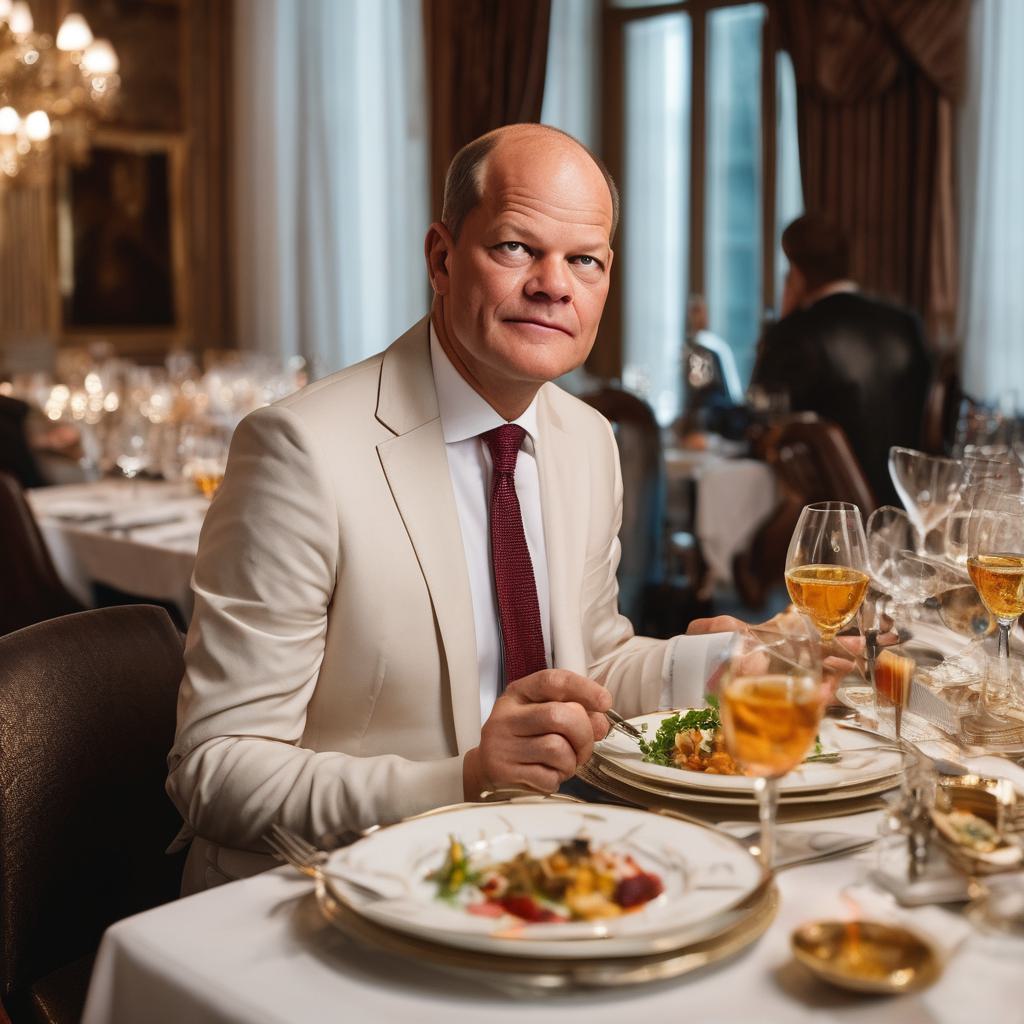}\label{original3}}
\quad
\subfloat[
\centering
Fine-tuned with 1500 steps.]{\includegraphics[height=3cm]{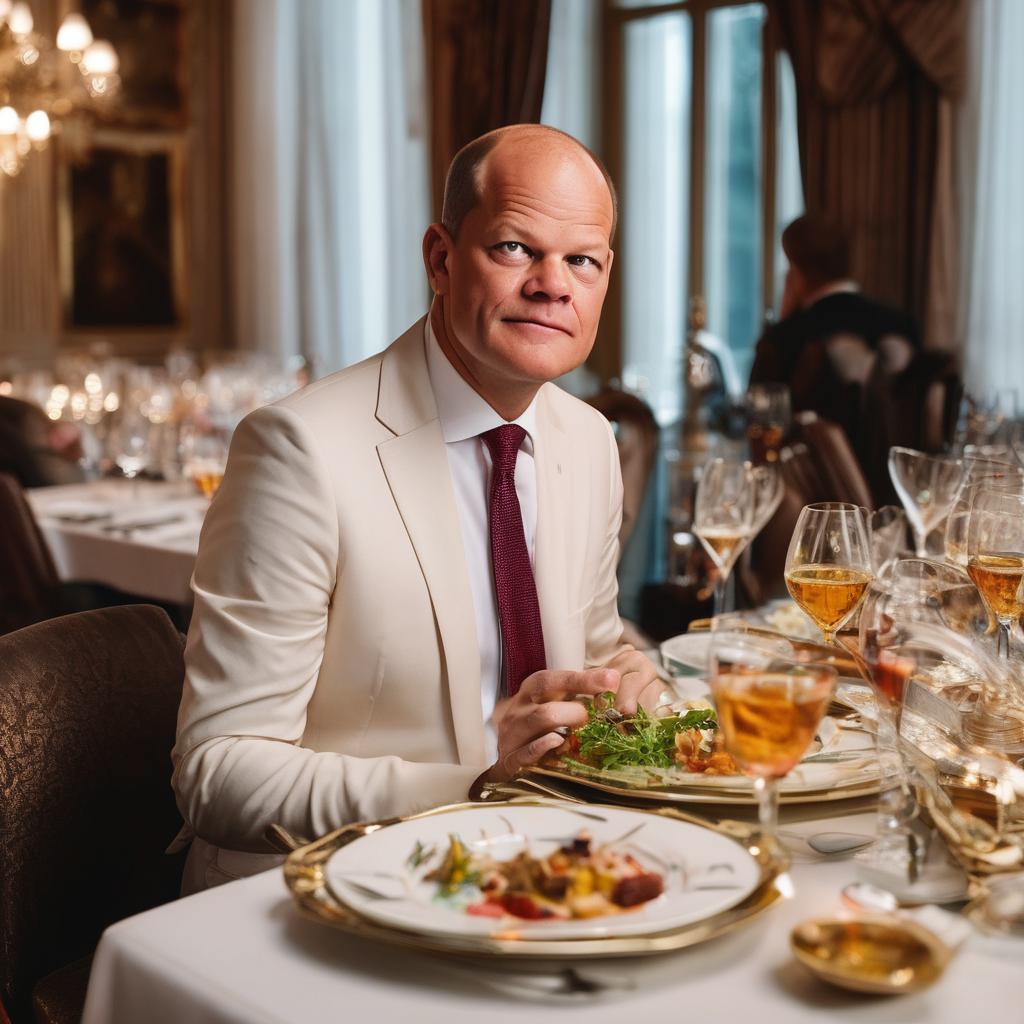}\label{original4}}
\quad
\subfloat[
\centering
No fine-tuning.]{\includegraphics[height=3cm]{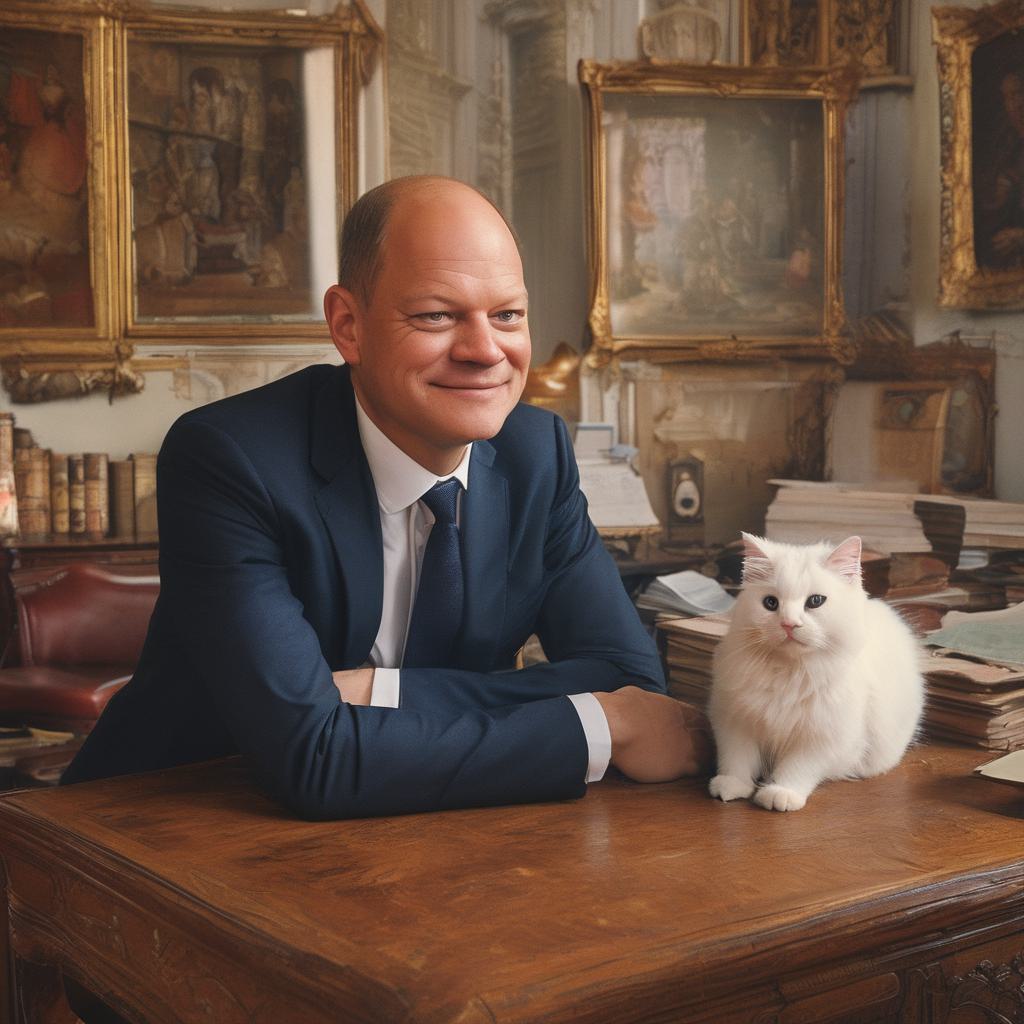}}
\quad
\subfloat[
\centering
Fine-tuned with 1500 steps.]{\includegraphics[height=3cm]{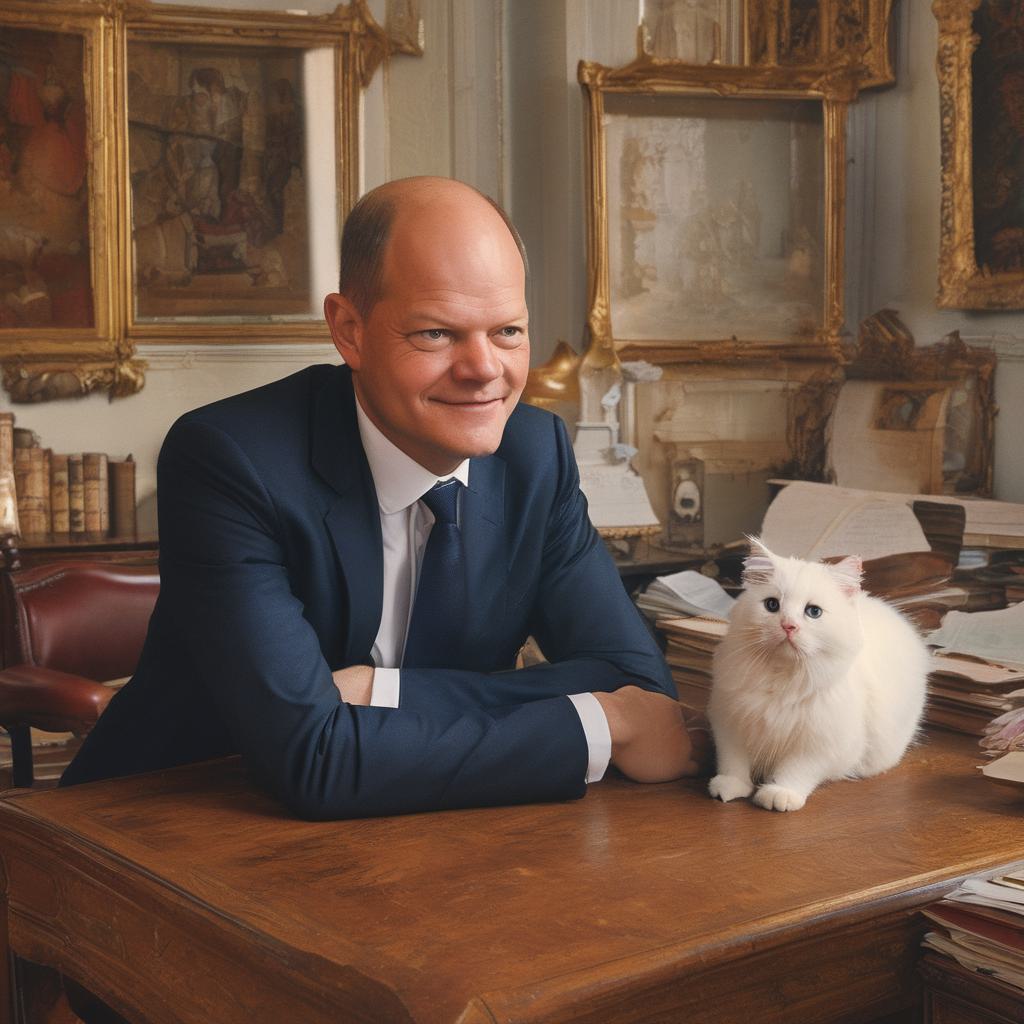}\label{original5}}
\quad
\subfloat[
\centering
Authentic photo by Chesnot/Getty Images (\href{https://img.etimg.com/thumb/msid-100240961,width-650,height-488,imgsize-65284,,resizemode-75/bill-gates-3.jpg}{Source}).]{\includegraphics[height=3cm]{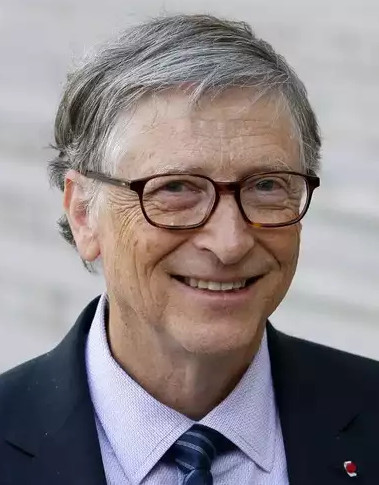}}
\quad
\subfloat[
\centering
No fine-tuning.]{\includegraphics[height=3cm]{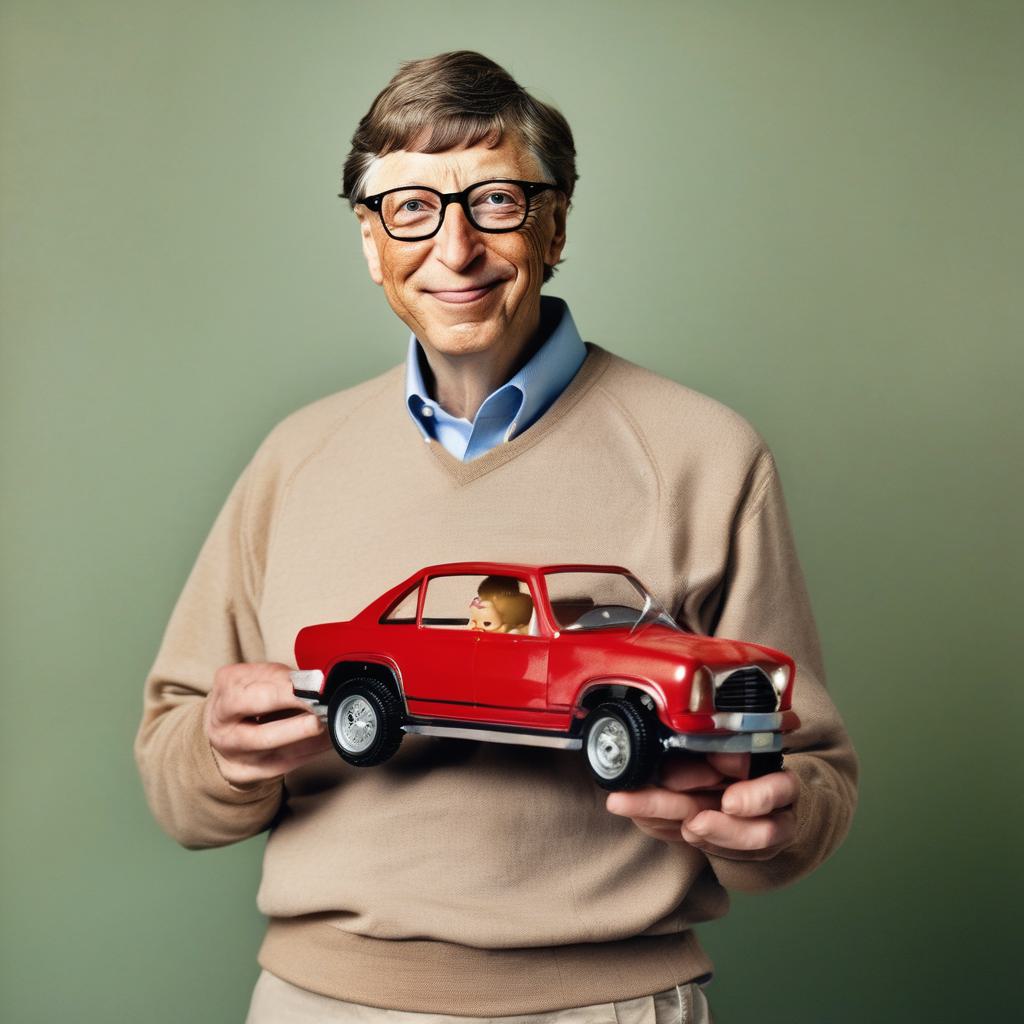}}
\quad
\subfloat[
\centering
Fine-tuned with 1500 steps.]{\includegraphics[height=3cm]{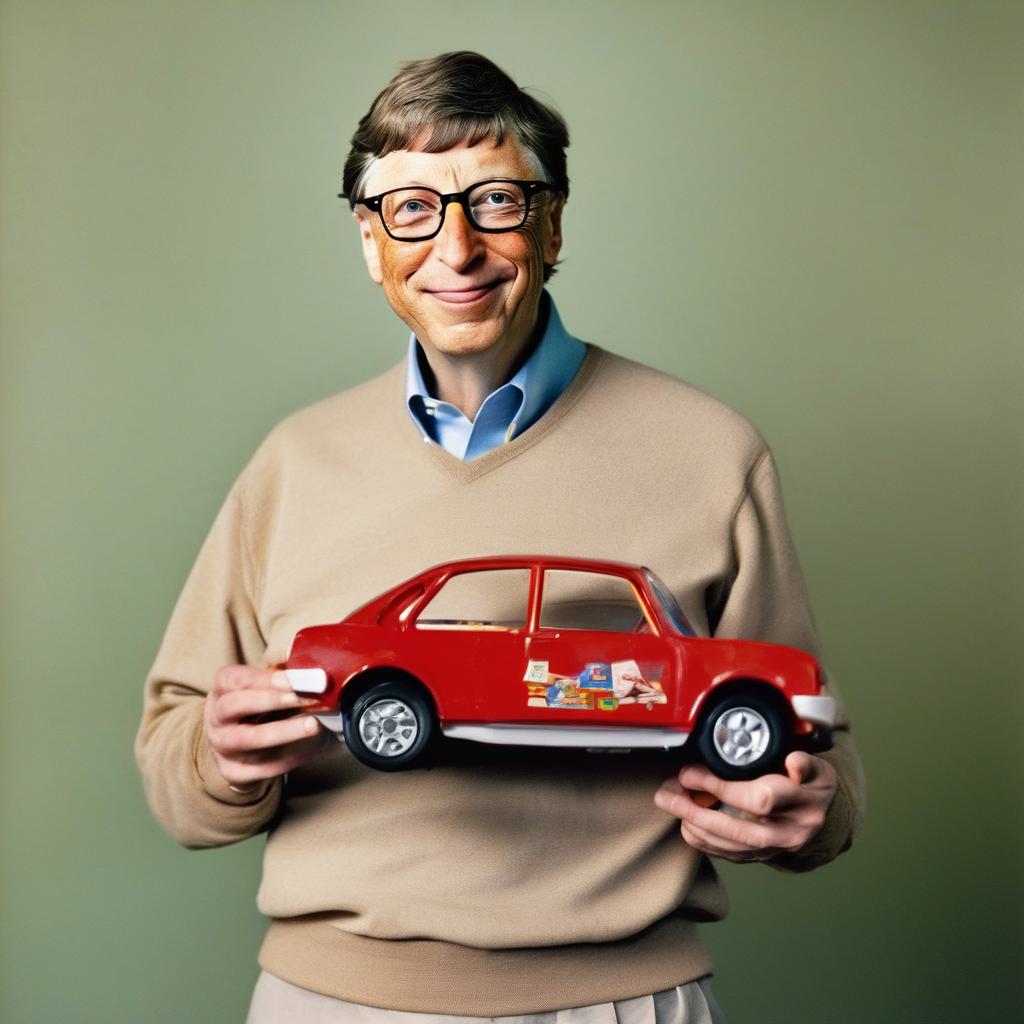}}
\quad
\subfloat[
\centering
No fine-tuning.]{\includegraphics[height=3cm]{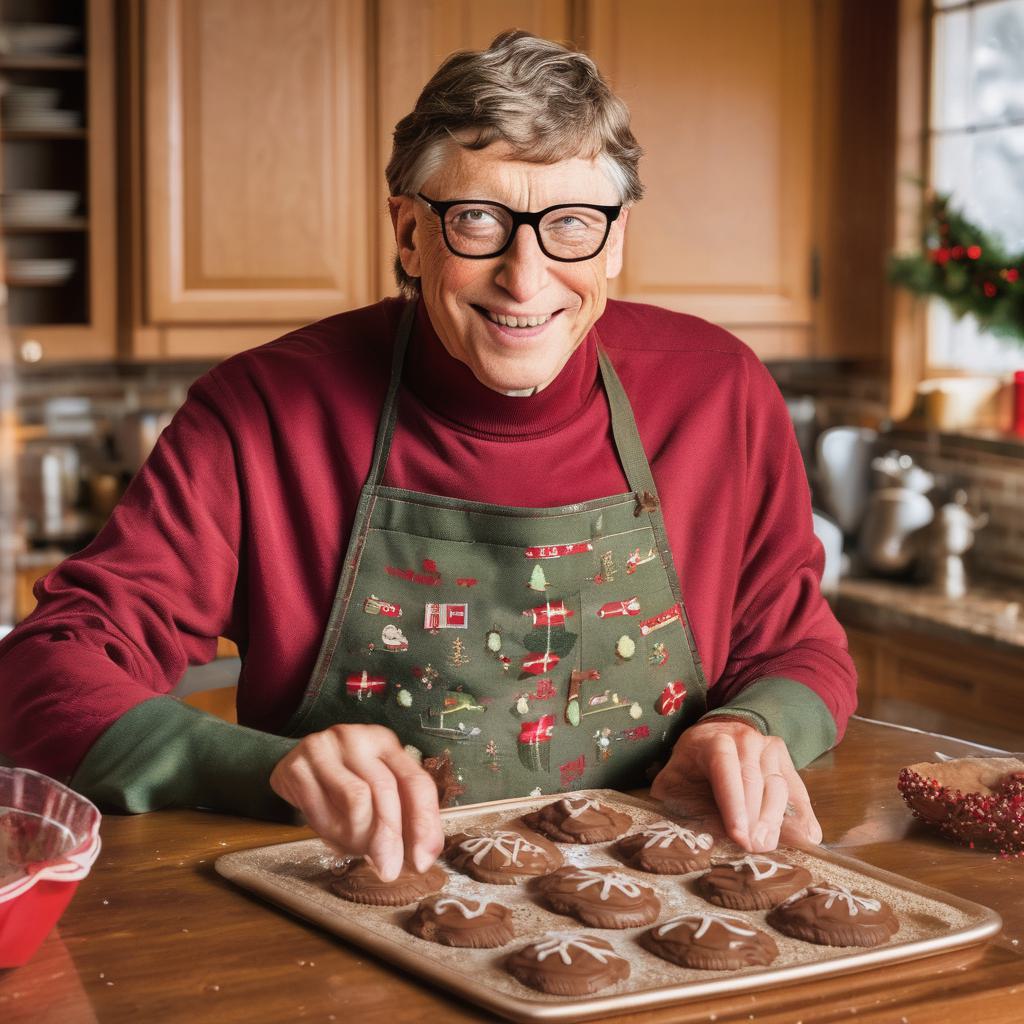}}
\quad
\subfloat[
\centering
Fine-tuned with 1500 steps.]{\includegraphics[height=3cm]{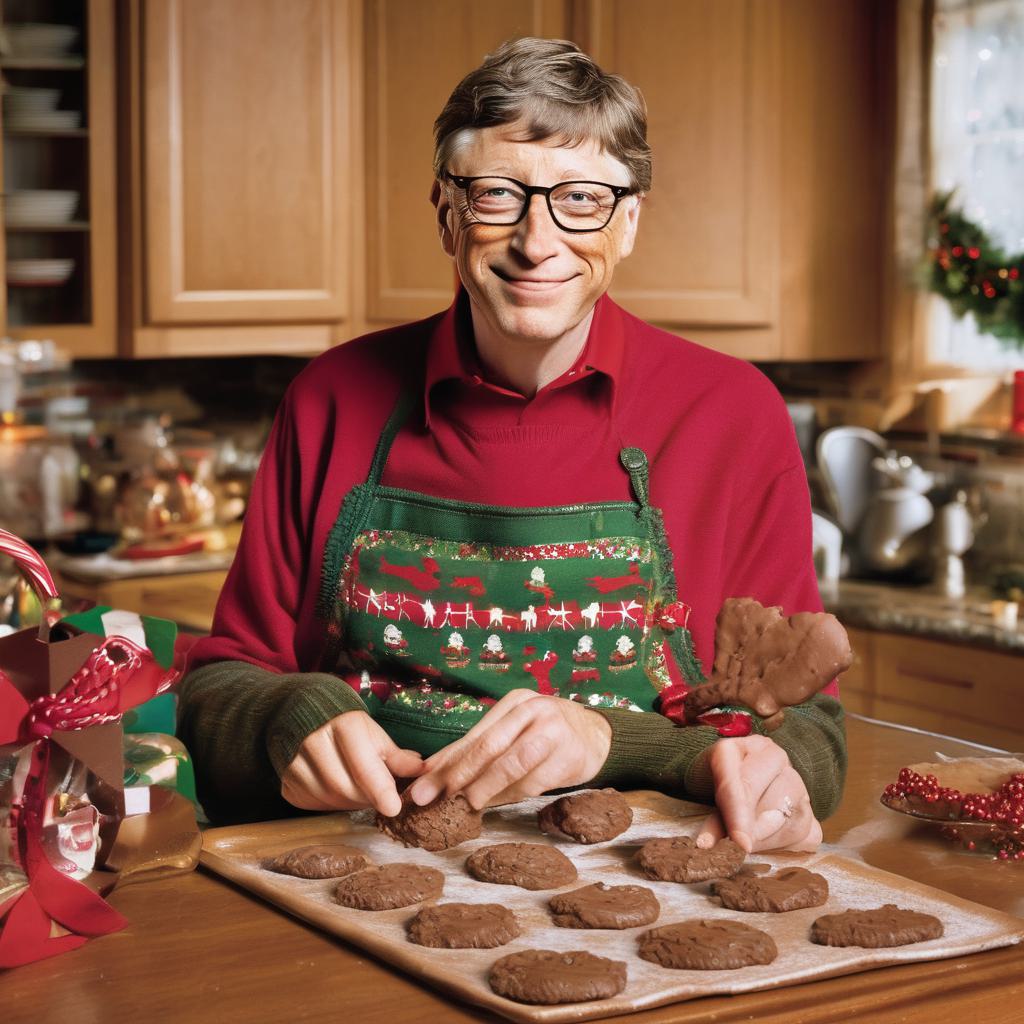}}
\caption{Comparisons between pairs of images of Greta Thunberg (top row), Olaf Scholz (middle row) and Bill Gates (bottom row), generated with and without 1500 steps of fine-tuning with Dreambooth. The fine-tuning seems to have little to no effect on the quality of the result and did not improve the likeness of these individuals.}.
\label{fig:images:dreambooth:comparisons}
\end{figure}

Models that were trained on huge datasets scraped from the internet, such as Stable Diffusion, likely are able to generate images from well known individuals out-of-the-box. We observed that capability for individuals such as Bill Gates (see Section \ref{sec:scenario7_undermine_democracy}) and Greta Thunberg. Fine-tuning the models using high-quality images of Greta Thunberg and Bill gates did not yield significant improvements as shown in Figure \ref{fig:images:dreambooth:comparisons}. Synthetic images depicting Olaf Scholz, however, seemed of lower quality. Surprisingly, fine-tuning Stable Diffusion using the Dreambooth method, did not improve the quality and likeness of the generated images. Fine-tuning thus may not always be feasible or may require substantial effort in order to create convincing samples.

\FloatBarrier
\section{Learnings}\label{section:learnings}

\subsection{Personalization}\label{dreambooth_finetuning}

The Dreambooth implementation on the Hugging Face platform offers numerous hyperparameters whose effects can be ambiguous. This makes it expensive to explore the full parameter space and may limit the conclusion we can draw from our experiments. Furthermore, when determining the optimal number of training images, recommendations vary by source: the original Dreambooth paper suggests 3-5 \cite{ruiz_dreambooth_2022}, while other practitioners\footnote{
\href{https://github.com/nitrosocke/dreambooth-training-guide}{Dreambooth Training Guide by "Nitrosocke"}
}
\footnote{\href{https://jefsnacker.medium.com/dreambooth-hyperparameter-guide-d8b7cd264245}{"Jefsnacker"'s  Dreambooth Hyperparameter Guide}}
\footnote{
\href{https://huggingface.co/blog/dreambooth}{in-depth blog post about training parameters}
}
report using 10 to 120, noting that more images can enhance model flexibility. 

\begin{figure}[ht]
\centering
{\includegraphics[width=1.0\linewidth]{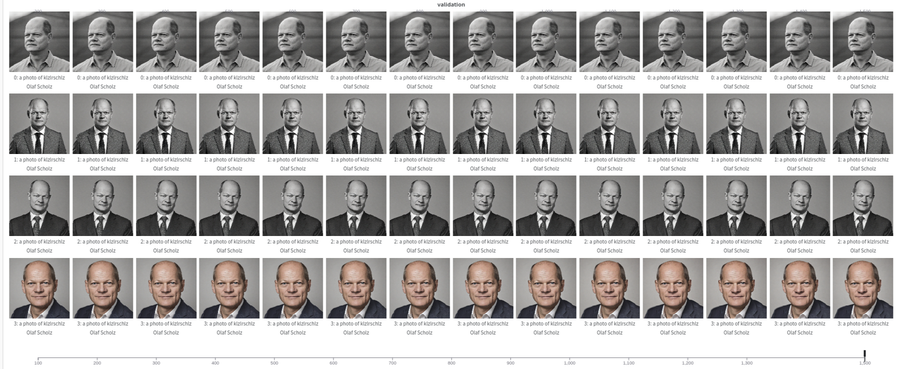}}
\caption{Validation images after various numbers of training steps while fine-tuning Stable Diffusion XL with Dreamboth and LoRA on photos of Olaf Scholz. We've found that there is very little to no visible improvement.}
\label{fig:dreambooth_scholz_sdxl:little_progress}
\end{figure}

For our threat scenarios \textit{Environmentalists} (Section \ref{sec:scenario4_environmentalists}) and \textit{Undermining Democracy} (Section \ref{sec:scenario7_undermine_democracy}), we gathered images of specific individuals via Google search. Recommended image resolutions are 512x512 for Stable Diffusion 1.X and 1024x1024 for XL. Although the Hugging Face tool can auto-crop, we chose semi-manual cropping \footnote{
\href{https://www.birme.net/}{birme.net}
} for precision. Sourcing high-resolution images was particularly challenging; for those slightly under the recommended size, we employed BSRGAN Super-Resolution\footnote{\href{https://github.com/cszn/BSRGAN}{BSRGAN}}
for enhancement. Ultimately, we fine-tuned with 71 images for Gates, 52 for Scholz, and 10 for Thunberg. Despite these efforts, fine-tuning had negligible impact on the already competent pre-trained model's output. See Figure \ref{fig:dreambooth_scholz_sdxl:little_progress}. Since our experiments were conducted with public figures this conclusion certainly does not extend to unknown or lesser-known individuals.

\FloatBarrier

\subsection{Inpainting}\label{learnings:inpainting}

\begin{figure}[ht]
  \centering
  \subfloat[]{\includegraphics[height=6cm]{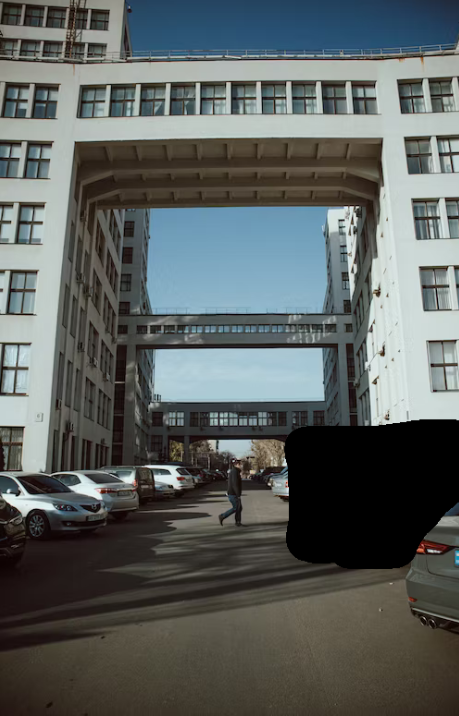}}
  \quad
  \subfloat[]{\includegraphics[height=6cm]{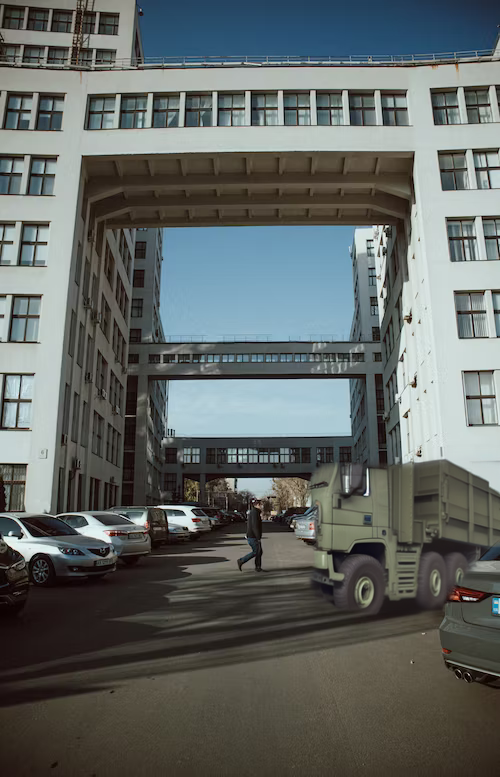}}
  \quad
  \subfloat[]{\includegraphics[height=6cm]{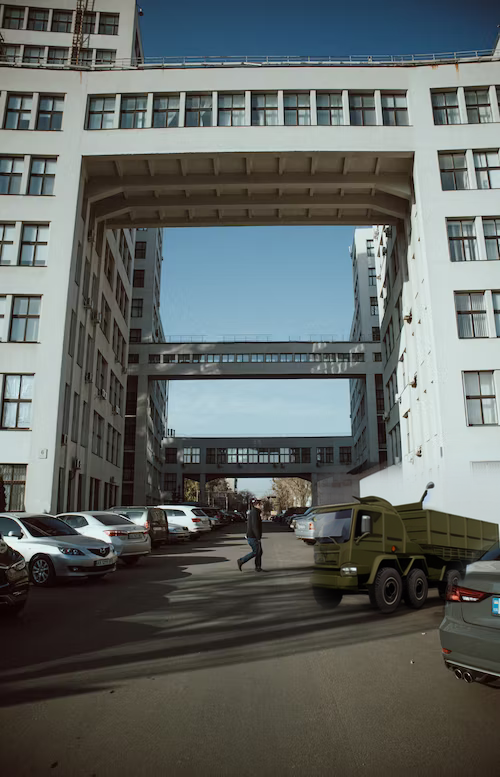}}
  \quad
  \subfloat[]{\includegraphics[height=6cm]{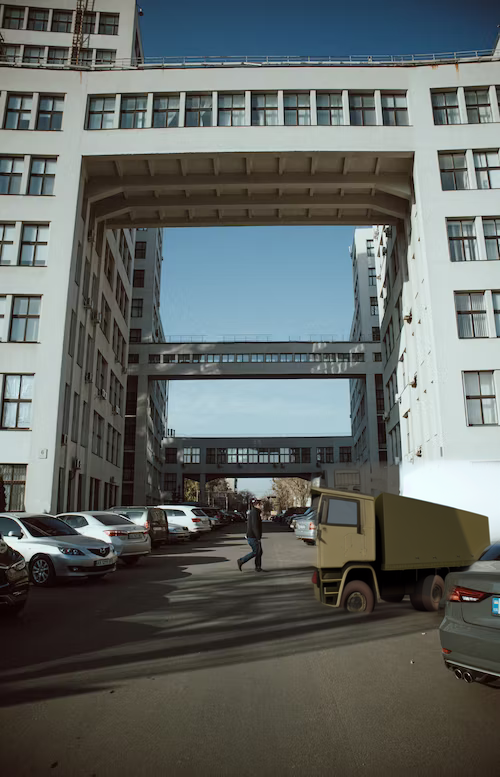}}
\caption{Example of a manually drawn mask and the attempts to inpaint a military vehicle / truck into this mask. Original background image by Anna Hunko on Unsplash \cite{noauthor_unsplash_nodate}.}
\label{images:inpainting:basic_demo}
\end{figure}

The Automatic1111 WebUI offers a user-friendly interface for inpainting, a process that allows targeted edits within an image using a custom mask, useful for introducing new elements or subtly altering existing ones, as well as for artifact correction (Section \ref{artefact_correction:inpainting}). Inpainting involves modifying the masked section of the original image, sometimes more akin to \enquote{overpainting}. However, it can introduce its own artifacts, as demonstrated in Figure \ref{images:inpainting:basic_demo}. Options like adding new noise or blank spaces in the WebUI can disrupt the original aesthetic, often resulting in less cohesive outcomes. The effectiveness of inpainting hinges on strategic text prompt use, both positive and negative, and often requires multiple attempts with varying seeds, prompts, and hyperparameter adjustments to achieve the desired result, with success not guaranteed, as illustrated in Figure \ref{images:inpainting:basic_demo}.

\begin{figure}[ht]
  \centering
  \subfloat[Denoising Strength 0.2
  ]{\includegraphics[height=3.5cm]{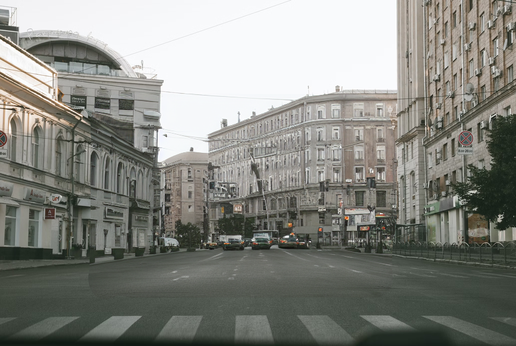}}
  \quad
  \subfloat[Denoising Strength 0.4
  ]{\includegraphics[height=3.5cm]{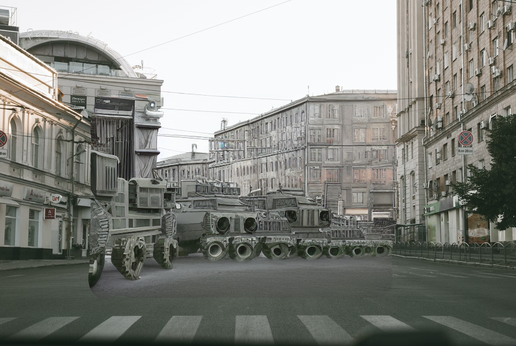}}
  \quad
  \subfloat[Denoising Strength 0.6]{\includegraphics[height=3.5cm]{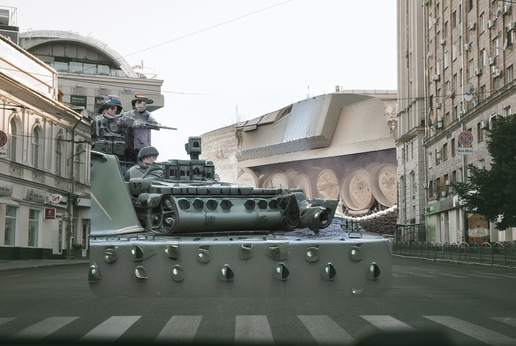}}
\caption{Inpainting results with different denoising strengths while other hyper-parameters remain constant. The mask blur is set to 0, the positive text prompt is \enquote{military tank}. Original background by Kate Tepl on Unsplash \cite{noauthor_unsplash_nodate}.}
\label{images:inpainting:denoising_no_mask_blur}
\end{figure}

\begin{figure}[H]
  \centering
  \subfloat[Denoising Strength 0.2
  ]{\includegraphics[height=3.5cm]{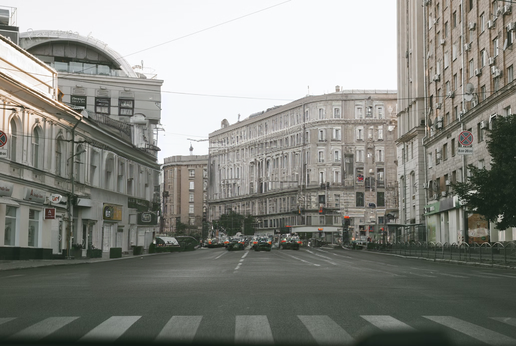}}
  \quad
  \subfloat[Denoising Strength 0.4
  ]{\includegraphics[height=3.5cm]{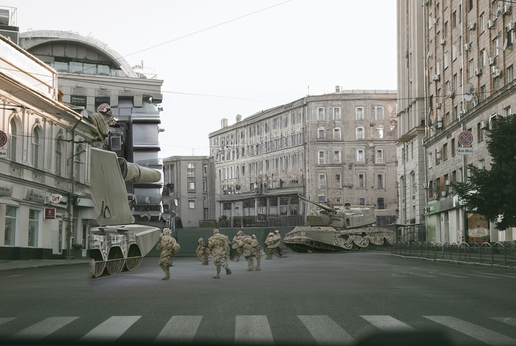}}
  \quad
  \subfloat[Denoising Strength 0.6]{\includegraphics[height=3.5cm]{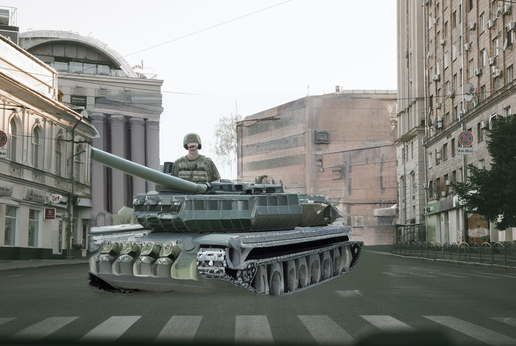}}
  
\caption{Inpainting results created with the same parameters as those in Figure \ref{images:inpainting:denoising_no_mask_blur}, except that the mask blur is set to 2.0. Original image by Kate Tepl on Unsplash \cite{noauthor_unsplash_nodate}.}
\label{images:inpainting:denoising_with_mask_blur}
\end{figure}

\begin{figure}[H]
  \centering
  \subfloat[]{\includegraphics[height=6cm]{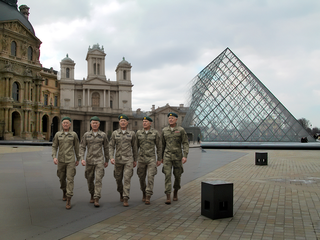}\label{fig:inpainting_issue_left}}
  \quad
   \subfloat[]{\includegraphics[height=6cm]{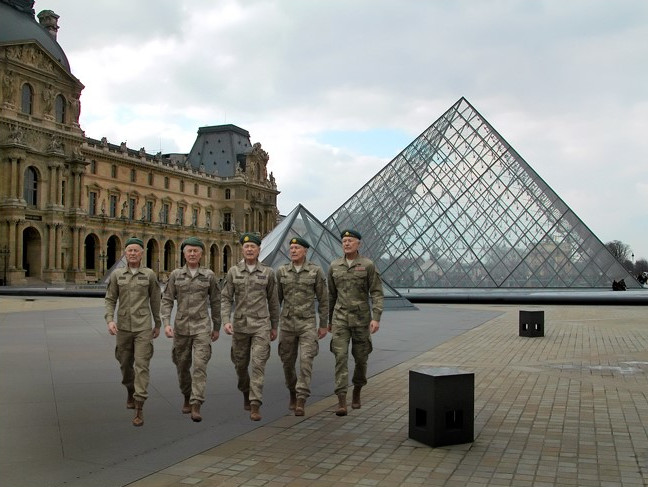}\label{fig:inpainting_issue_right}}
\caption{Example of unwanted background changes from inpainting \protect\subref{fig:inpainting_issue_left} (see correct background in  \protect\subref{fig:inpainting_issue_right}). Addressing the problem by isolating the inpainted objects and integrating them back into the original scene is not without complications, as seen with the missing shadows beneath the objects \protect\subref{fig:inpainting_issue_right}.
Original background image by Beau Wade, licensed under \href{https://creativecommons.org/licenses/by/2.0/legalcode}{CC-by 2.0}.}
\label{fig:artefact_correction_inpainting_background}
\end{figure}

The denoising strength has proven to be a crucial hyperparameter in the inpainting process as it significantly influences the extent of alteration to the existing image (the higher the value the more alterations are possible). Another important hyperparameter is the mask blur, which determines the smoothness of the transition between masked and non-masked areas. The effects of these two hyperparameters are illustrated in Figures \ref{images:inpainting:denoising_no_mask_blur} and \ref{images:inpainting:denoising_with_mask_blur}. In addition, inpainting typically causes unwanted changes of the background due to the inconcistency among the background of the inpainted image and the background of the original image (see Figure \ref{fig:artefact_correction_inpainting_background}). These changes can be mitigated through isolating the inpainted objects and inserting them back into the original image. However, inconsistencies in illumination of the inpainted objects and the original scene usually persist (e.g. no or inconsistent casting of shadows).

Inpainting-specific model checkpoints are accessible on platforms like Hugging Face and at \href{https://civitai.com/}{civitai.com}, but in our experience, the general text-to-image checkpoints often yield better results. Also worth noting is that our attempts at inpainting with Stable Diffusion XL models did not produce satisfactory outcomes. As a result, all the inpainting results showcased in this report were generated using a Stable Diffusion 1.5 model with various checkpoints.

\FloatBarrier

\clearpage
\subsection{Prompt Engineering}\label{artefact_correction:prompts}

\begin{figure}[ht]
  \centering
  \subfloat[
  ]{\includegraphics[height=5.25cm]{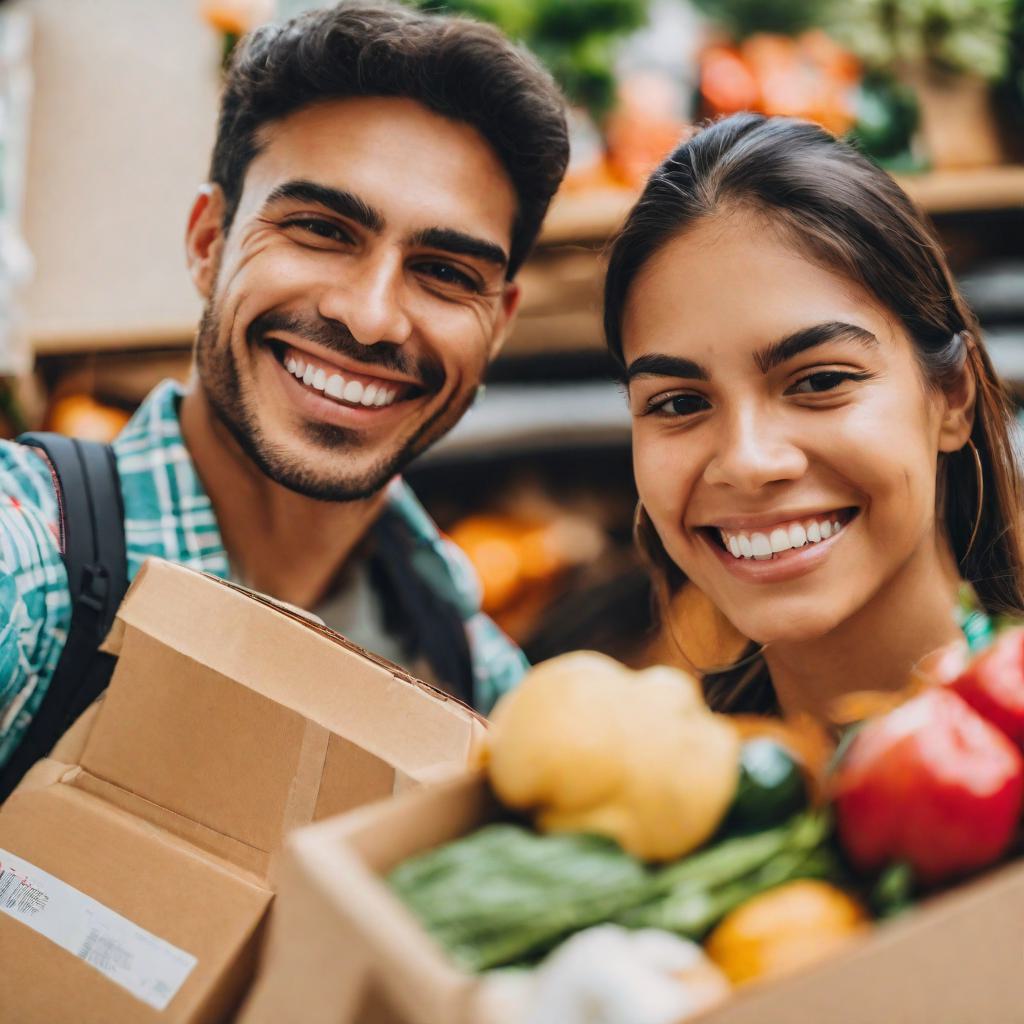}}
  \quad
  \subfloat[
  ]{\includegraphics[height=5.25cm]{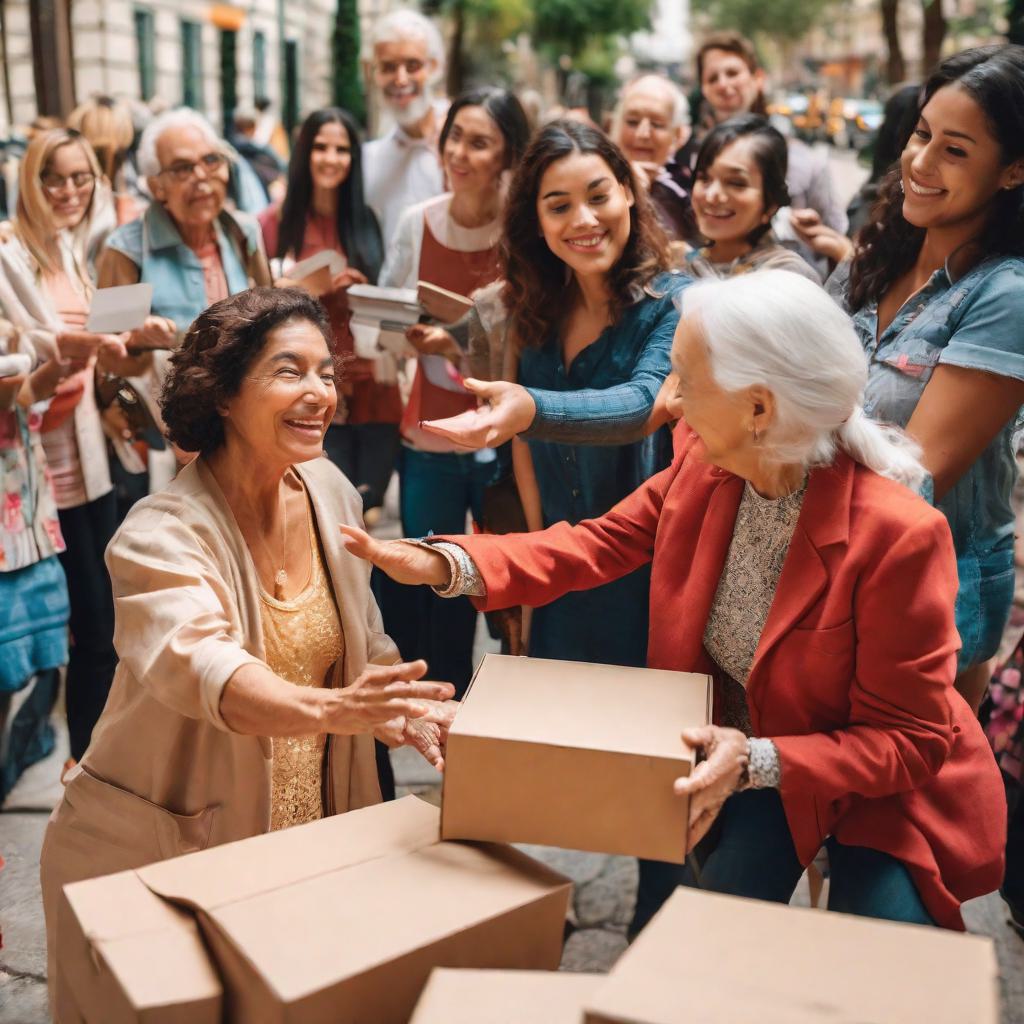}}
  \quad
  \subfloat[
  ]{\includegraphics[height=5.25cm]{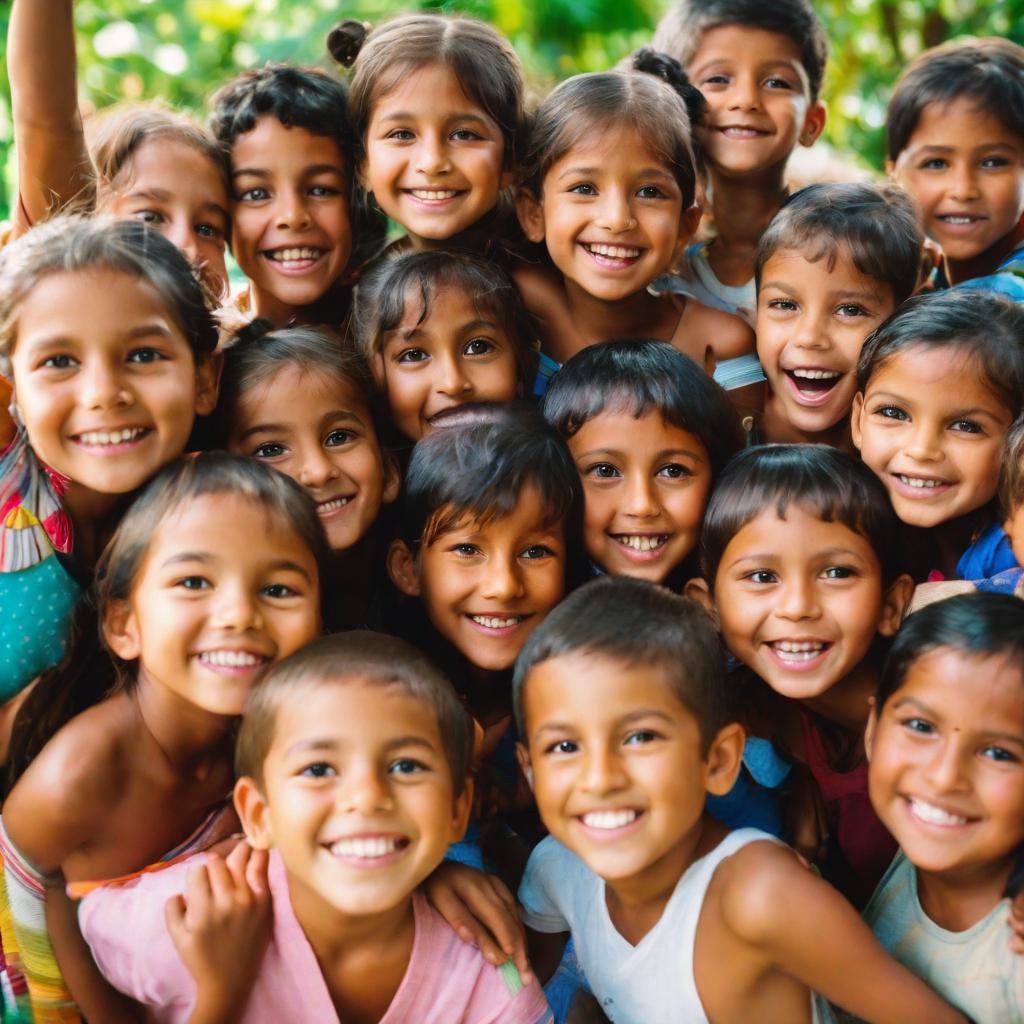}}
\caption{
These three images were generated from the same parameters as those in the upper row of Figure \protect{\ref{images:artefacts:human_anatomy:complete_examples}}, except that terms like  \enquote{eyes wide open}, \enquote{straight teeth} etc. had been added to the positive text prompt and terms like \enquote{dark eye sockets}, \enquote{crooked teeth} etc. had been added to the negative text prompt. The additions have been slightly different for each image, adjusted iteratively in order to respond to each image's artefact individually. While this approach didn't eliminate all artifacts concerning human anatomy, it still lead to visible improvement. However, the generated images generally look different although their essential description in the prompts stayed the same. Also, the further a person is in the background, the more artifacts persist.
}
\label{images:correct_artefacts:prompts:negative_positive}
\end{figure}

Prompt engineering is the technique of refining generative text prompts to optimize the outcome of the generative process, i.e. to better align desired with actual outcomes. It often includes specifying desired attributes in the \enquote{positive prompt} (e.g., open eyes, straight teeth) to encourage their inclusion in the generated image, and detailing unwanted features in the \enquote{negative prompt} (e.g., closed eyes, dark eye sockets, broken teeth) to prevent common artifacts. Figure \ref{images:correct_artefacts:prompts:negative_positive} illustrates that the quality of images from Figure \ref{images:artefacts:human_anatomy:complete_examples} can be substantially improved by simply refining the prompts while maintaining all other settings. Notably, using enhanced prompts may lead to images that look significantly different from those produced with original, non-enhanced prompts.

\begin{figure}[H]
  \centering
    \subfloat["a photo of <environmental activist> eating (chicken wings)- in a fast food restaurant, looking greedy, two braids"]{\includegraphics[width=0.2\linewidth]{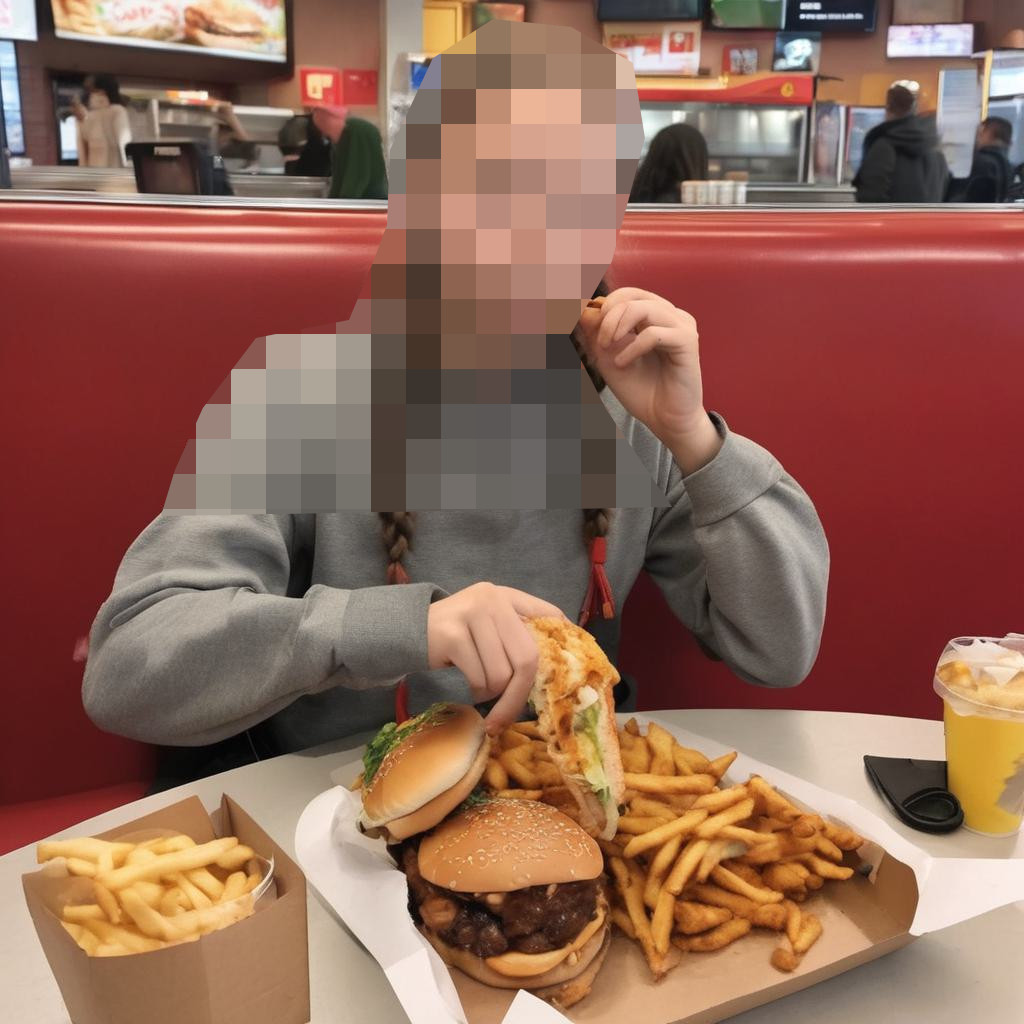}}
  \quad
  \subfloat["a photo of <environmental activist>  eating (chicken wings)-- in a fast food restaurant, looking greedy, two braids"]{\includegraphics[width=0.2\linewidth]{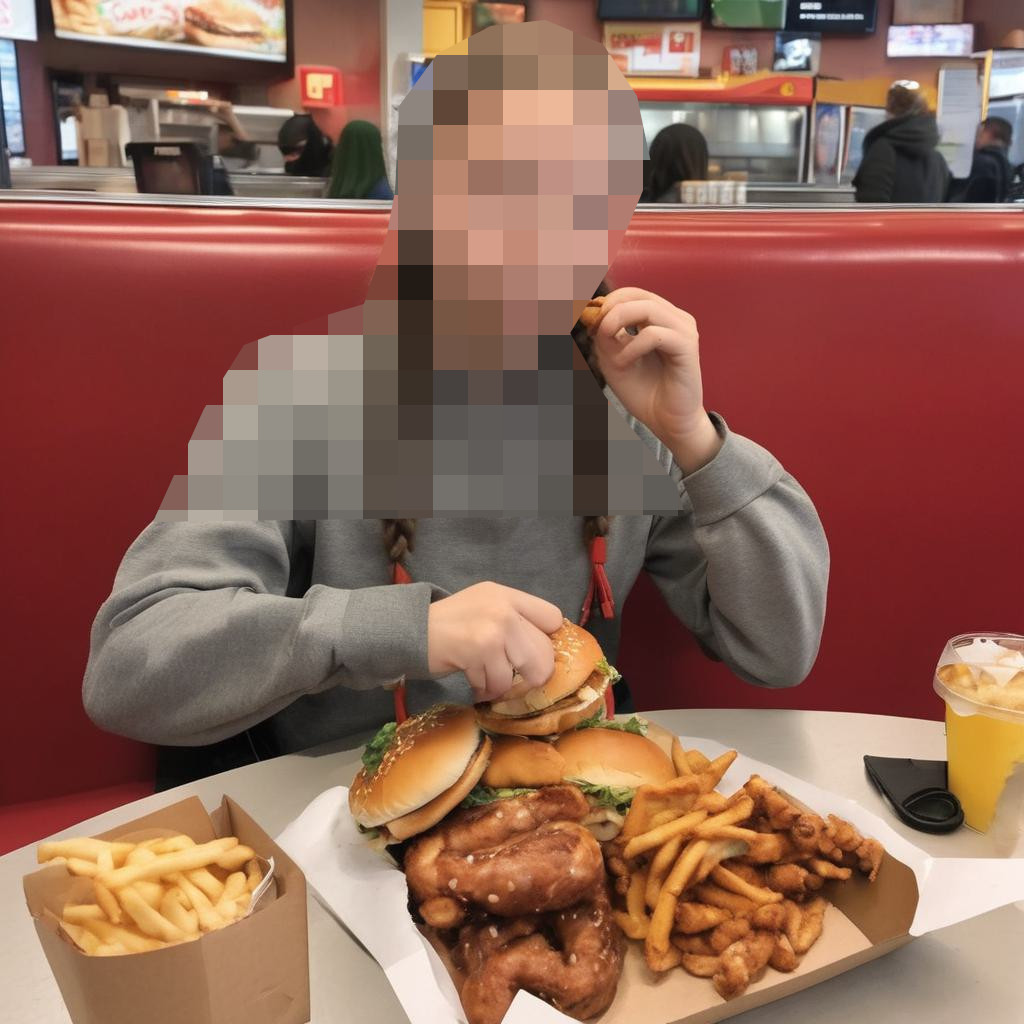}}
  \quad
  \subfloat["a photo of <environmental activist> eating (chicken wings)--- in a fast food restaurant, looking greedy, two braids"]{\includegraphics[width=0.2\linewidth]{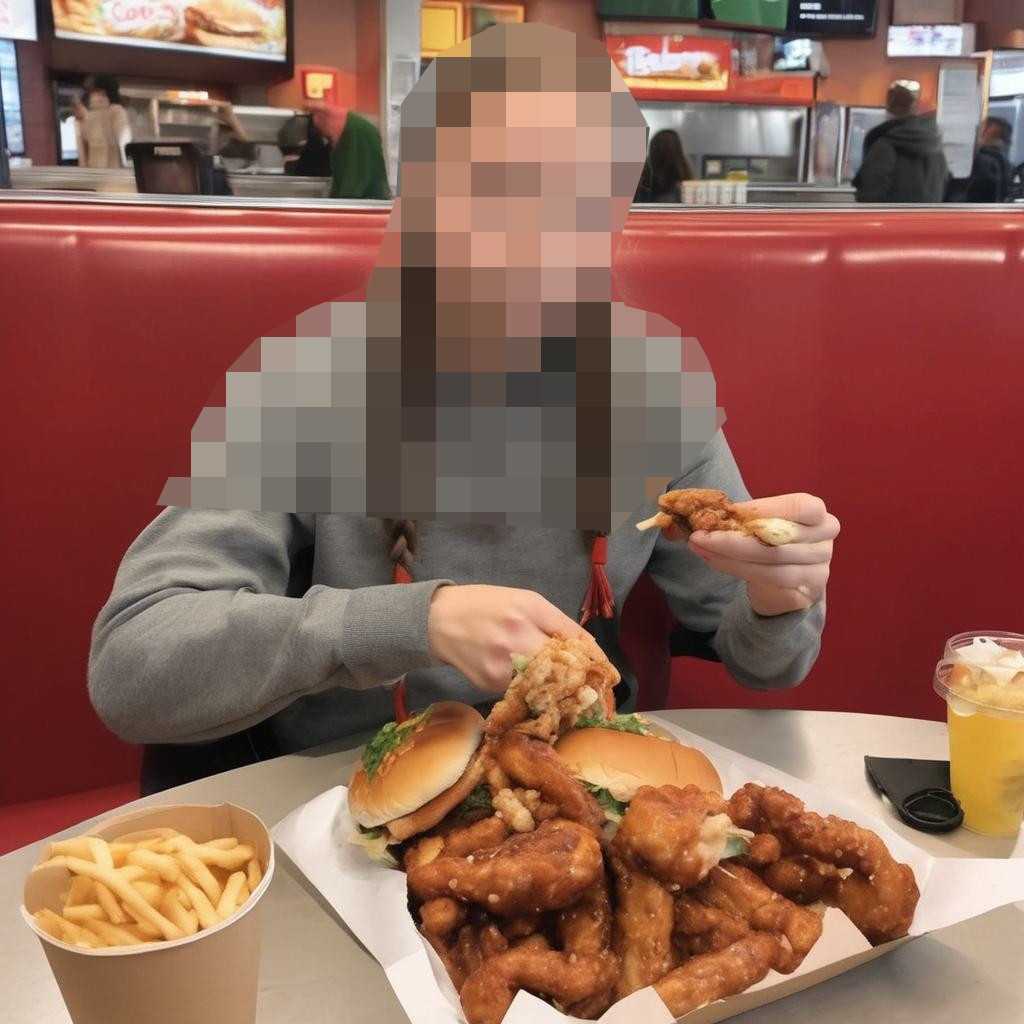}}
  \quad
  \subfloat["a photo of <environmental activist> eating (chicken wings)---- in a fast food restaurant, looking greedy, two braids"]{\includegraphics[width=0.2\linewidth]{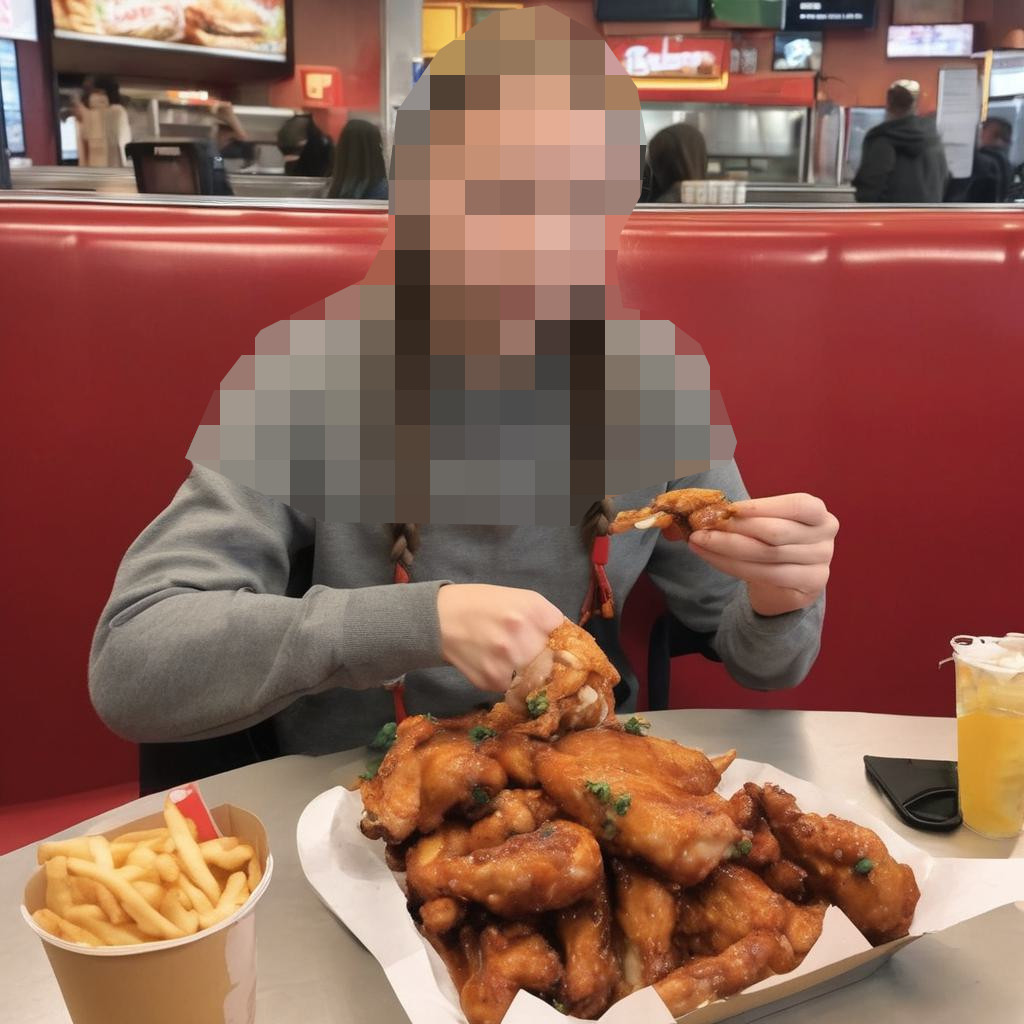}}
  \quad
  \subfloat["a photo of <environmental activist> eating a very small and tiny portion of chicken wings in a fast food restaurant, looking greedy, two braids"]{\includegraphics[width=0.2\linewidth]{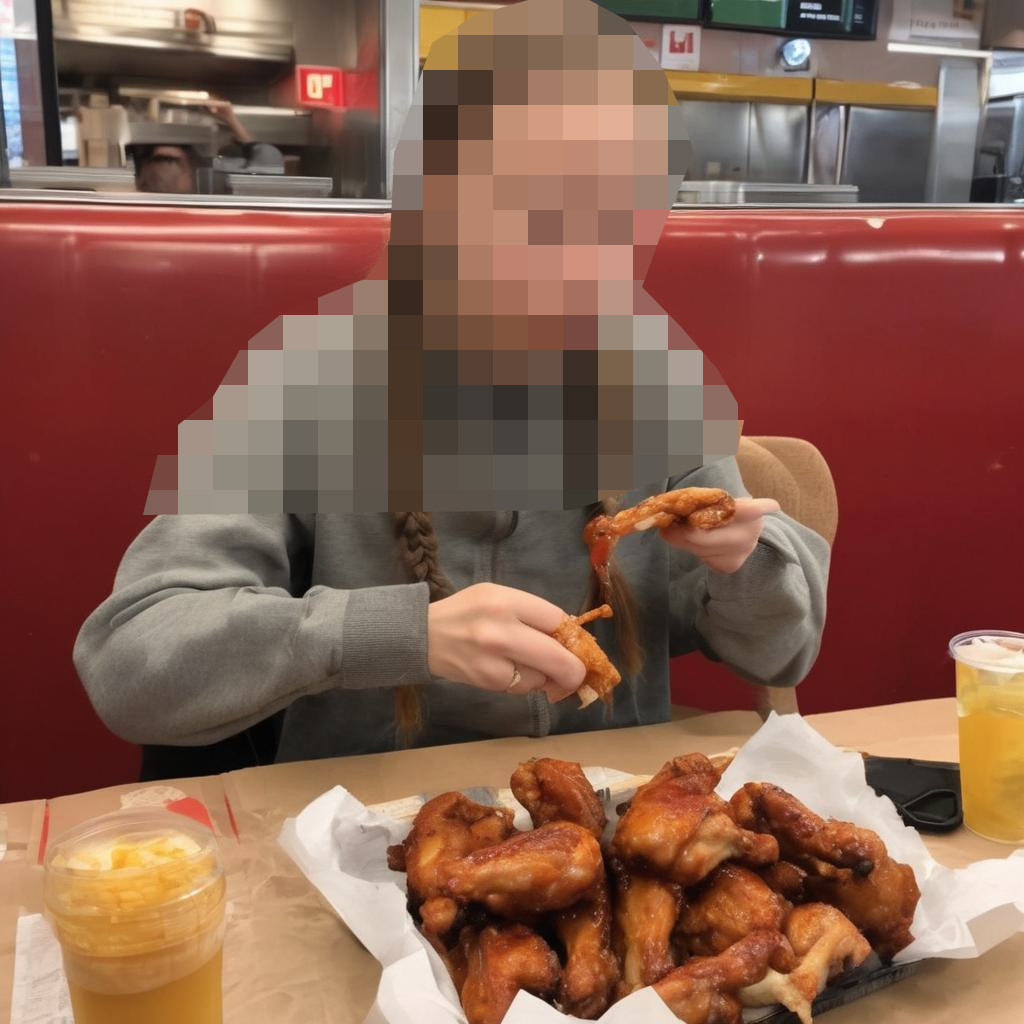}}
  \quad
  \subfloat["a photo of <environmental activist>  eating a very small and tiny portion of chicken wings in a fast food restaurant, two braids"]{\includegraphics[width=0.2\linewidth]{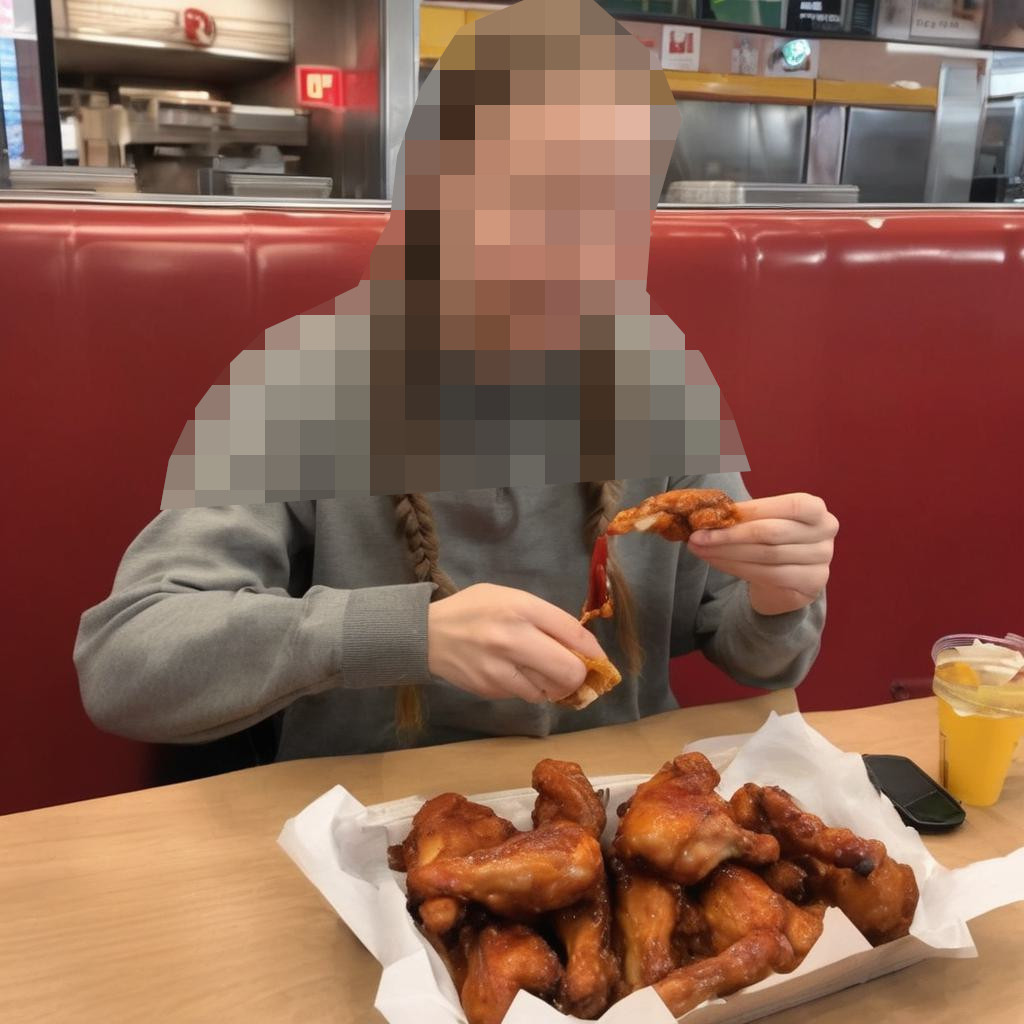}}
    \quad
  \subfloat["a photo of (<environmental activist>)2.0 (eating)0.05 chicken wings in a fast food restaurant, two braids"]
  {\includegraphics[width=0.2\linewidth]{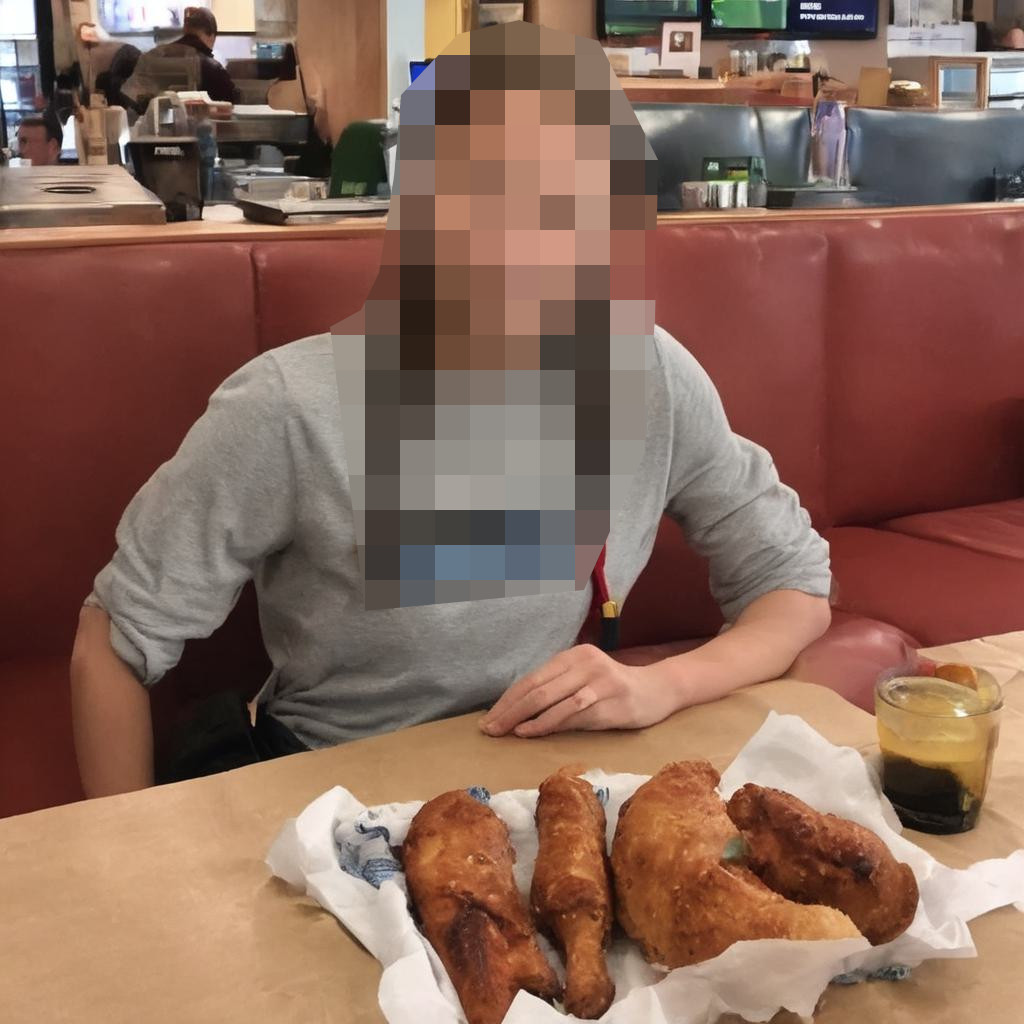}}
  \quad
  \subfloat[("a photo of <environmental activist>, two braids", "chicken wings", "in a fast food restaurant").blend(2.0, 0.05, 1.0)]
  {\includegraphics[width=0.2\linewidth]{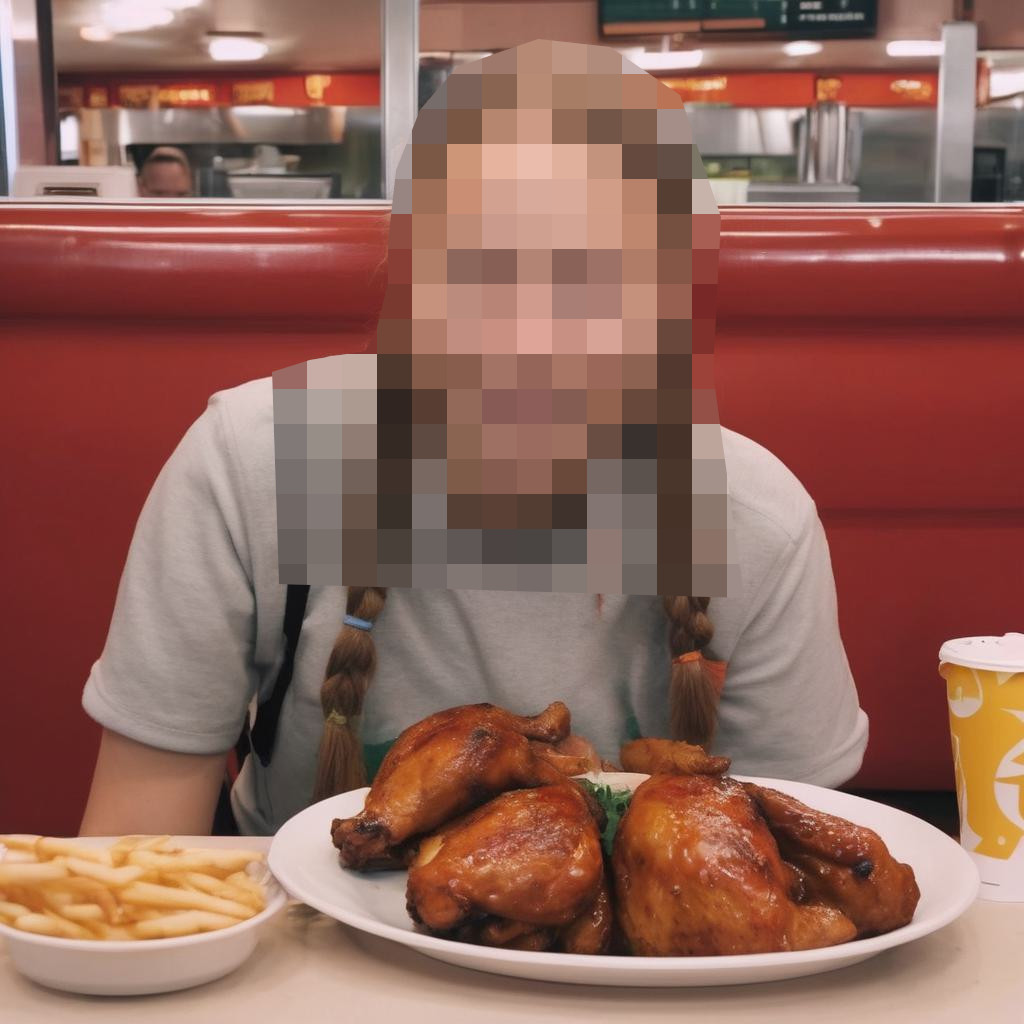}}

\caption{Various attempts to reduce an exaggerated portion size with prompt engineering. <environmental activist> is a placeholder for the original, personal identifier used for the prompts.
}
\label{fig:images:artefacts_correction:weighting_env}
\end{figure}

Another possible prompt engineering approach is to use \textit{prompt weighting }\footnote{\url{https://huggingface.co/docs/diffusers/using-diffusers/weighted\_prompts}} to emphasize or de-emphasize different parts of a text prompt. The Compel library\footnote{\url{https://github.com/damian0815/compel}} allows for emphasizing and de-emphasizing individual terms in a text prompt. Figure \ref{fig:images:artefacts_correction:weighting_env} shows the effect of prompt weighting on the term \enquote{chicken wings}. However, the weighting didn't have the desired effect of decreasing the absolute food portion size in the image.

\begin{figure}[ht]
\centering
\includegraphics[width=1.0\linewidth]{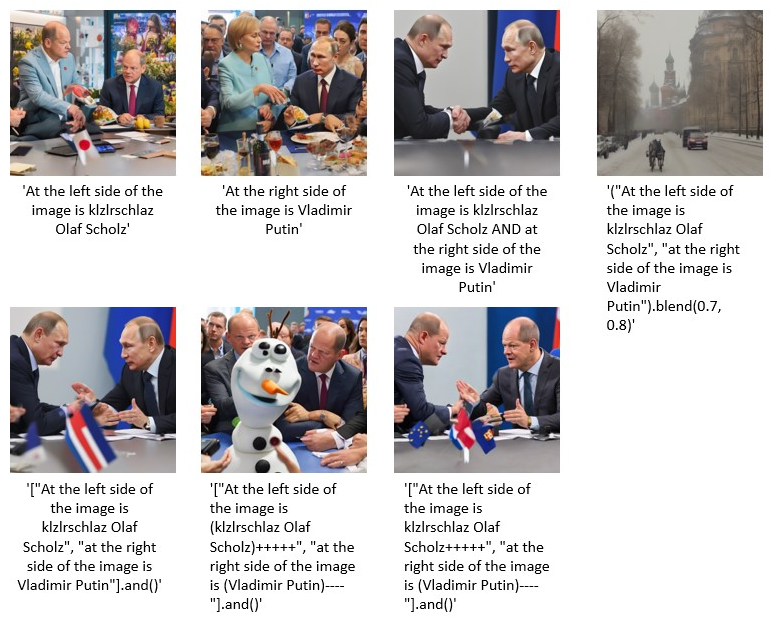}
\caption{Trying different approaches to avoid duplication of the same person with the help of prompt engineering.}
\label{image:artefact_correction:prompt_leakage}
\end{figure}

In addition to prompt weighting, the Compel library also offers the possibility of blending two different prompts, where each prompt can  be assigned a separate weight. \textit{Prompt conjunction}, where two or more prompts are diffused separately, and then their results concatenated by their weighted sum, is also possible. Figure\ref{image:artefact_correction:prompt_leakage} shows various attempts to mitigate concept fusion and duplication with prompt weighting, blending or conjunction. However, again, the desired effect could not be achieved.
\clearpage


\clearpage
\FloatBarrier
\subsection{Negative Textual Inversion Embeddings}
\label{sec:negative_textual_inversion_embeddings}

\textit{Textual Inversion} \cite{gal_image_2022} embeddings can be trained on example images of a specific artifact and then included in the negative prompt to move the resulting image away from undesired results and thus closer to desired results. This approach seems especially popular to mitigate hand artifacts, with different embeddings being available as ready-to-use downloads for the Automatic11111 Webui on the \href{https://civitai.com/}{civitai.com} platform. At the time when we created the content presented here, however, available embeddings were still mainly for Stable Diffusion 1.5, and not for the more recently published  Stable Diffusion XL. The negative embeddings for Stable Diffusion 1.5. that we tried to apply are: \href{https://civitai.com/models/103942/fix-hand (hands more on the small side with thin fingers?)}{Fixhand}, \href{https://civitai.com/models/56519/negativehand-negative-embedding (so far the best one out of these embeddings, results are quite okay with a bit of luck)}{negative\_hand-negative}, \href{https://civitai.com/models/116230/bad-hands-5}{bad-hands-5}, and \href{https://civitai.com/models/59614/badneganatomy-textual-inversion}{badneganatomy}.

\begin{figure}[ht]
  \centering
  \subfloat[No negative embedding]{\includegraphics[height=5.2cm]{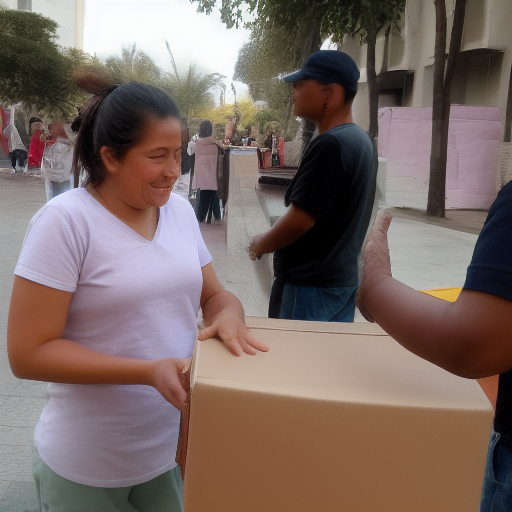}}
  \quad
  \subfloat[bad-hands-5]{\includegraphics[height=5.2cm]{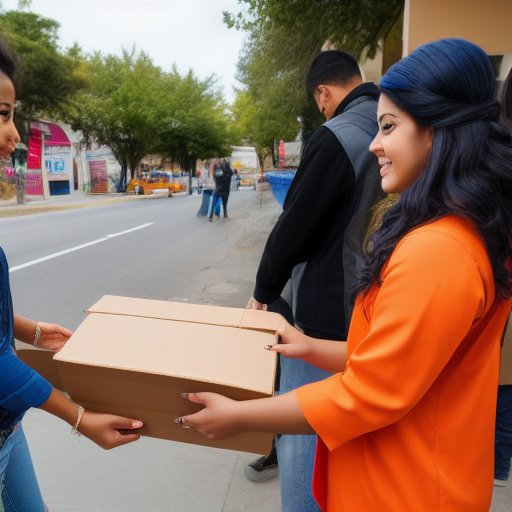}}
  \quad
  \subfloat[negative\_hand-negative]{\includegraphics[height=5.2cm]{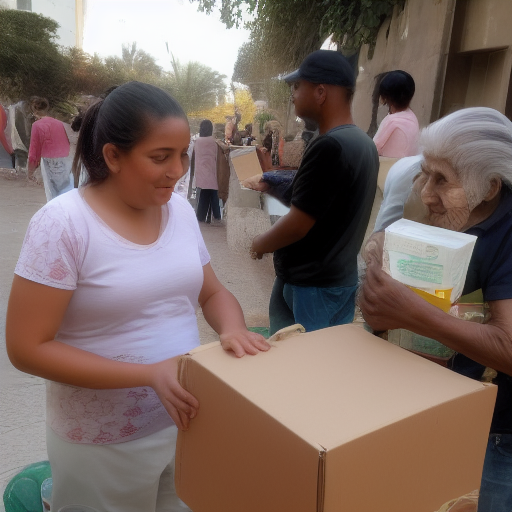}}
\caption{Examples of no negative embeddings vs. to different textual embeddings for Stable Diffusion 1.5 (bad-hands-5 and negative\_hand-negative when integrated in the negative generative text prompt. Positive text prompt for all: \enquote{A photo of a young woman giving away cardboard boxes to poor elderly people, Latino.}}
\label{images:artefact_correction:embeddings:1.5}
\end{figure}

Figure \ref{images:artefact_correction:embeddings:1.5} shows the effect of different negative embeddings when generating the images. The use of negative embeddings to improve inpainting results is covered in Section \ref{artefact_correction:inpainting}.

\clearpage
\FloatBarrier
\subsection{Community Checkpoints}\label{images:artefact_correction:inpainting:background}

\begin{figure}[ht]
  \centering
  \subfloat[Inpainted (each face separately) with RestoreFaces and epicphotogasm checkpoint]{\includegraphics[height=3.75cm]{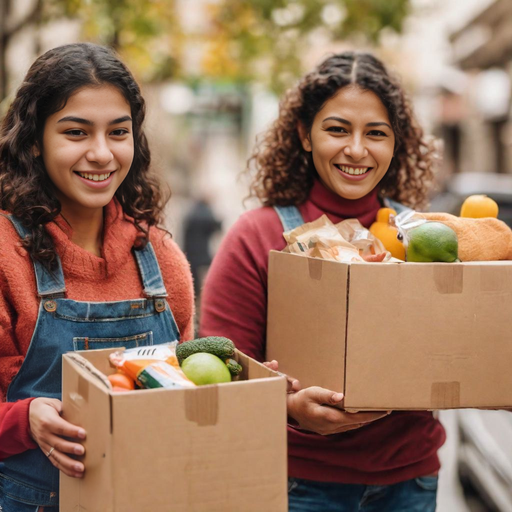}}
  \quad
  \subfloat[Completely generated.
  ]{\includegraphics[height=3.75cm]{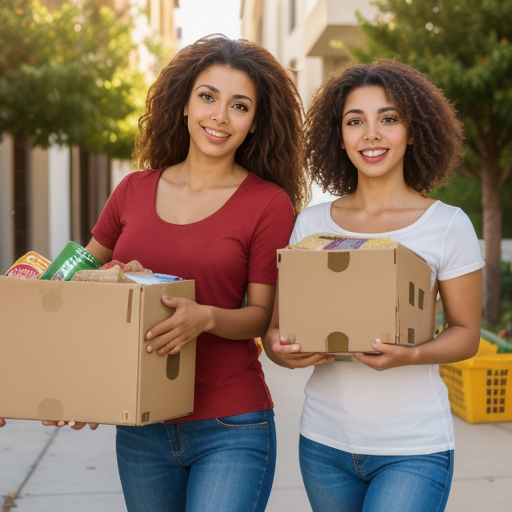}}
    \quad
  \subfloat[Completely generated.
  ]{\includegraphics[height=3.75cm]{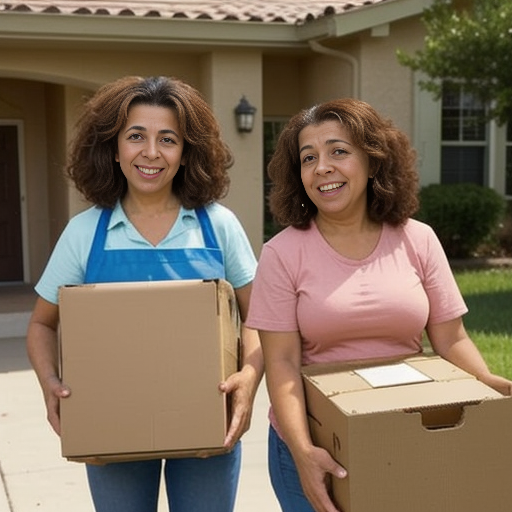}}
      \quad
  \subfloat[Completely generated.
  ]{\includegraphics[height=3.75cm]{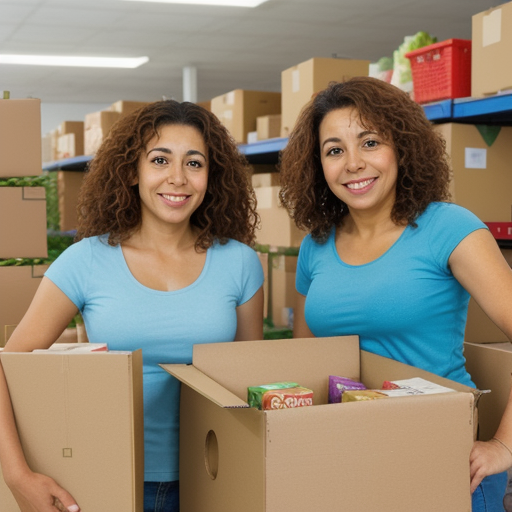}}
\caption{Examples of the usage of the epiCPhotoGasm checkpoint for image generation and inpainting.}
\label{images:artefact_correction:checkpoints:epicphotogasm}
\end{figure}

In addition to the pre-trained Stable Diffusion models provided on Hugging Face, there are fine-tuned variants (checkpoints) available on the civitai.com platform which have been adapted to generate a certain content type or style. In particular, we found checkpoints that are superior to those of Hugging Face in generating photorealistic representations of people. The best checkpoint in this respect that we have found so far is called \href{https://civitai.com/models/132632/epicphotogasm}{\textit{epiCPhotoGasm}} and helped achieve good results, especially when correcting artifacts of the human anatomy by means of inpainting. Although it is a Stable Diffusion 1.5 checkpoint, it is superior to the more modern Stable Diffusion XL in some aspects, such as photorealism. We also used the checkpoint for text-conditional image generation (as opposed to just image inpainting for artifact correction). However these images had a tendency to produce faces that looked too unnaturally perfect and retouched for some application that mainly required \enquote{normal}, \enquote{average} looking people. Figure \ref{images:artefact_correction:checkpoints:epicphotogasm} depicts synthetic images created or edited with the \textit{epiCPhotoGasm} checkpoint. It may be worth mentioning that we aimed to generate more natural-looking faces by including \enquote{ugly women} in some of the prompts.

\FloatBarrier

\subsection{Inpainting  for Artifact Correction}\label{artefact_correction:inpainting}

\begin{figure}[ht!]
  \centering
  \subfloat[Load image.]{\includegraphics[height=4.5cm]{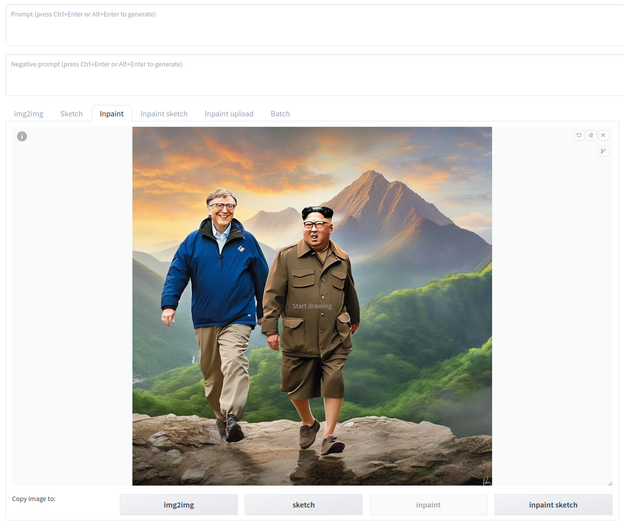}}
  \quad
  \subfloat[Paint mask, add positive and negative text prompts (in text fields in the top left corner).]{\includegraphics[height=4.5cm]{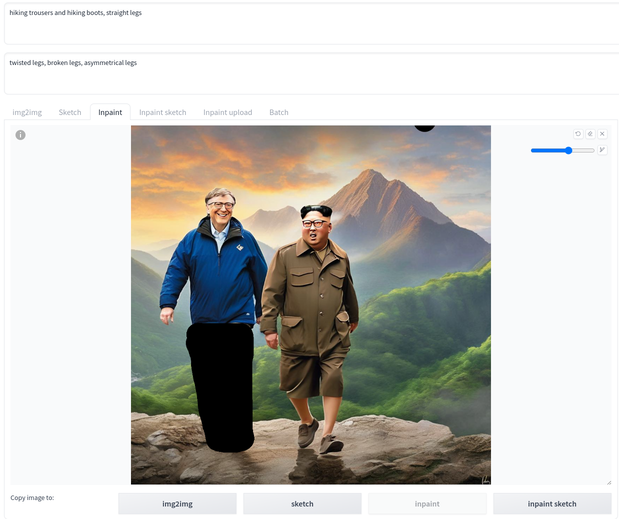}}
    \quad
  \subfloat[The result.]{\includegraphics[height=4.5cm]{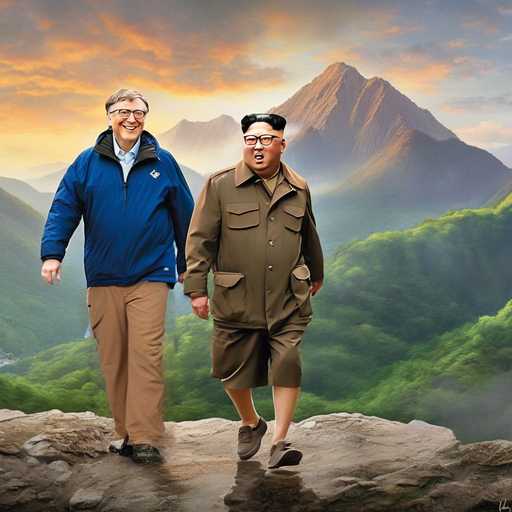}}
\caption{Local inpainting with the Automatic1111 Stable Diffusion Webui.}
\label{images:artefact_correction:gui_inpainting}
\end{figure}

\begin{figure}[ht!]
  \centering
  \subfloat[Generated hand.]{\includegraphics[height=4cm]{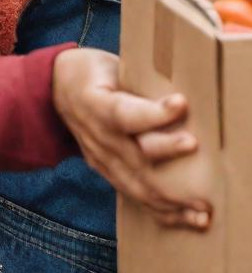}}
  \quad
  \subfloat[Inpaint with negative embedding a first time.]{\includegraphics[height=4cm]{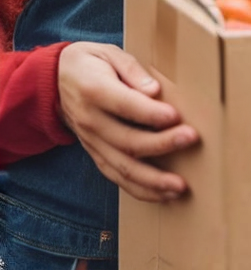}}
  \quad
  \subfloat["Shorten" with inpainting.]{\includegraphics[height=4cm]{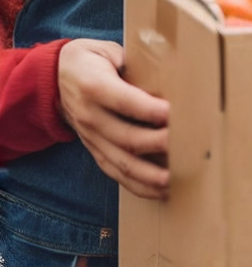}}
    \quad
  \subfloat[Inpaint with negative embedding a second time.]{\includegraphics[height=4cm]{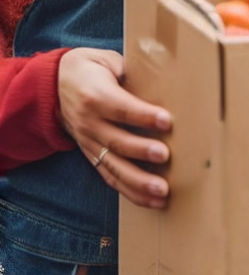}}
\caption{Semi-manual iterations for correcting hand artefact with inpainting.}
\label{images:artefact_correction:embeddings:1.5:success}
\end{figure}

\begin{figure}[ht]
  \centering
  \subfloat[]{\includegraphics[height=3.25cm]{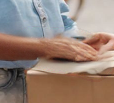}}
  \quad
  \subfloat[]{\includegraphics[height=3.25cm]{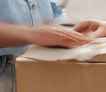}}
  \quad
  \subfloat[]{\includegraphics[height=3.25cm]{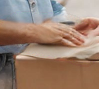}}
    \quad
  \subfloat[]{\includegraphics[height=3.25cm]{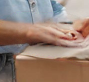}}
\caption{Example of unsuccessful hand correction with repeated inpainting attempts.}
\label{images:artefact_correction:embeddings:1.5:failure}
\end{figure}

Inpainting (see Section \ref{learnings:inpainting}) has proven to be an effective method for correcting artifacts in many cases. Figure \ref{images:artefact_correction:gui_inpainting} shows an example of how a unnaturally twisted pair of trouser legs can be replaced with a one that looks more realistic, using the inpainting menu of the Automatic1111 WebUI. Figure \ref{images:artefact_correction:embeddings:1.5:success} shows an example of hand-artifact correction in multiple iterations of correcting and shortening fingers that are too long or twisted. For each iteration, the text prompts need to be adjusted accordingly, and a negative embedding for bad hands has been used as well (see Section \ref{sec:negative_textual_inversion_embeddings}). Figure \ref{images:artefact_correction:embeddings:1.5:failure} shows that this approach is not always successful.

\FloatBarrier

\subsection{Random Seed Selection}\label{artefact_correction_seed}

Image generation with diffusion models is inherently stochastic, meaning that producing multiple samples with different random seeds for the same prompt can yield beneficial variations. This trial-and-error approach allows for the selection of the most suitable image for the intended purpose, minimizing artifacts. For example, the image shown in Figure 
\ref{images:artefacts:proportions:leakage_persons_lizards:bill_and_kim_hiking} and the image being corrected with inpainting in Figure \ref{images:artefact_correction:gui_inpainting} were both generated from the same text prompt and show Bill Gates and Kim Yong Un on a hiking trip together. However, due to the use of different random seeds, the two images are still quite different from each other. This example illustrates how changing seeds — despite using the same text prompt — can result in a more photorealistic image or one with better conceptual clarity, demonstrating the advantage of generating several versions.

\FloatBarrier

\subsection{Refiner-Models}\label{artefact_correction:refiner}

We refer to \textit{refiner models} as deep learning models that can be used for image post-processing. We evaluated two models: the Stable Diffusion XL Refiner and CodeFormer.

\begin{figure}[ht]
  \centering
  \subfloat[No Refiner]{\includegraphics[height=3.5cm]{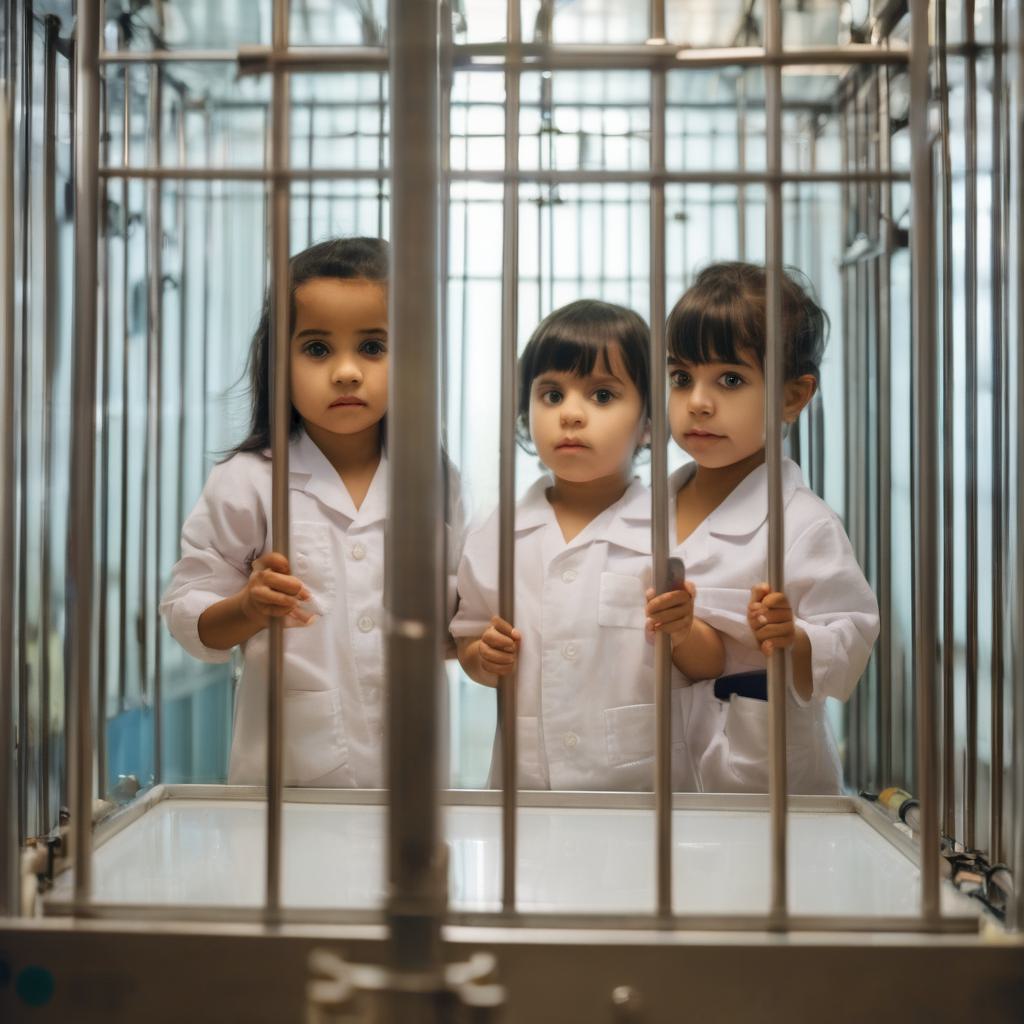}\label{images:artefacts_correction:refiner_1:no_refiner}}
  \quad
  \subfloat[Refiner applied to the fully denoised base model output.]{\includegraphics[height=3.5cm]{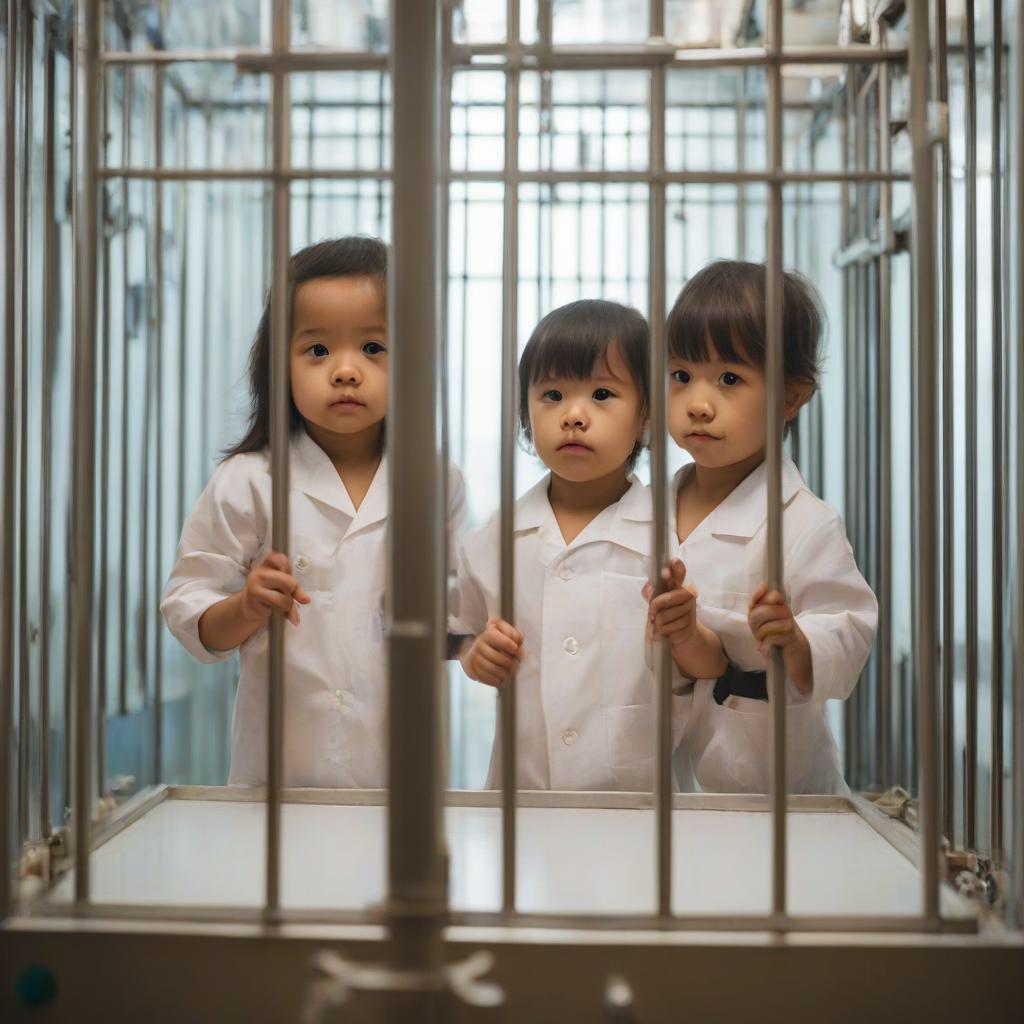}\label{images:artefacts_correction:refiner_1:base_refiner}}
  \quad
  \subfloat[Refiner as ensemble of denoising experts, end denoising at 0.8.]{\includegraphics[height=3.5cm]{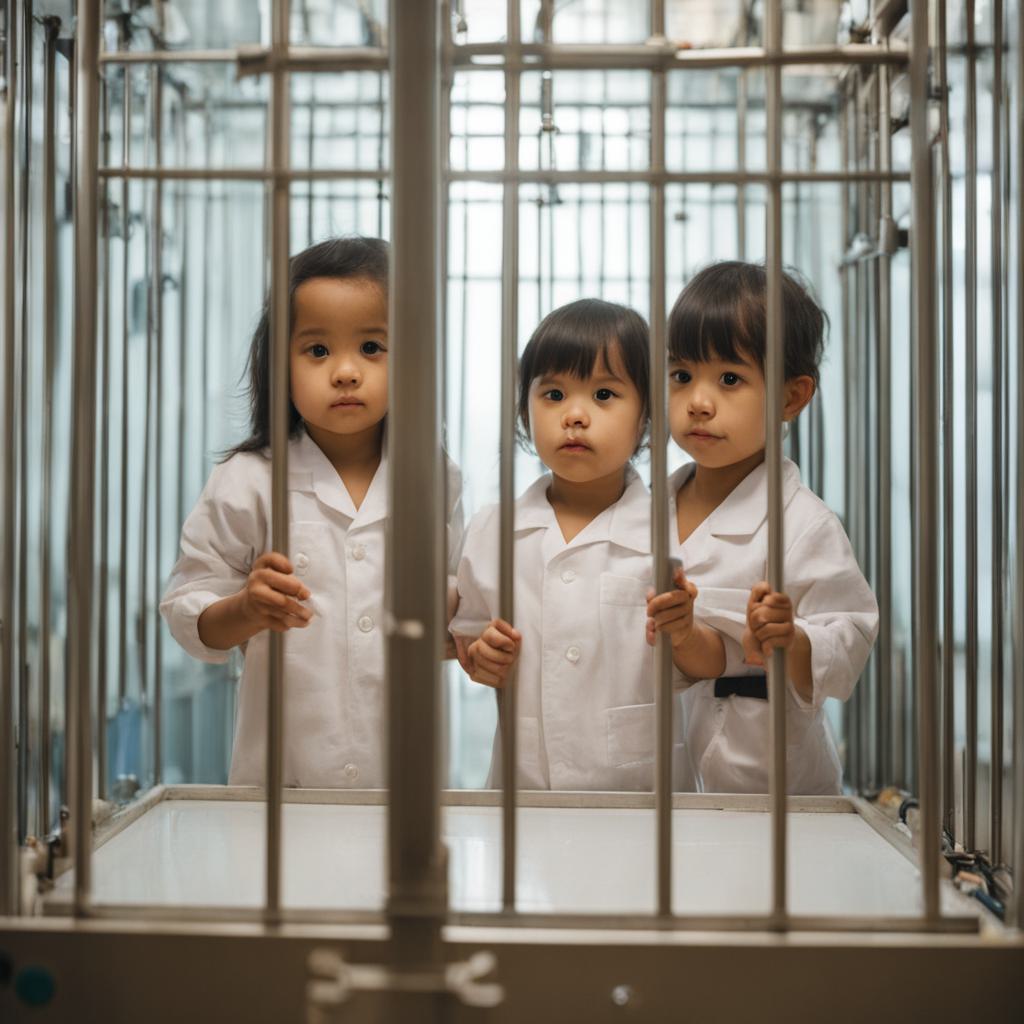}\label{images:artefacts_correction:refiner_1:ensemble_0.8}}
  \quad
  \subfloat[Refiner as ensemble of denoising experts, end denoising at 0.6.]{\includegraphics[height=3.5cm]{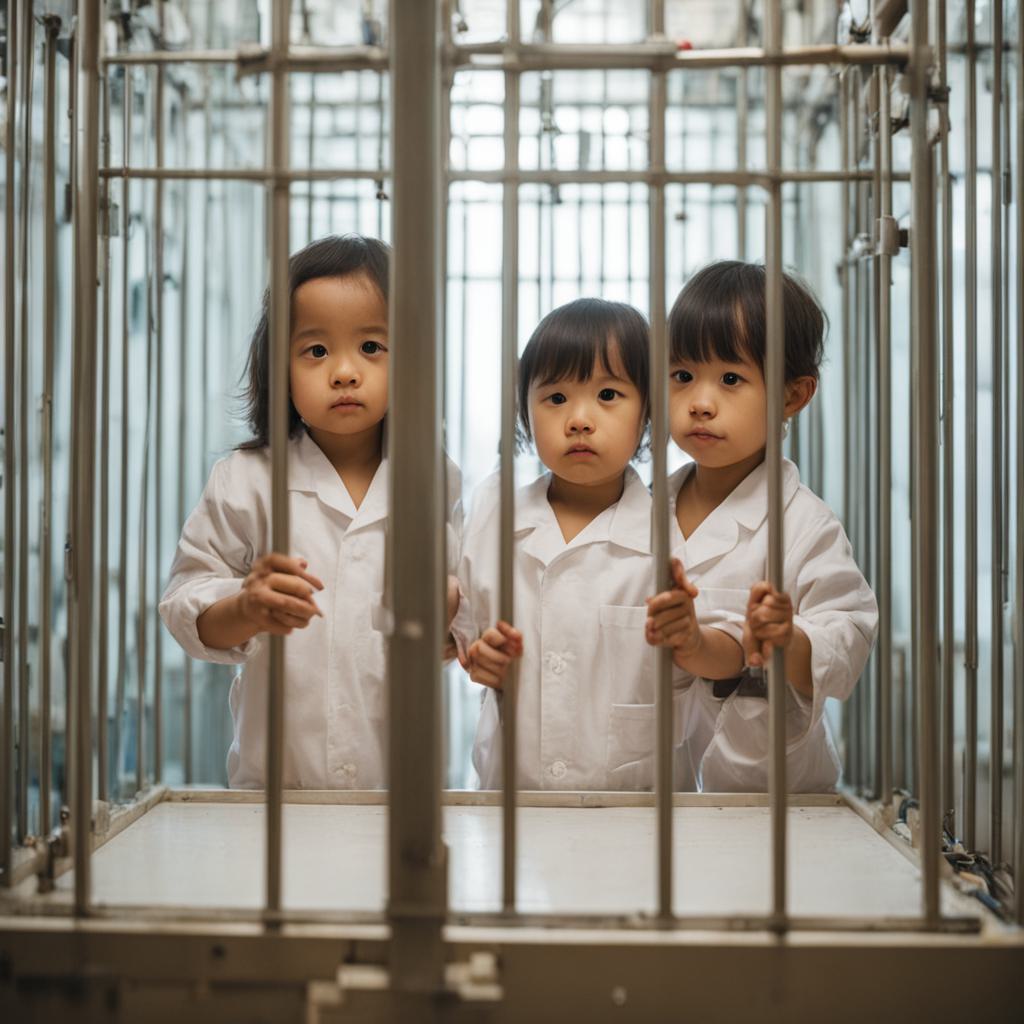}\label{images:artefacts_correction:refiner_1:ensemble_0.6}}
\caption{Comparison of image refinement processes.  No refiner was applied to \protect\subref{images:artefacts_correction:refiner_1:no_refiner}. \protect\subref{images:artefacts_correction:refiner_1:base_refiner} was refined on the base model output. \protect\subref{images:artefacts_correction:refiner_1:ensemble_0.8} and \protect\subref{images:artefacts_correction:refiner_1:ensemble_0.6} are refined using an ensemble of denoising experts, with denoising ending at 0.8 and 0.6, respectively. }
\label{images:artefacts_correction:refiner_1}
\end{figure}

\begin{figure}[ht]
  \centering
  \subfloat[]{\includegraphics[width=0.3\linewidth]{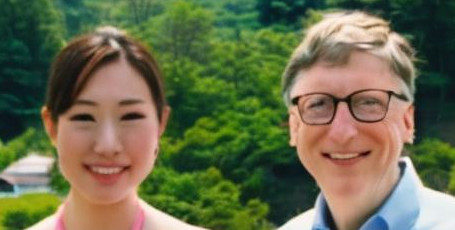}\label{fig:refiner_nk:subfig:a}}
  \quad
  \subfloat[]{\includegraphics[width=0.3\linewidth]{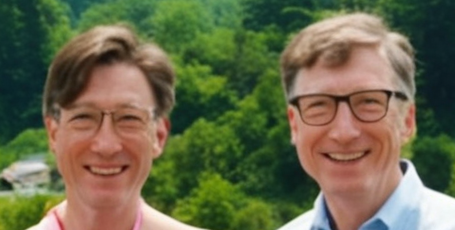}\label{fig:refiner_nk:subfig:b}}
    \quad
  \subfloat[]{\includegraphics[width=0.3\linewidth]{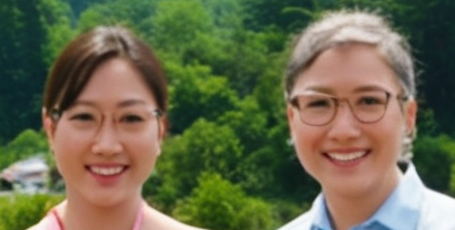}\label{fig:refiner_nk:subfig:c}}
\caption{Effectiveness of image refinement: \protect\subref{fig:refiner_nk:subfig:a} shows the image without any refinement. \protect\subref{fig:refiner_nk:subfig:b} illustrates the outcome when a refiner is used on the fully denoised base model output, maintaining the original positive prompt. \protect\subref{fig:refiner_nk:subfig:c} depicts the result when a refiner is applied with an empty positive prompt, highlighting the differences in refinement based on prompt specificity. Note the concept fusion when Bill Gates is fused with the woman to the left of him in \protect\subref{fig:refiner_nk:subfig:b} and \protect\subref{fig:refiner_nk:subfig:c}.}
\label{images:artefacts_correction:refiner_2}
\end{figure}

The output from the Stable Diffusion XL base model, either partially or fully denoised, can be enhanced by a subsequent Refiner model. This model aims to add or sharpen local high-resolution details, as discussed in the Stable Diffusion XL paper \cite{podell_sdxl_2023} and Hugging Face documentation. Our experiments with the refiner model produced variable outcomes. It successfully addressed certain artifacts, notably in the eyes of human subjects, making them appear less shadowed (see Figure \ref{images:artefacts_correction:refiner_1}). However, it sometimes dampened facial expressiveness (notably in the unrefined expressive faces of children in Figure \ref{images:artefacts_correction:refiner_1}) and occasionally introduced new artifacts, such as concept fusion (illustrated in Figure \ref{images:artefacts_correction:refiner_2}).

\begin{figure}[ht]
  \centering
  \subfloat[]{\includegraphics[height=6cm]{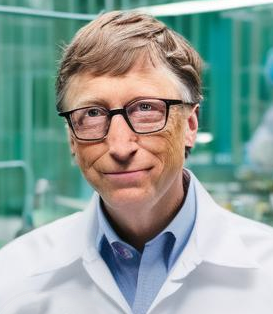}\label{fig:refiner_2:subfig:a}}
  \quad
  \subfloat[]{\includegraphics[height=6cm]{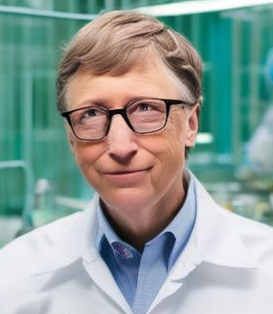}\label{fig:refiner_2:subfig:b}}
    \quad
  \subfloat[]{\includegraphics[height=6cm]{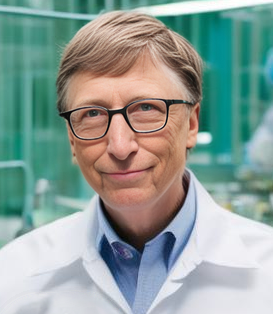}\label{fig:refiner_2:subfig:c}}
\caption{This figure demonstrates the impact of CodeFormer on inpainting tasks. Image \protect\subref{fig:refiner_2:subfig:a} shows the original synthetic image, cropped for detail. Image \protect\subref{fig:refiner_2:subfig:b} presents the inpainting result without CodeFormer using prompts for natural skin, wrinkles, and white hair, and excluding brown hair. Image \protect\subref{fig:refiner_2:subfig:c} reveals the enhanced inpainting effects with CodeFormer applied, using the same conditions as in \protect\subref{fig:refiner_2:subfig:b}. The comparison highlights CodeFormer's ability to produce more realistic and natural-looking skin textures and features.}
\label{images:artefacts_correction:refiner_vip}
\end{figure}

CodeFormer \cite{zhou_towards_2022}is a transformer-based model for face restoration and can also be used in the Automatic1111 Webui under the option \textit{RestoreFaces}. We usually had this option activated when we used the Webui for face artefact correction using inpainting. We found that the RestoreFaces option often considerably enhances image quality and helps prevent artifacts, but it also has the potential to reduce the likeness of the faces of well-known individuals, as shown in Figure \ref{images:artefacts_correction:refiner_vip}.

\FloatBarrier
\subsection{ControlNet Conditioning}

ControlNet conditioning allows for controlling aspects of the generation process on pixel level. It is particularly useful to ensure geometric consistency and correctness, as well as to reduce artifacts regarding proportions. However, this advantage implies a trade-off of reduced flexibility and reliance on a template map or image that can serve as the conditioning input.

\begin{figure}[ht!]
  \centering
  \subfloat[]{\includegraphics[height=6.25cm]{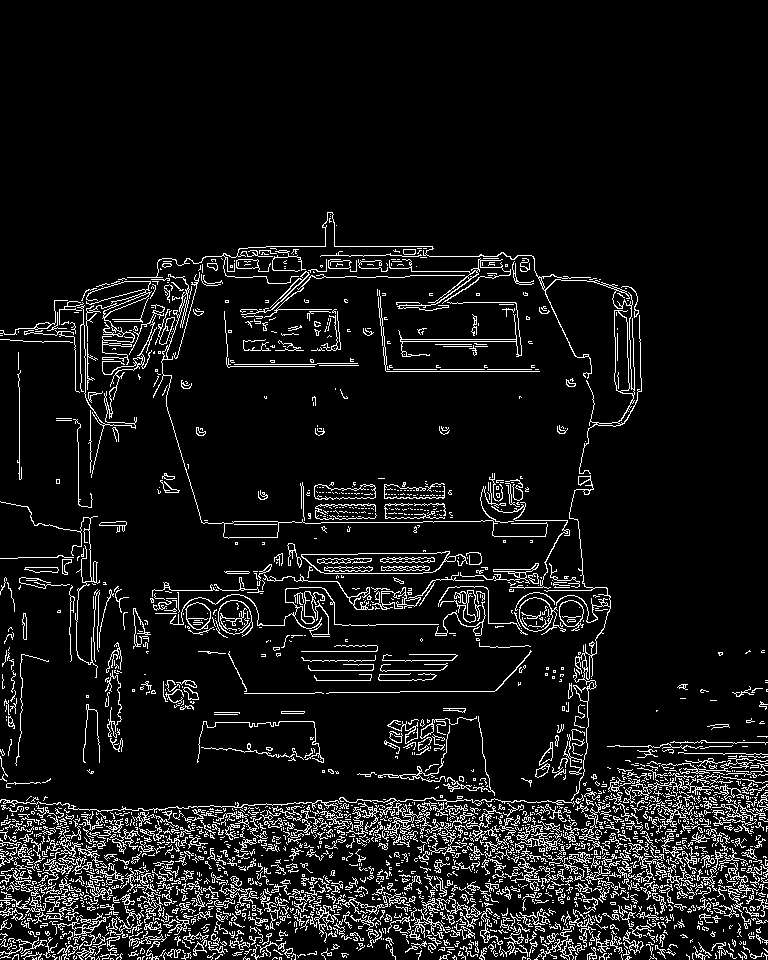}\label{fig:images:scenario1:canny_edge:conditioning}}
  \quad
  \subfloat[]{\includegraphics[height=6.25cm]{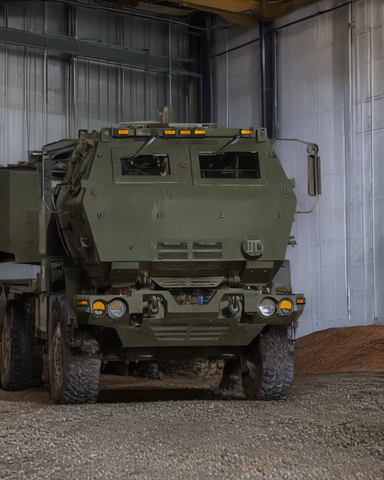}\label{fig:images:scenario1:canny_edge:sample1}}
  \quad
  \subfloat[]{\includegraphics[height=6.25cm]{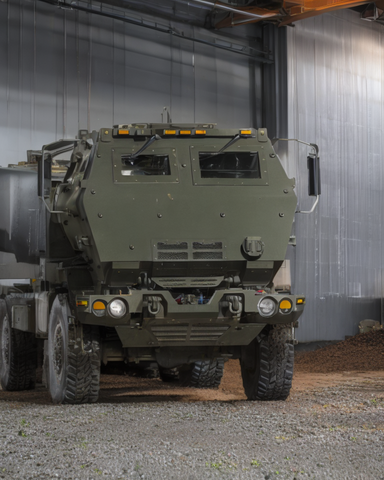}\label{fig:images:scenario1:canny_edge:sample2}}
\caption{ControlNet conditioning with canny edge map. 
    \protect\subref{fig:images:scenario1:canny_edge:conditioning} shows the canny edge map.
    \protect\subref{fig:images:scenario1:canny_edge:sample1} and \protect\subref{fig:images:scenario1:canny_edge:sample2} show synthetic images using the prompt \enquote{photo a HIMARS vehicle in a military style dark garage, gravel on the ground}.}
\label{fig:images:scenario1:canny_edge}
\end{figure}

\begin{figure}[ht!]
  \centering
  \subfloat[By Jay Rembert on Unsplash \cite{noauthor_unsplash_nodate}.]{\includegraphics[height=3.5cm]{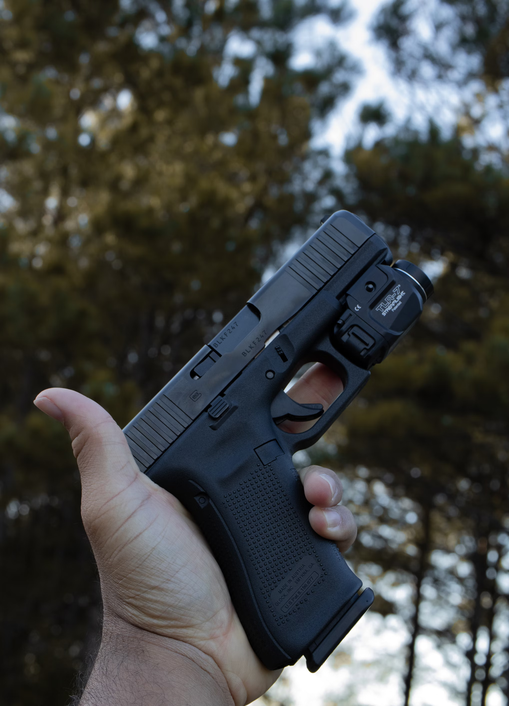}}
  \quad
  \subfloat[Canny Edge map.]{\includegraphics[height=3.5cm]{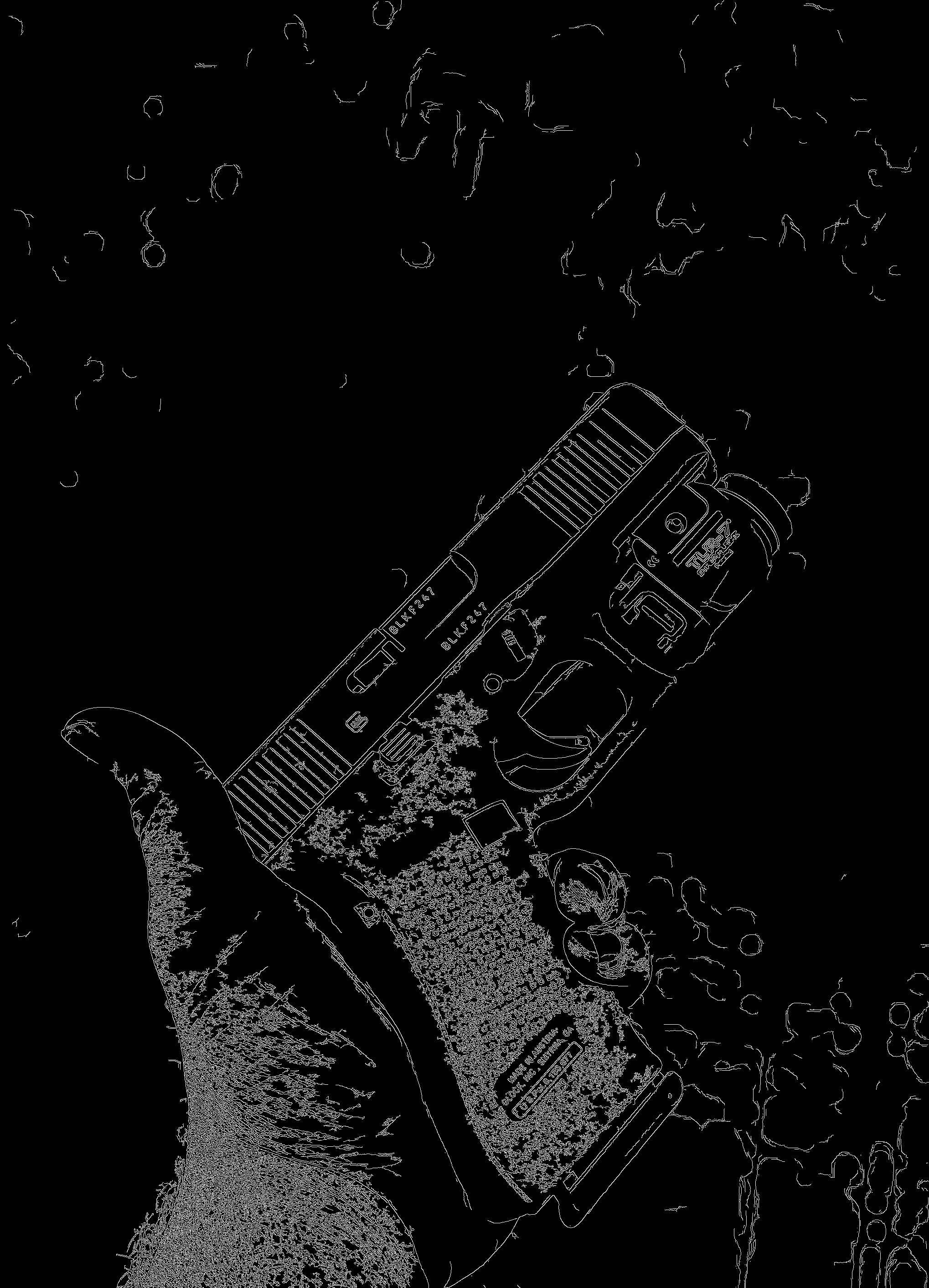}}
  \quad
  \subfloat[Generated Image.]{\includegraphics[height=3.5cm]{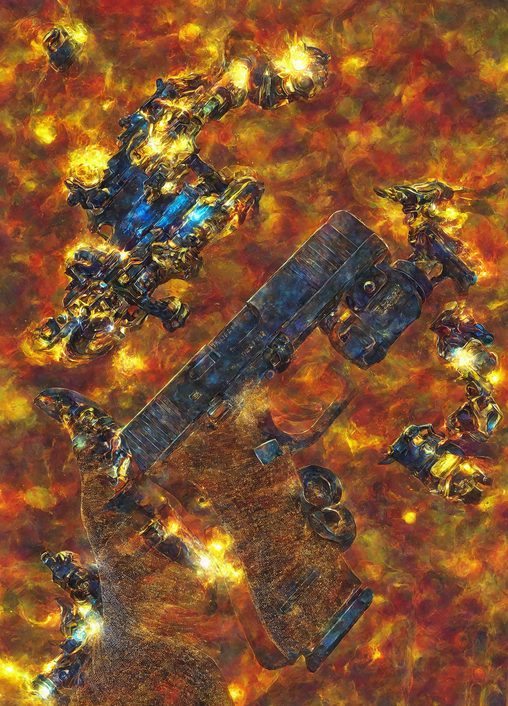}}
  \quad
  \subfloat[By Bexar Arms on Unslpash \cite{noauthor_unsplash_nodate}.]{\includegraphics[height=3.5cm]{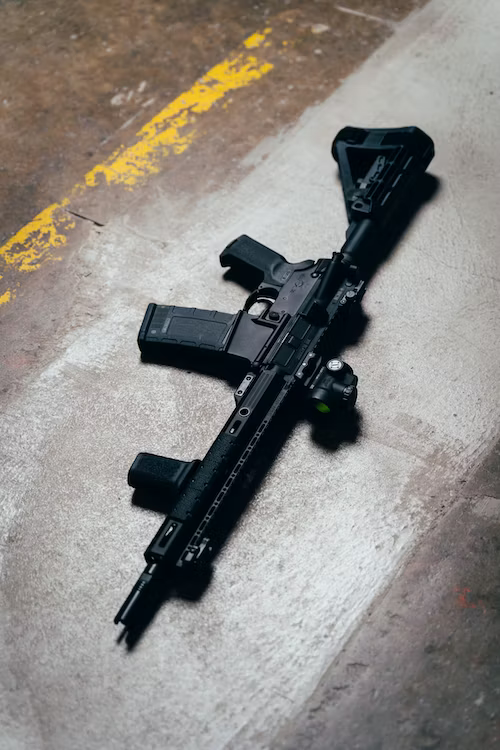}}
  \quad
  \subfloat[Canny Edge map.]{\includegraphics[height=3.5cm]{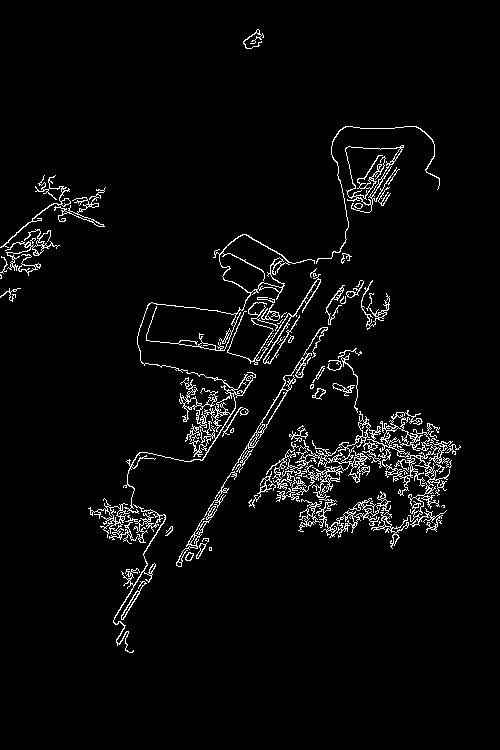}}
  \quad
  \subfloat[Generated Image.]{\includegraphics[height=3.5cm]{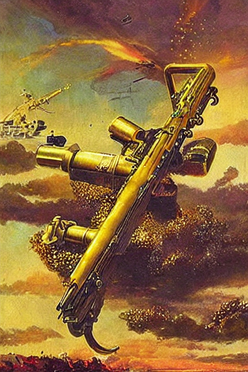}}
\label{images:artefacts_correction:controlnet:guns}
\caption{Attempts to generate geometrically plausible firearms from Stable Diffusion 1.5 using a canny edge conditioned ControlNet. }
\end{figure}

\begin{figure}[ht!]
  \centering
  \subfloat[By Lance Asper on Unsplash \cite{noauthor_unsplash_nodate}.]{\includegraphics[height=4cm]{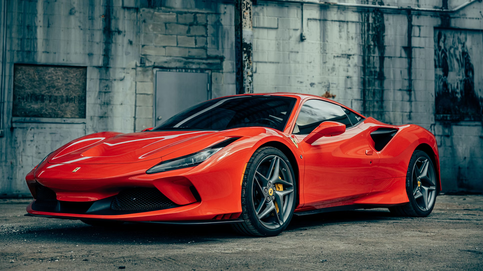}}
\quad
  \subfloat[Canny Edge map.]{\includegraphics[height=4cm]{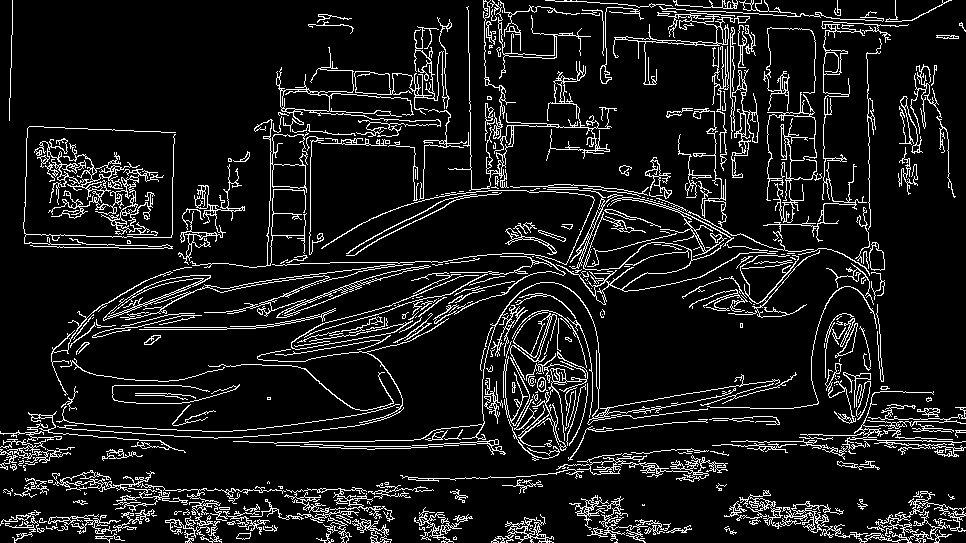}}
\hfill
  \subfloat[Generated from ControlNet for Stable Diffusion 1.5.]
{\includegraphics[height=4cm]{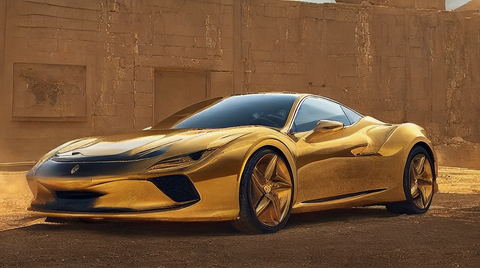}}
\quad
  \subfloat[Generated from ControlNet for Stable Diffusion XL]
{\includegraphics[height=4cm]{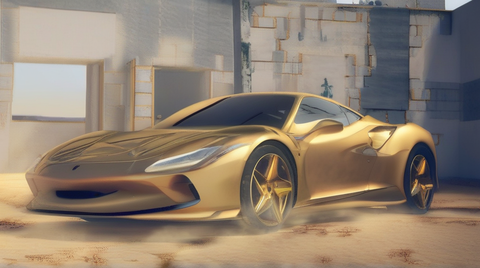}}
\label{images:artefacts_correction:controlnet:car}
\caption{Images of a car generated using Stable Diffusion models with a Canny Edge ControlNet and an original photograph.}
\end{figure}

\FloatBarrier

\subsection{From drawings to images}

\begin{figure}[ht!]
  \centering
    \subfloat[]{\includegraphics[height=2.4cm]{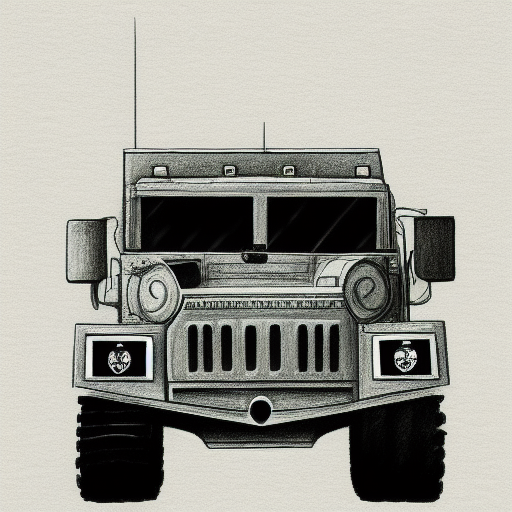}}
    \quad
    \subfloat[]{\includegraphics[height=2.4cm]{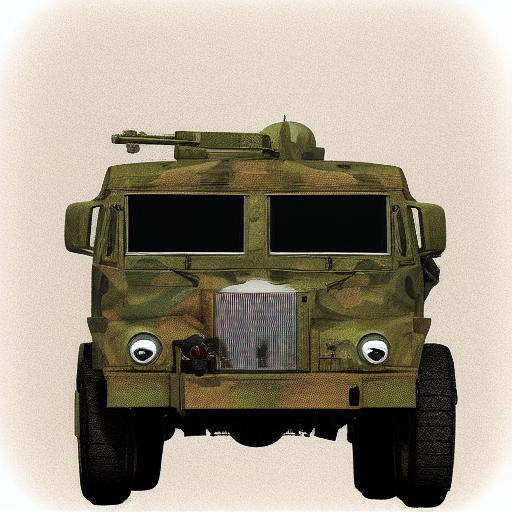}}
    \quad
    \subfloat[]{\includegraphics[height=2.4cm]{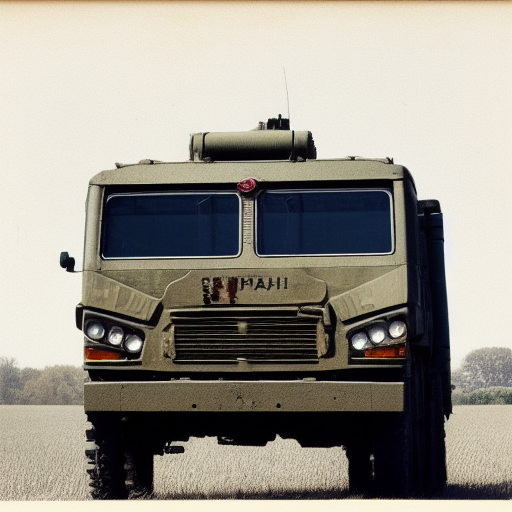}}
    \quad
    \subfloat[]{\includegraphics[height=2.4cm]{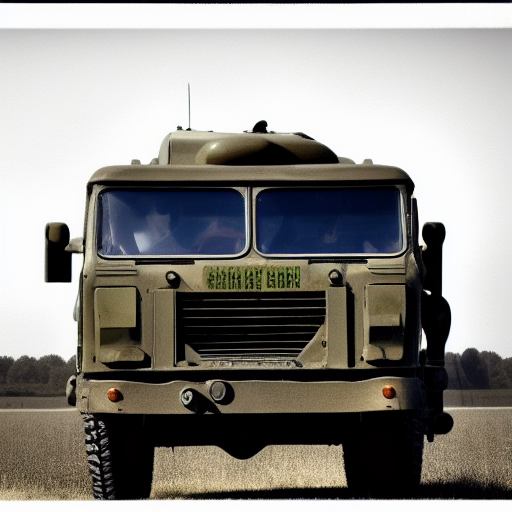}}
    \quad
    \subfloat[]{\includegraphics[height=2.4cm]{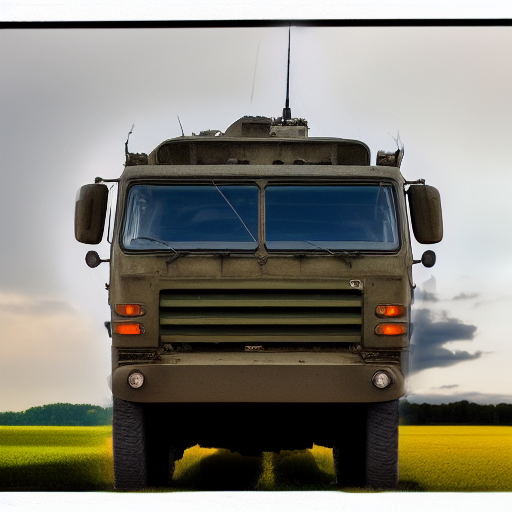}}
    \quad
    \subfloat[]{\includegraphics[height=2.4cm]{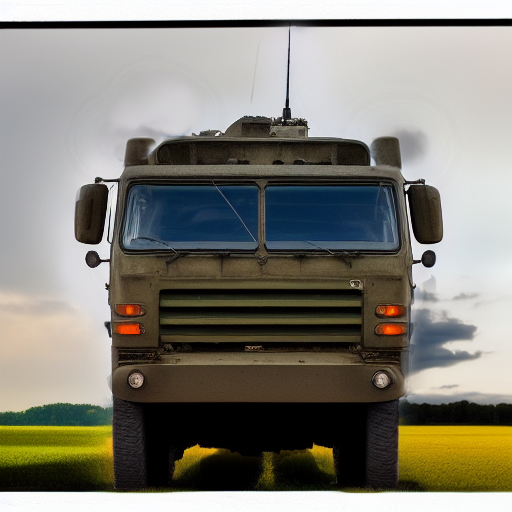}}
\caption{Approach to generate geometrically consistent synthetic images of objects by first generating an image as a simple drawing, then using a image-to-image model for step-wise adjustments until photorealism is achieved. Shown samples were generated with Stable Diffusion 1.5 and the inpainting feature of the Automatic1111 WebUI.}
\label{fig:drawing_first:truck}
\end{figure}

Stable Diffusion tends to produce images with fewer line and symmetry artifacts when tasked with generating straightforward, minimalist drawings. For instance, symmetrical and photorealistic vehicle images have been successfully generated by initially creating them as drawings and subsequently adapting them into photographs through varied text prompts by conditioning on the initial result as demonstrated in Figure \ref{fig:drawing_first:truck}. This method, however, has not been thoroughly investigated.

\FloatBarrier
\subsection{Workflow to minimize Artifacts}

\begin{figure}[ht!]
\centering
\subfloat[Generated Image.]{\includegraphics[height=3.5cm]{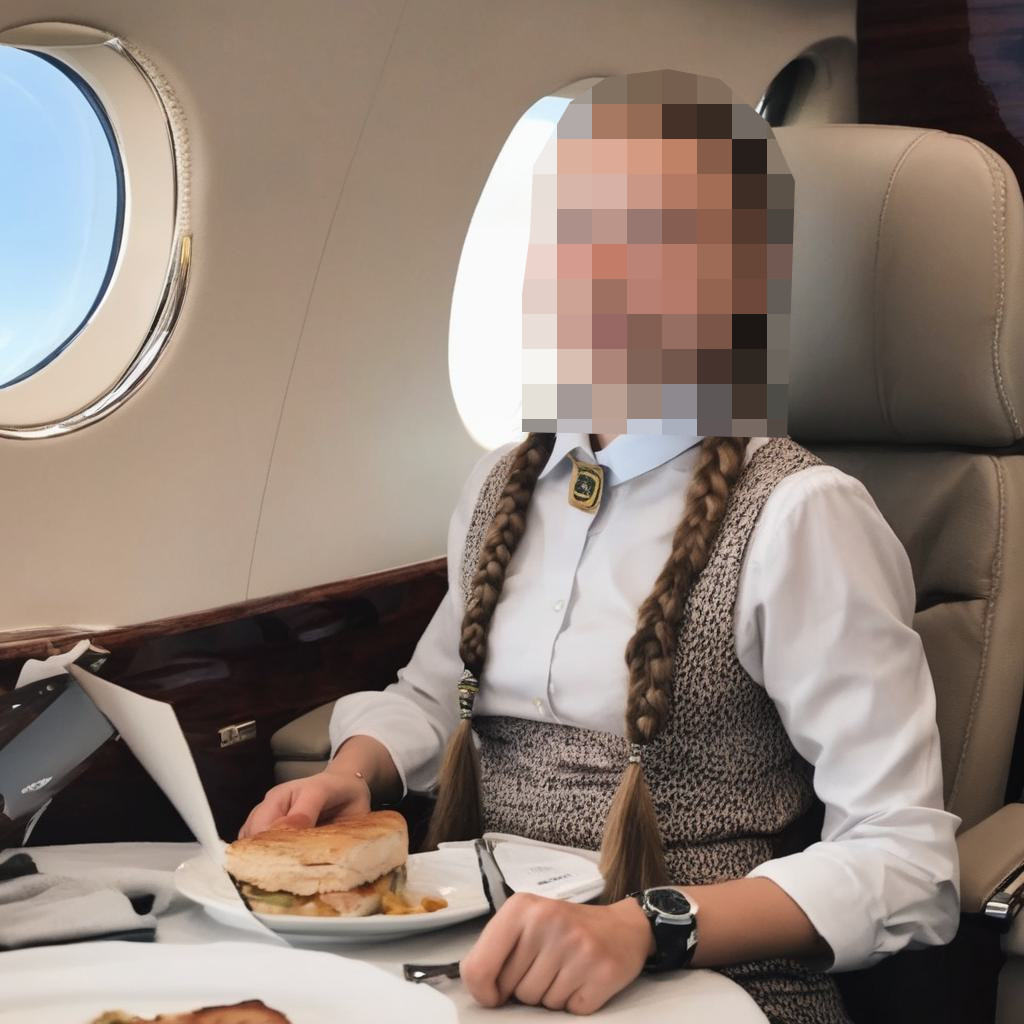}}
\quad
\subfloat[Correct shirt front with inpainting]{\includegraphics[height=3.5cm]{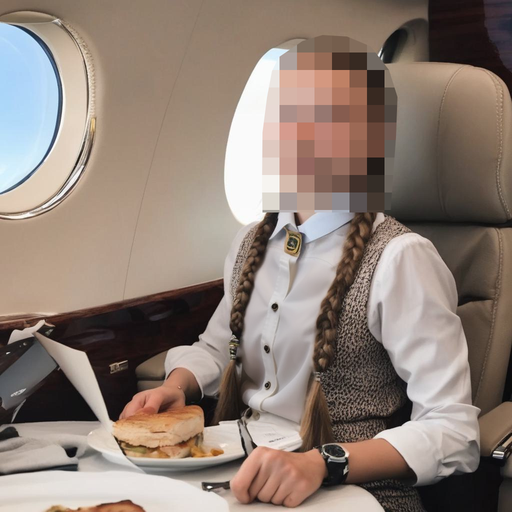}}
\quad
\subfloat[Correct left arm and shoulder with inpainting.]{\includegraphics[height=3.5cm]
{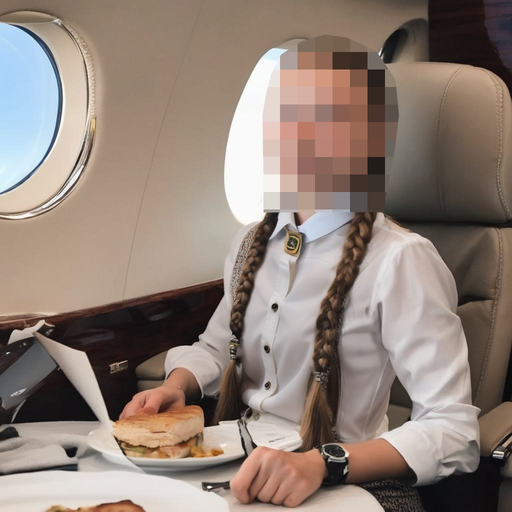}}
\quad
\subfloat[Correct right shoulder with inpainting.]{\includegraphics[height=3.5cm]{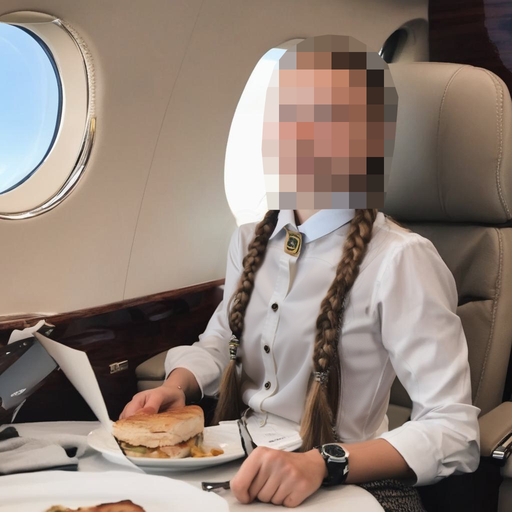}}
\quad
\subfloat[Add a different food with inpainting.]{\includegraphics[height=3.5cm]{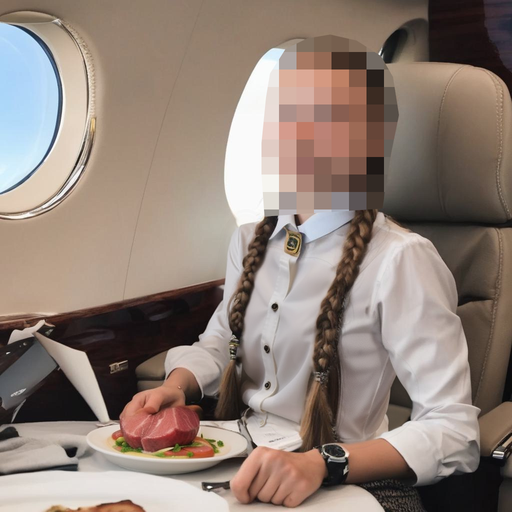}}
\quad
\subfloat[Color the brown wall paneel manually.]{\includegraphics[height=3.5cm]{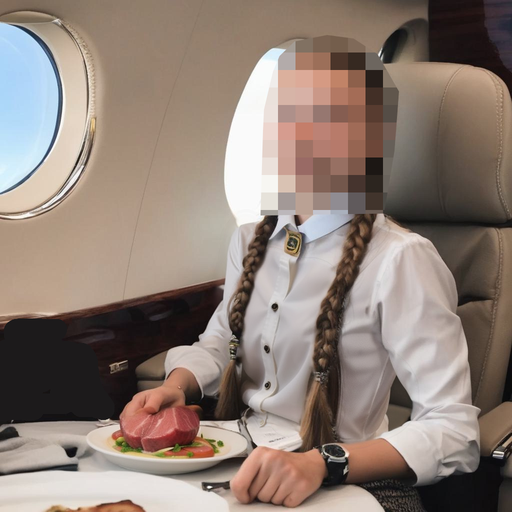}}
\quad
\subfloat[Smooth out the manual paintwork with very light inpainting.]{\includegraphics[height=3.5cm]{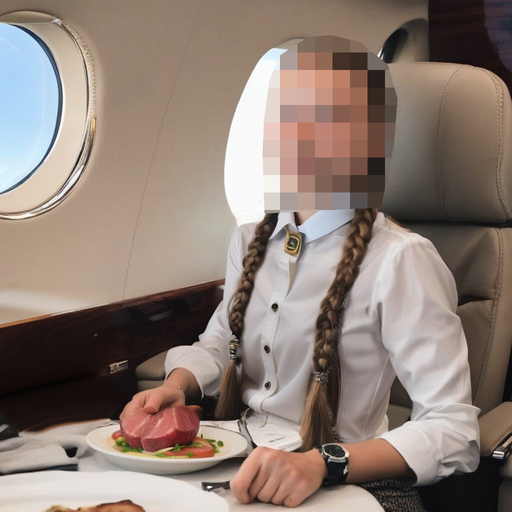}}
\caption{Sequence of synthetic images with each showing a different step in artifact removal (from left to right). However, some artifacts, such as the number of fingers on the left hand remain.}
\label{fig:images:artefact_correction:workflow_env}
\end{figure}

\begin{figure}[ht]
\centering
\subfloat[Generated Image.]{\includegraphics[height=6.2cm]{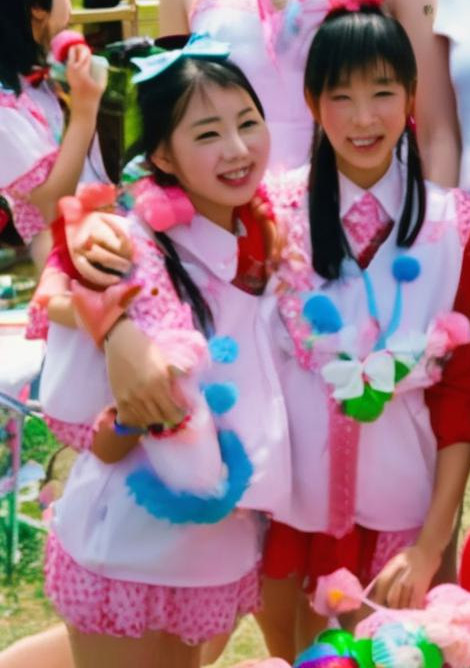}}
\quad
\subfloat[Correct the girls with very heavy inpainting with a different model checkpoint.]{\includegraphics[height=6.2cm]{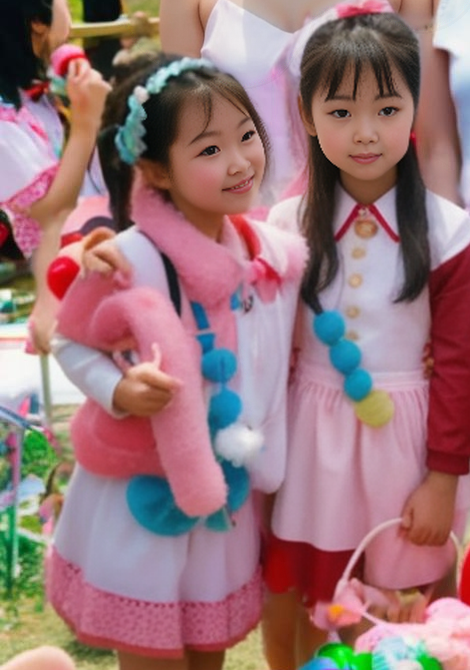}}
\quad
\subfloat[Crop out persons or object that are not needed and probably difficult to correct with inpainting.]{\includegraphics[height=6.2cm]{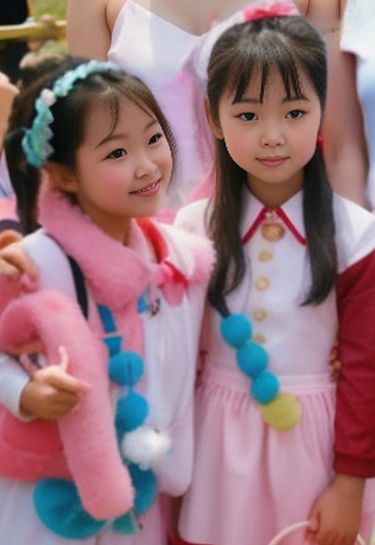}}
\caption{Sequence of synthetic images with each showing a different step in artifact removal (from left to right).}
\label{fig:images:artefact_correction:workflow_vip_1}
\end{figure}

Artifacts are quite common in synthetic images. The following examples demonstrate how various post-processing steps can be used to minimize artifact occurrence. Figures \ref{fig:images:artefact_correction:workflow_env} and \ref{fig:images:artefact_correction:workflow_vip_1} each demonstrate the intermediate results of post-processing a synthetic image for artifact removal. At times, simply cropping the image can be a feasible approach to exclude regions affected by artifacts that are not integral to what the image intends to convey (applied in Figure \ref{fig:images:artefact_correction:workflow_vip_1}).

\clearpage

\section{Practical Considerations}

When considering the practical aspects of employing generative models from a threat actor's perspective, especially in the context of cyber influence operations, several factors come into play. These factors influence the decision-making process regarding the utilization of deep learning-generated image content.

Accessibility of the methods is a crucial consideration. For the Stable Diffusion models that we assessed, there exists a wide array of open-source resources, including libraries, pre-trained models, training scripts, tutorials, and applications such as the Automatic1111 Stable Diffusion WebUI. These resources are readily available to anyone without the need for registration or payment. Furthermore, there is a vibrant community surrounding these methods, where members actively assist each other, share tips, and collaborate on training, improvement, and model sharing. However, it is important to note that many of these resources may not work flawlessly right out of the box, often requiring some level of programming knowledge and a basic grasp of the underlying methods to address common bugs and errors.

Utilizing a pre-trained model without any additional training demands substantial computing power. Even the lightest pipelines we employed necessitated a GPU with at least 8 GB VRAM, while most of our models or model pipelines were run on graphics cards with 24 GB VRAM. Consequently, investing in suitable hardware can be costly, although it remains within reach for the average citizen of a moderately affluent country. Furthermore, procurement of the required hardware is well within reach for nation state actors, state-sponsored actors, and well-resourced criminal organizations.

It is important to acknowledge that Stable Diffusion XL continues to face challenges in generating physically realistic, high-quality images. Creating images that convincingly maintain their authenticity even under scrutiny seems unattainable at the moment.

One of the significant strengths of Stable Diffusion and most other generative models, as previously highlighted in Section \ref{illustrations_and_cartoons}, lies in their remarkable ability to generate illustrations, drawings, and cartoons without requiring specialized artistic or drawing skills. Moreover, they perform this task much faster than a human artist or graphic designer could manually create such images.

However, it is important to recognize that generating these images still demands time and human intervention. Automating the image generation process is feasible to only a limited extent, if at all. It involves manual effort, countless iterations, numerous trials and errors, and human judgment to determine which text prompts yield the closest approximation to the desired output. Fully exploring the entire parameter space is nearly impossible, and the cause-and-effect relationships between configuration settings and outcomes are unpredictable.

The time needed to generate a single image can vary, but based on our experience, it can take up to half an hour. This duration includes post-processing and waiting time during the lengthy denoising process.

\FloatBarrier

\section{Implications for Threat Landscape and Cyber Defense}

The consolidation of the previous findings on synthetic image generation provides a good starting point to speculate about the future evolution of the threat landscape in cyber influence operations and potential countermeasures useful for cyber defense units.

In our \textit{Synthetic Force Multiplier} (Section \ref{sec:scenario1_synthetic_force_multiplier}) and \textit{Targeted Military Disinformation}  (Section \ref{sec:scenario2_tailored_military_disinfo}) threat scenarios, the lack of photorealism in generated images was a significant challenge. Artifacts were noticeable upon close inspection, indicating synthetic generation. Creating convincing samples required extensive training, specific scenario research, and trial-and-error, with no guarantee of success. Synthetic image generation demands substantial resources (e.g. dedicated personnel, hardware, and training) from threat actors, often yielding less-than-ideal results. Consequently, threat actors might prefer purchasing these capabilities rather than developing them. Therefore, cyber defense units should monitor the landscape of private entities offering \enquote{influence-for-hire} services involving synthetic media generation. Knowing the capabilities of private actors in this space may be important for the assessment and attribution of future threats.

As outlined in the previous section, the semi-automated creation process may take up to half an hour followed by a time-consuming selection process to choose the most suitable set of images for a given scenario. The automation of these steps, in particular the assessment of the images' overall quality and suitability, seems unattainable in the near future. They necessitate an understanding of the real world, as well as a degree of empathy unavailable to current AI models. However, Large Language Models (LLMs) have recently shown basic ability of cognitive empathy (e.g. recognition of emotions in text) \cite{sorin_large_2023}. Thus, advances in computational empathy should be closely monitored by cyber defense units since threat actors could employ these methods in a dual-use manner to further automate the creation and selection process of synthetic media. A threat actor may also try to reduce the number of manual interventions by choosing quantity over quality. In cases of leakage of sensitive information, a threat actor may choose to flood information channels with synthetic content in an attempt to bury the leaked information. Furthermore, large quantities of synthetic images with relatively low fidelity may be sufficient to increase uncertainty in the target audience about the veracity of online information in general. Hence, cyber defense units should actively explore and prepare guidelines and countermeasures for cyber influence operations involving large quantities of synthetic media content. 

Synthetic image generation may play a more important role for threat actors which do not rely on photo-realistic content to conduct their cyber influence operations. The implemented threat scenarios \textit{Environmentalists} and \textit{Undermining democracy} (Sections \ref{sec:scenario4_environmentalists} and \ref{sec:scenario7_undermine_democracy}) showed that stylized or deliberately exaggerated image content, as well as drawings can be generated with convincing quality. In particular, threat actors may use synthetic images for propaganda or the promotion of conspiracy theories. In general, synthetic image generation may serve as an amplifier for content creation.

Synthetic images exhibit distinct artifacts, some of which have been presented in this report. Intelligence analysts in cyber defense units should be educated about the features of synthetic images and particularly about the presence of such artifacts. This will enable them to more confidently assess the authenticity of image content. Threat actors may over time develop a preference for particular generative models. Experienced intelligence analysts could use information on the supposed generative model behind a synthetic image for soft attribution of cyber influence operations. Therefore, continuous technology monitoring should be implemented to keep track of the newest generative models and to catalogue their respective artifacts and weaknesses.

Depending on the hyperparameters and chosen models, a generative pipeline can yield wildly different results featuring different artifacts (as e.g. shown for Refiner-Models in Section \ref{artefact_correction:refiner}). We hypothesize that this renders the performance of methods to detect synthetic images inherently unstable and necessitates their frequent re-training. Until proven otherwise, automatic detection methods should be used with caution. Threat actors may exploit suboptimal detection performances of publicly-available detectors to generate uncertainty about the authenticity of image content. In fact, real evidence of the war between Israel and Hamas has recently been falsely deemed AI-generated with cherry-picked false positive results of a publicly-available synthetic image detection method \cite{maiberg_ai_2023}, leading major media outlets to doubt its credibility. The repurposing of synthetic image detectors by threat actors alongside the release of synthetic imagery to muddy the line between real and synthetic appears to be a new variation of the \textit{liar's dividend} \cite{chesney_deep_2019}. It will make it easier for threat actors to discredit authentic image content online as AI-generated and deny responsibility for factual events (e.g. atrocities in a war).

Finally, it is important to note that synthetic image generation is just one more tool for threat actors to enhance their cyber influence operations. Established media manipulation techniques, e.g. the use of recontextualized media, will stay relevant\footnote{For definitions of established techniques, cf. the media manipulation casebook: \url{https://mediamanipulation.org/definitions}.}. 
Cyber defense units should not be tempted to overemphasize the threat caused by synthetic image generation methods while neglecting the significance of already established techniques.

\FloatBarrier

\section{Conclusion \& Recommendations}

The potential threat of synthetic image generation methods to augment cyber influence operations is significant but varies across particular scenarios. As we have explored in this report, the accessibility and availability of these methods for threat actors, along with the required computational resources, play a crucial role in determining their feasibility. Additionally, the quality and authenticity of generated image content remain challenging aspects for threat actors to address, especially when aiming to deceive or manipulate target audiences effectively.

While these methods excel in generating illustrations, drawings, and cartoons quickly and without the need for artistic skills, their application for creating photo-realistic, convincing images faces limitations. The inherent complexity of creating images that can withstand scrutiny remains a formidable challenge.

Furthermore, the human factor cannot be discounted. The process of generating image content, selecting text prompts, and refining outputs still heavily relies on human intervention, judgment, and iterative refinement. The automation of this process is only achievable to a limited extent, given the unpredictable nature of cause-and-effect relationships within the parameter space.

We identify two primary categories of threat scenarios likely to utilize synthetic image generation in the near future: 1) scenarios employing synthetic images for the amplification of propaganda or conspiracy, where photo-realism is not essential, and 2) scenarios using numerous low-fidelity synthetic images to distract or sow doubt about the overall veracity of online information. Cyber defense units should prioritize monitoring developments in generative models and private entities in the \textit{influence-for-hire} industry offering synthetic image generation services, develop guidelines and countermeasures against cyber influence operations with synthetic images, and continuously train intelligence analysts in forensic analysis of synthetic content.

To further this field responsibly and safeguard against the misuse of synthetic image generation methods as a society, we recommend the following:

\begin{itemize}
    \item  Continued Research: There is a clear need for ongoing research into the capabilities and limitations of synthetic image generation models, focusing on the study of maximum achievable realism of outputs while understanding the mechanics behind artifact generation. This is vital to stay ahead of threat actors and develop the knowledge base necessary for any effective mitigation strategy.
    \item Defense Mechanisms: Develop sophisticated detection tools that leverage machine learning to identify generated content, and make them available in a responsible way that mitigates their potential abuse by threat actors. In particular, novel and creative approaches that combine automatic detection tools with human supervision are needed. Educate the public about the hallmarks of synthetic media. Initiate awareness campaigns to inform the public about the existence and capabilities of such technologies, including a transparent reporting and education on how synthetic media was employed by threat actors for propagandistic purposes and cyber influence operations in the past. This serves the goal of increasing societal resilience against influence-related activities involving synthetic media.
    \item Collaboration: Encourage collaboration between technologists, policymakers, and media experts to stay ahead of malicious applications and formulate coherent responses. Consider the consequences of increased penetration of information space with synthetic media on public discourse, authenticity verification, and perceived credibility of digital content.
    \item Policy and Regulation: Consider the development of policies and regulations that control the distribution of synthetic image generation technology to prevent its use in harmful applications. Collaboration among nation states and technology providers will be necessary to prevent misuse of communication channels and platforms. To be effective, regulations must address that boundaries between private and state action in cyber influence operations are often blurred.

\end{itemize}

By acknowledging the power and risks associated by synthetic image generation methods and taking proactive steps to address them, we can mitigate its threats to information integrity.


\printbibliography[heading=bibintoc] %

\end{document}